\documentclass[pdftex,bookmarks,hidelinks]{article}
\pdfoutput=1
\usepackage{jheppub}
\usepackage{graphicx,epstopdf,amsmath,amsfonts,amssymb,appendix,comment}
\usepackage{color,slashed,subfigure,braket,multirow}
\usepackage{epsfig,mathrsfs,latexsym,color,url,etoolbox}
\usepackage{bbm}
\usepackage{ulem}
\usepackage{mathtools}
\usepackage{xspace}
\usepackage{footnote}
\usepackage[usenames,dvipsnames,table]{xcolor}

\definecolor{nicered}{rgb}{0.7,0.1,0.1}
\definecolor{nicegreen}{rgb}{0.1,0.5,0.1}
\DeclareMathAlphabet{\mathbbold}{U}{bbold}{m}{n}

\newcommand{\beq}{\begin{equation}}
\newcommand{\eeq}{\end{equation}}
\newcommand\bout{\bgroup\markoverwith{\textcolor{blue}{\rule[0.5ex]{4pt}{0.8pt}}}\ULon}

\newcommand{\nusi}{$\nu$SI\xspace}
\usepackage{hyperref}
\hypersetup{colorlinks,citecolor= nicegreen,linkcolor= nicered}

\newcommand\snowmass{\begin{center}\rule[-0.2in]{\hsize}{0.01in}\\\rule{\hsize}{0.01in}\\
\vskip 0.1in Submitted to the  Proceedings of the US Community Study\\ 
on the Future of Particle Physics (Snowmass 2021)\\ 
\rule{\hsize}{0.01in}\\\rule[+0.2in]{\hsize}{0.01in} \end{center}}

\begin{document}

\pagestyle{titlepage}

\preprint{CERN-TH-2022-024, DESY-22-035, FERMILAB-PUB-22-099-T}
\date{\today}

\title{{\huge Neutrino Self-Interactions: A White Paper}}

\collaboration{Editors: Nikita Blinov, Mauricio Bustamante, Kevin J. Kelly, Yue Zhang\vspace{0.25cm}}

\def\UCBerkeley{Department of Physics, University of California, Berkeley, CA 94720, USA}
\def\INT{Institute for Nuclear Theory, University of Washington, Seattle, WA 98195, USA}
\def\Victoria{Department of Physics and Astronomy, University of Victoria, Victoria, BC V8P 5C2, Canada}
\def\Ferrara{Dipartimento di Fisica e Scienze della Terra, Universit\'a degli Studi di Ferrara, via Giuseppe Saragat 1, 44122 Ferrara, Italy}
\def\INFN{Istituto Nazionale di Fisica Nucleare (INFN), Sezione di Ferrara, Via Giuseppe Saragat 1, 44122 Ferrara, Italy}
\def\NBIA{Niels Bohr International Academy, Niels Bohr Institute, University of Copenhagen, Copenhagen, Denmark}
\def\DARK{DARK, Niels Bohr Institute, University of Copenhagen, Copenhagen, Denmark}
\def\SLAC{SLAC National Accelerator Laboratory, 2575 Sand Hill Road, Menlo Park, CA 94025, USA}
\def\BNL{High Energy Theory Group, Physics Department, Brookhaven National Laboratory, Upton, NY 11973, USA}
\def\WUSTL{Department of Physics and McDonnell Center for the Space Sciences, Washington University, St. Louis, MO 63130, USA}
\def\ND{Department of Physics, University of Notre Dame, South Bend, IN 46556, USA}
\def\HCRI{Regional Centre for Accelerator-based Particle Physics, Harish-Chandra Research Institute, A CI of Homi Bhabha National Institute, Chhatnag Road, Jhunsi, Prayagraj (Allahabad) 211019, India}
\def\NCSU{Department of Physics, North Carolina State University, Raleigh, NC 27695, USA.}
\def\BU{Department of Physics \& Astronomy, Bowdoin College, Brunswick ME, USA}
\def\MU{Department of Physics, Moravian University, Bethlehem PA, USA}
\def\FNAL{Fermi National Accelerator Laboratory, Batavia IL, USA}
\def\CERN{CERN, Esplande des Particules, 1211 Geneva 23, Switzerland}
\def\DESY{Deutsches Elektronen-Synchrotron DESY, Notkestr. 85, 22607 Hamburg, Germany}
\def\TAMU{Mitchell Institute for Fundamental Physics and Astronomy, Department of Physics and Astronomy, Texas A\&M University, College Station, TX 77843, USA}
\def\UMN{School of Physics and Astronomy, University of Minnesota, Minneapolis, MN 55455, U.S.A.}
\def\MPIK{Max-Planck-Institut für Kernphysik, Saupfercheckweg 1, 69117 Heidelberg, Germany}
\def\LUC{Loyola University Chicago, Chicago, IL, USA}
\def\Carleton{Department of Physics, Carleton University, Ottawa, ON, Canada}
\def\PITT{PITT PACC, Department of Physics and Astronomy, University of Pittsburgh, 3941 O’Hara St., Pittsburgh, PA 15260, USA}
\def\NU{Department of Physics and Astronomy, Northwestern University, 2145 Sheridan Road, Evanston, IL 60208, USA}
\def\VT{Center for Neutrino Physics, Department of Physics,
Virginia Tech University, Blacksburg, VA 24601, USA}
\def\CCAPP{Center for Cosmology and AstroParticle Physics
  (CCAPP), Ohio State University, Columbus, Ohio 43210, USA}
\def\OSU{Department of Physics, Ohio State University, Columbus, Ohio 43210, USA}
\def\IIT{Department of Physics, Indian Institute of Technology Indore, India}
\def\UNM{Department of Physics and Astronomy, University of New Mexico, Albuquerque, NM 87106, USA}
\def\PI{Perimeter Institute for Theoretical Physics, Waterloo, ON N2J 2W9, Canada}
\def\SSM {Scuola Superiore Meridionale, Universit\`a degli studi di Napoli ``Federico II'', Largo San Marcellino 10, \mbox{80138 Napoli, Italy}}
\def\INFNNap {INFN, Sezione di Napoli, Complesso Universitario, Monte S. Angelo, I-80126 Napoli, Italy}
\def\UNINA{Dipartimento di Fisica ”Ettore Pancini”, Universita` degli studi di Napoli “Federico II”, Complesso Univ. Monte S. Angelo, I-80126 Napoli, Italy}
\def\Aarhus{Department of Physics and Astronomy, Aarhus University, DK-8000 Aarhus C, Denmark}
\def\SEU{School of Physics, Southeast University, Nanjing 211189, China}
\def\UW{Department of Physics, University of Washington, Seattle, WA 98195, USA}

\emailAdd{nblinov@uvic.ca}
\emailAdd{mbustamante@nbi.ku.dk}
\emailAdd{kjkelly@cern.ch}
\emailAdd{yzhang@physics.carleton.ca}

\author[1, 2]{Jeffrey M.~Berryman,}
\affiliation[1]{\INT}
\affiliation[2]{\UCBerkeley}
\author[3]{Nikita Blinov,}
\affiliation[3]{\Victoria}
\author[4, 5]{Vedran Brdar,}
\affiliation[4]{\FNAL}
\affiliation[5]{\NU}
\author[6, 7]{Thejs Brinckmann,}
\affiliation[6]{\INFN}
\affiliation[7]{\Ferrara}
\author[8]{Mauricio Bustamante,}
\affiliation[8]{\NBIA}
\author[9]{Francis-Yan Cyr-Racine,}
\affiliation[9]{\UNM}
\author[10]{Anirban Das,}
\affiliation[10]{\SLAC}
\author[5]{Andr\'e de Gouv\^ea,}
\author[11]{Peter B.~Denton,}
\affiliation[11]{\BNL}
\author[12]{P. S. Bhupal Dev,}
\affiliation[12]{\WUSTL}
\author[13]{Bhaskar Dutta,}
\affiliation[13]{\TAMU}
\author[14, 15]{Ivan Esteban,}
\affiliation[14]{\CCAPP}
\affiliation[15]{\OSU}
\author[8, 16, 17]{Damiano Fiorillo,}
\affiliation[16]{\UNINA}
\affiliation[17]{\INFNNap}
\author[7]{Martina Gerbino,}
\author[18]{Subhajit Ghosh,}
\affiliation[18]{\ND}
\author[19]{Tathagata Ghosh,}
\affiliation[19]{\HCRI}
\author[20]{Evan Grohs,}
\affiliation[20]{\NCSU}
\author[21]{Tao Han,}
\affiliation[21]{\PITT}
\author[22]{Steen Hannestad,}
\affiliation[22]{\Aarhus}
\author[23, 24]{Matheus Hostert,}
\affiliation[23]{\PI}
\affiliation[24]{\UMN}
\author[25]{Patrick Huber,}
\affiliation[25]{\VT}
\author[26, 27]{Jeffrey Hyde,}
\affiliation[26]{\MU}
\affiliation[27]{\BU}
\author[4, 28]{Kevin J. Kelly,}
\affiliation[28]{\CERN}
\author[29]{Felix Kling,}
\affiliation[29]{\DESY}
\author[24]{Zhen Liu,}
\author[7]{Massimiliano Lattanzi,}
\author[30]{Marilena Loverde,}
\affiliation[30]{\UW}
\author[31]{Sujata Pandey,}
\affiliation[31]{\IIT}
\author[17, 32]{Ninetta Saviano,}
\affiliation[32]{\SSM}
\author[33]{Manibrata Sen,}
\affiliation[33]{\MPIK}
\author[25]{Ian M. Shoemaker,}
\author[34]{Walter Tangarife,}
\affiliation[34]{\LUC}
\author[35]{Yongchao Zhang,}
\affiliation[35]{\SEU}
\author[36]{Yue Zhang}
\affiliation[36]{\Carleton}

\abstract{Neutrinos are the Standard Model (SM) particles which we understand the least, often due to how weakly they interact with the other SM particles. Beyond this, very little is known about interactions among the neutrinos, i.e., their self-interactions. The SM predicts neutrino self-interactions at a level beyond any current experimental capabilities, leaving open the possibility for beyond-the-SM interactions across many energy scales. In this white paper, we review the current knowledge of neutrino self-interactions from a vast array of probes, from cosmology, to astrophysics, to the laboratory. We also discuss theoretical motivations for such self-interactions, including neutrino masses and possible connections to dark matter. Looking forward, we discuss the capabilities of searches in the next generation and beyond, highlighting the possibility of future discovery of this beyond-the-SM physics.

\begin{figure}[hb]
    \centering
    \includegraphics[width=0.9\linewidth]{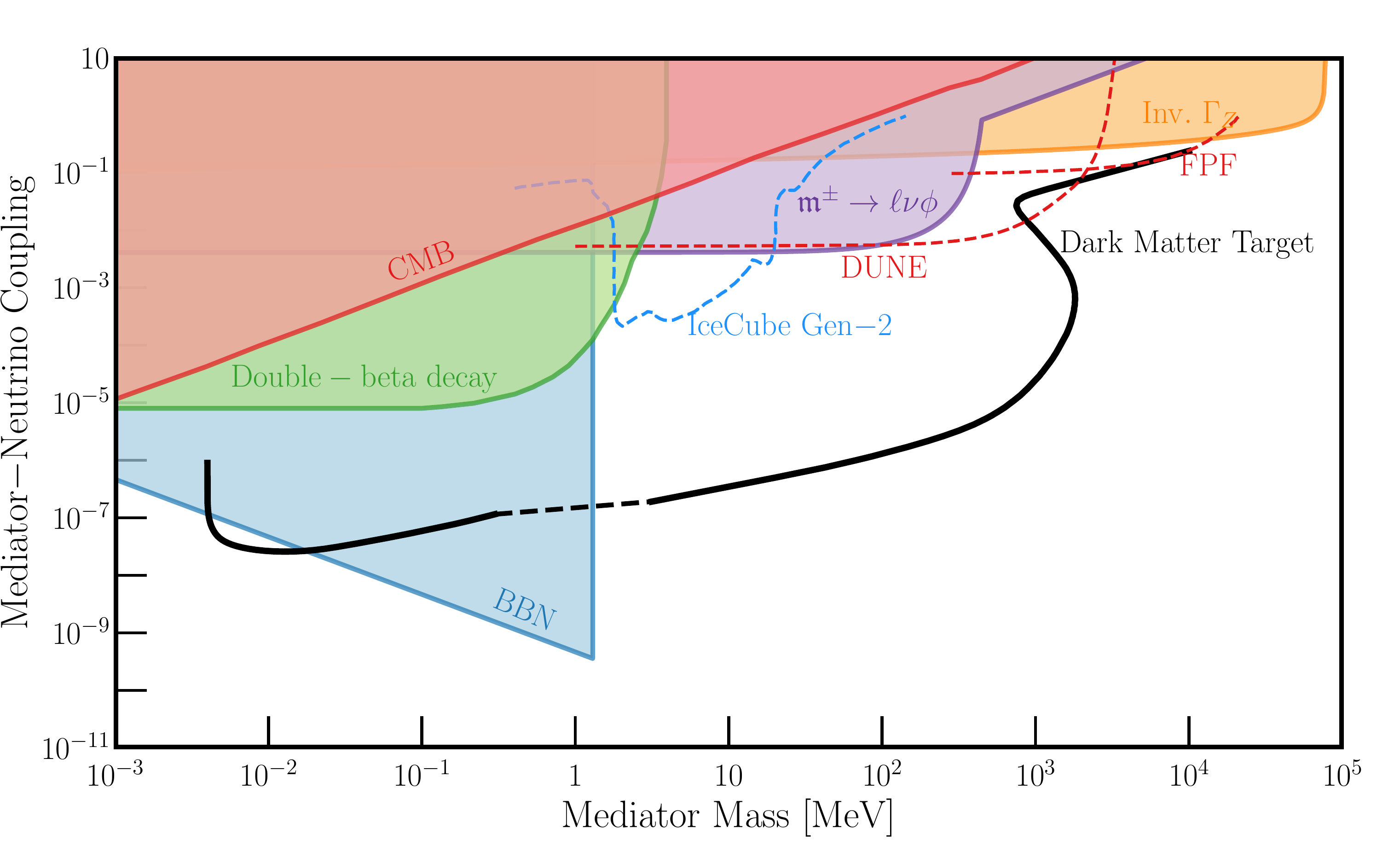}
    \caption{A schematic summary of the various searches for neutrino self-interactions discussed in this white paper. The self-interaction is generated by a new mediator particle, whose mass (coupling to neutrinos) is shown on the horizontal (vertical) axis. 
    The relative importance of laboratory, astrophysical, and cosmological observations depends on the flavor structure of the mediator-neutrino coupling (assumed here to be universal). Cosmological constraints arise from Big Bang Nucleosynthesis (BBN) and Cosmic Microwave Background (CMB). Laboratory bounds come from searches for neutrinoless double beta decay; rare meson, $\tau$ and $Z$ decays; collider searches for new neutrino scattering (DUNE, FPF) and missing energy channels. The presence of the new self-interactions can also modify the observed spectrum of high-energy neutrinos at IceCube.
    The neutrino self-interaction can also play a key role in producing the dark matter of the Universe via a freeze-in mechanism, leading to a theoretically motivated target in the mediator mass-coupling parameter space. This target for a representative model is shown by the solid black line.}
    \label{fig:summary}
\end{figure}

\snowmass
}

\maketitle

\setcounter{tocdepth}{2}
\flushbottom

\section{Introduction \& Executive Summary}\label{sec:intro}
Neutrinos are a fundamental ingredient of the Standard Model (SM) of particle physics and cosmology. Due to their neutrality and weak interactions, however, their properties are the least understood out of the SM particles. In particular, neutrinos can serve as a portal to beyond-Standard Model physics, imbuing them with new interactions. Remarkably, these interactions can be significantly larger than those provided by the electroweak (EW) gauge bosons. This observation suggests a multitude of new phenomena that can be enabled by such interactions. 
While non-standard interactions with charged SM fermions are in some cases straightforwardly tested with traditional $\nu$ scattering and oscillation experiments, it is remarkable that neutrino \textit{self}-interactions, \nusi, can also be tested by a variety of methods. In this review we focus on the latter class of non-standard neutrino interactions, and discuss both the theoretical motivation and the physical systems in which they can be observed. Our goals are (i) to define a set of benchmark models, (ii) to comprehensively review the relevant physical phenomena within these scenarios, and (iii) to highlight upcoming short- and long-term prospects for discovering \nusi. 

We begin our discussion in Sec.~\ref{sec:theory} by outlining the main contexts in which neutrino self-interactions arise. These include the important problems of neutrino mass generation, gauge extensions of the SM, and the production of dark matter in the early Universe. Below the scale of EW symmetry breaking, \nusi can usually be described by a schematic interaction like $\nu\nu\phi$ where $\phi$ is a scalar or vector mediator particle. However, since $\nu$ are part of an EW doublet, an ultraviolet (UV) completion is usually required to embed such an interaction consistently within the SM. We therefore also consider different UV-complete scenarios. We emphasize that, unlike for non-standard interactions in the charged fermion sector, in many cases the mediator $\phi$ cannot be integrated out. This becomes particularly important when we explore the complementarity of experimental and observational probes of \nusi that are sensitive to a wide range of energy scales. 

Sec.~\ref{sec:cosmo} then focuses on the imprints of neutrino-self interactions on cosmological observables, including light element abundances, the Cosmic Microwave Background (CMB), and the matter distribution in the Universe. These observables probe self-interactions at comparatively low scales of ${\sim}\mathrm{eV}$ to ${\sim}\mathrm{MeV}$. Supernovae and other astrophysical sources of neutrinos discussed in Sec.~\ref{sec:astro} can test characteristic self-interaction scales up to $\mathcal{O}(100\;\mathrm{MeV})$.  Finally we consider laboratory experiments in Sec.~\ref{sec:lab}, which can access the broadest range of self-interaction scales all the way up to $\mathcal{O}(100\;\mathrm{GeV})$. We demonstrate how several future experiments will further probe \nusi, testing important theoretically motivated targets. 

Remarkably, all of the aforementioned probes (cosmological, astrophysical, and laboratory) provide important constraints across a broad range of parameter space -- Fig.~\ref{fig:summary} presents a selection of the probes discussed in this white paper across this broad range. In order for us to thoroughly test the motivated scenarios (including those connected to neutrino masses, dark matter production, etc.), it is imperative that the community considers all of these domains simultaneously. Only then can we maximize the potential for discovering new physics via neutrino self-interactions.

\section{Theoretical Motivations}\label{sec:theory}
Self-interactions among active neutrinos, mediated by a new scalar or vector boson, occur in many theories beyond the SM. Given the elusive nature of neutrinos, their SM self-interaction has never been directly measured in the laboratory. The $Z$-boson invisible width measured at the LEP experiment~\cite{ALEPH:2005ab} is only an indirect measurement and does not exclude the presence of additional mediators. 

Motivations for introducing new neutrino self interactions include explaining the origin of neutrino mass, connections to dark matter and addressing its relic abundance, as well as other puzzles (e.g., the Hubble tension, muon $g-2$). If the mediator couples universally to a left-handed lepton doublet, \nusi are often accompanied by non-standard neutrino interactions with charged SM fermions, leading to other complementary probes. It is possible to make the mediator neutrinophilic given that the EW symmetry is broken. A famous example of this type is the Majoron.

The vast majority of phenomenological studies considered in this work can be completely described in terms of the minimal coupling of a new mediator $\phi$ to neutrinos as 
\begin{equation}\label{eq:LScalar}
    \mathcal{L} \supset g_{\alpha\beta}\phi \nu_\alpha \nu_\beta
\end{equation}
if $\phi$ is a scalar or pseudoscalar, and as 
\begin{equation}\label{eq:LVector}
    \mathcal{L} \supset g_{\alpha\beta}\phi_\mu \nu^\dagger_\alpha \bar{\sigma}^\mu \nu_\beta
\end{equation}
if $\phi$ is a vector. We consider $\alpha$ and $\beta$ to be flavor or mass eigenstate indices and have used two-component spinor notation. 
Here, $\phi$ can be a new particle in the usual sense, or a manifestation of strongly coupled physics. 

It is clear that the interactions of Eqs.~\eqref{eq:LScalar} and~\eqref{eq:LVector} are not invariant under EW symmetry transformations ($\nu$ is part of a $SU(2)_L$ doublet with a  hypercharge), so there must be additional beyond-SM matter at a higher scale. All existing astrophysical and cosmological observables, as well as many laboratory probes probe \nusi with $\sqrt{s} \lesssim \mathrm{GeV}$, such that the simplified model with only the new mediator particle is sufficient. The exceptions to this observation are high-energy collider experiments that access $\sqrt{s}$ at the EW scale and above, rendering the UV completions of Eqs.~\eqref{eq:LScalar} and~\eqref{eq:LVector} relevant. 

Additionally, unless $\phi$ carries lepton number, Eq.~\eqref{eq:LScalar} violates lepton-number conservation in scenarios where this is a global symmetry. While this distinction is important for the theoretical motivations of the following discussion, it rarely yields distinct phenomena in the searches for \nusi discussed in Sections~\ref{sec:cosmo}--\ref{sec:lab}; we will highlight these distinctions wherever relevant.

Neutrino self-interactions are also an inevitable aspect of neutrino non-standard interactions (NSI) \cite{Wolfenstein:1977ue,Farzan:2017xzy,Proceedings:2019qno} which are new interactions between neutrinos and electrons, up quarks, and/or down quarks. In addition, since the coupling to charged particles is often more tightly constrained than that to neutrinos, NSI may lead to sizable neutrino self-interactions.

\subsection{Neutrino Masses}\label{sec:theory:NuMass}
Neutrino self-interactions can be generated in theories that explain the origin of neutrino mass. In models where lepton number symmetry is spontaneously broken~\cite{Gelmini:1980re,Chikashige:1980ui,Aulakh:1982yn}, 
the corresponding (pseudo-)Goldstone boson (Majoron) $J$ can play the role of a light neutrinophilic scalar boson.
Because a Majorana neutrino mass carries both $SU(2)_L$ and $U(1)_Y$ charges, if the Majoron originates from a scalar multiplet charged under these symmetries,
it suffers from strong constraints from $Z\to \sigma J$ and $Z\to Z^* JJ \to \nu\bar\nu JJ$ decays that contribute to the $Z$-boson invisible width.
In the first decay channel, $\sigma$ stands for the ``Higgs'' boson for lepton-number violation,. Such a decay mode is kinematically allowed if the lepton-number breaking scale occurs below the EW scale.\footnote{In the model where $J$ and $\sigma$ belong to an $SU(2)_L$ triplet with hypercharge 2 whose vacuum expectation value, which breaks both lepton number and $SU(2)_L$, is constrained to be lower than a few GeV for the $\rho$ parameter to pass the EW precision tests~\cite{Joshipura:1992hp}.}
To avoid such limitations, the minimal setup is to have the Majoron from a SM gauge singlet complex scalar $\phi$ whose vacuum expectation value serves the sole purpose of breaking lepton number~\cite{Gelmini:1980re}.
A gauge-invariant neutrino mass operator is constructed using Higgs field insertions.
The corresponding effective interacting Lagrangian takes the form
\begin{equation}\label{eq:LHSquaredPhi}
\mathcal{L} = \frac{1}{\Lambda^2} (LH)^2 \phi \ ,
\end{equation}
where $\phi$ carries $-2$ units of lepton number. For clarity, we suppress the lepton flavor indices in this subsection.
Possible UV completions for this operator will be addressed in Section~\ref{sec:theory:UV}.

Below the EW scale, the above Lagrangian contributes to a Yukawa-like interaction between $\phi$ and neutrinos,
\begin{equation}
\mathcal{L} = \frac{v^2}{2\Lambda^2} \nu\nu \phi \ .
\end{equation}
Assuming $\phi$ has a potential which makes it pick up a vacuum expectation value at scale $f/\sqrt2$, this term contributes to Majorana neutrino mass
\begin{equation}
M_\nu = - \frac{v^2 f}{2\sqrt2\Lambda^2} \ .
\end{equation}
Below the $f$ scale, the $\phi$ field decomposes into $\phi = (f + \sigma + i J)/\sqrt2$.
In this very simple model, the interactions between $\sigma, J$ and, neutrino are proportional to the above neutrino mass
\begin{equation}
\mathcal{L}_{\rm int} = \frac{M_\nu}{f} \nu\nu (\sigma + iJ) \ .
\end{equation}
In perturbative theories, the mass of $\sigma$ is tied to the symmetry breaking scale $f$, whereas $J$ is massless if the Lagrangian we start with respects the lepton-number global symmetry. 
A nonzero mass of $J$ can be generated in the presence of explicit lepton-number violation.

Both $\sigma$ and $J$ can mediate self-interaction among the active neutrinos. If the physical processes involving neutrino self-interactions occur at an energy scale $E$ below $m_\sigma$ but above $m_J$, the Majoron plays the dominant role over $\sigma$. However, if $E$ lies above the lepton number breaking scale $f$, both $\sigma$ and $J$ contribute.
For $E\gg f$, symmetry restoration is expected if all mass scales are negligible at leading order. In this case, it is more effective to perform calculations using the whole complex field $\phi$~\cite{Kelly:2019wow}.

\subsection{Gauge Extensions of the Standard Model}
\label{sec:gauge_extensions}
While new gauge bosons are a key ingredient of high-scale theories like GUTs~\cite{Georgi:1974sy,Pati:1974yy,Mohapatra:1974hk,Fritzsch:1974nn,Georgi:1974my} and left-right symmetric models~\cite{Pati:1974yy, Mohapatra:1974gc, Senjanovic:1975rk}, an extended gauge sector with direct couplings to SM particles can also appear low energies. In this context, there is a special class of minimal models, corresponding to gauging of anomaly-free global currents of the SM. These scenarios are minimal because they do not require additional light matter to regulate pathological high-energy behavior in amplitudes involving these bosons.\footnote{Equivalently, this behavior can be used to set very strong 
constraints on anomalous gauge symmetries~\cite{Dror:2017ehi,Dror:2017nsg}.} 
Examples of anomaly-free SM currents include $B-L$, $L_i - L_j$, $B - 3L_i$ ($B-L$ requires the presence of an additional right-handed neutrino), where $B$ ($L$) stands for the baryon (lepton) number current with possible flavor-dependence indicated by the subscript index~\cite{Foot:1990mn, He:1990pn, He:1991qd} (see also Ref.~\cite{Allanach:2018vjg} for a comprehensive discussion of all anomaly-free $U(1)$ currents in SM and simple extensions). The key observation is that many of these currents involve couplings to the lepton doublets; the new gauge bosons therefore induce new interactions between neutrinos of the form in Eq.~\eqref{eq:LVector}.

In all of these gauge extensions, the beyond-SM interactions are not purely neutrinophilic, leading to interactions among the charged SM fermions and between them and the neutrinos. More often than not, this feature leads to the most powerful searches for these extensions emerging from experimental scenarios involving protons and electrons, rather than neutrinos. We refer the reader to Ref.~\cite{Bauer:2018onh} for a thorough discussion on searches for these types of new mediators.

\subsection{Connections to Dark Matter}\label{sec:theory:DM}
Neutrino oscillations allow for a direct connection between dark matter (DM) and neutrino physics when the DM candidate is a fermionic singlet with mass at the keV scale and a tiny mixture with neutrinos. Dodelson and Widrow (DW) showed that the entire content of DM in the Universe can be produced from the oscillation of active neutrinos into these {\it sterile} neutrinos through a  ``freeze-in'' mechanism~\cite{Dodelson:1993je}. In the absence of any other new physics, the freeze-in of the DM can be achieved for mixing angles $\sin^22\theta\sim 10^{-12}-10^{-9}$ in conjunction with sterile neutrino masses in the range $~1 - 50$ keV~\cite{Dodelson:1993je}. This mechanism is well motivated since oscillations among active and sterile neutrinos are typical of beyond-standard-model neutrino physics~\cite{Boyarsky:2009ix}. Adding motivation to this scenario, the small mixing allows for X-ray signals from the radiative decay of the sterile neutrino to $\nu_1$ plus a photon, which renders this mechanism testable.

The sterile neutrino DM scenario assumes the existence of a fourth mass eigenstate that is a linear combination of active and sterile neutrinos, $\nu_4 \equiv \nu_s \cos\theta + \nu_a \sin\theta$. The sterile neutrino population is negligible very early in the Universe. It then evolves as a function of time, for fixed neutrino energy $E\equiv x\,T$, where $T$ is the temperature of active neutrinos, governed by the Boltzmann equation~\cite{Dodelson:1993je, Abazajian:2005gj}
\begin{eqnarray}
\label{masterequation}
\frac{d f_{\nu_s}}{d z} & = & \frac{\Gamma \sin^22\theta_{\rm eff}}{4H z}  f_{\nu_a} \ ,  \qquad z\equiv {\rm MeV}/T,\\
\sin^22\theta_{\rm eff} & \simeq &  \frac{\Delta^2 \sin^22\theta}{\Delta^2 \sin^22\theta + \Gamma^2/4 + (\Delta \cos2\theta - V_T)^2} \ ,
\end{eqnarray}
where $f_{\nu_s}(x, z)$ is the sterile neutrino phase-space distribution function and $f_{\nu_a}$ is the usual thermal distribution function for the active neutrinos, $\Gamma$ is the total interaction rate among the active neutrino, and $H$ is the Hubble rate. $\Delta \equiv m_4^2/(2E)$ is the neutrino oscillation frequency in vacuum, where $m_4\gg m_{1,2,3}$, and $V_T$ is the thermal potential experienced by the active neutrino.  

This mechanism, in its simplest form proposed by Dodelson and Widrow, is however strongly constrained by non-observations of X-rays from DM-rich galaxies by X-ray telescopes~\cite{Watson:2011dw, Horiuchi:2013noa, Perez:2016tcq, Dessert:2018qih, Ng:2019gch}. If $\nu_4$ (with a mass larger than $2$~keV) is responsible for all the DM in the Universe, 
it requires a mixing angle $\theta$ large enough that X-ray radiation should have been observed by X-ray telescopes. In order to make this scenario compatible with observations, new-physics solutions are needed; for instance, the introduction of a large lepton-number asymmetry~\cite{Shi:1998km}. However, the amount of lepton asymmetry required for this mechanism to work is currently not testable by any cosmological surveys.

An alternative and simple solution comes from the introduction of new self-interactions among the active neutrinos mediated by scalar~\cite{ Hansen:2017rxr,DeGouvea:2019wpf} or vector particles~\cite{Kelly:2020pcy}, which allow for an enhancement of the effective mixing angle $\theta_{\rm eff}$ in Eq.~\eqref{masterequation} while maintaining the vacuum mixing angle $\theta$ well below the current experimental constraints. These models with new ``secret'' interactions have the advantage of being testable in a variety of current and future neutrino and dark photon experiments. 

\begin{figure}
\includegraphics[width=0.46\textwidth]{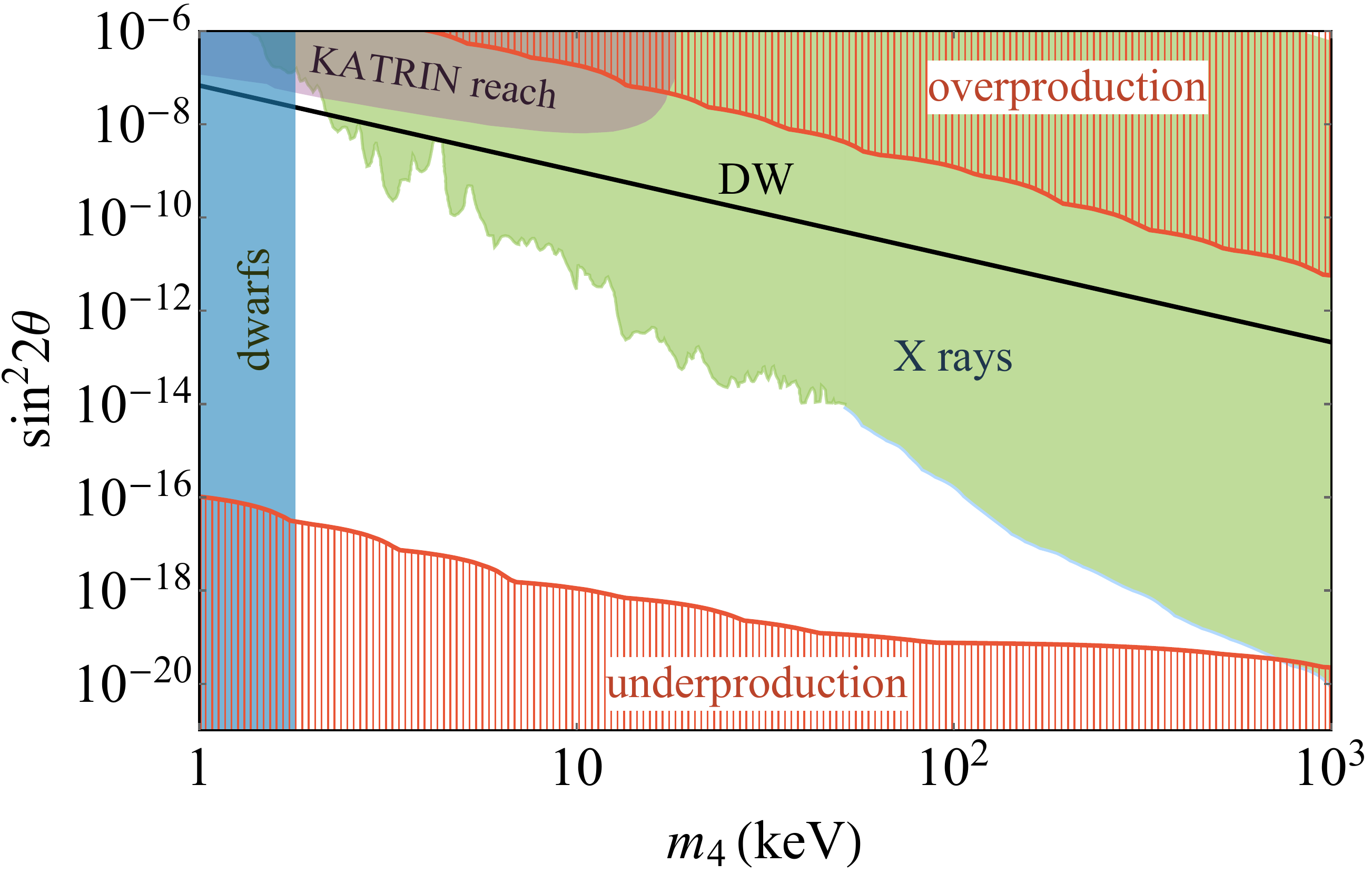}~~\includegraphics[width=0.45\textwidth]{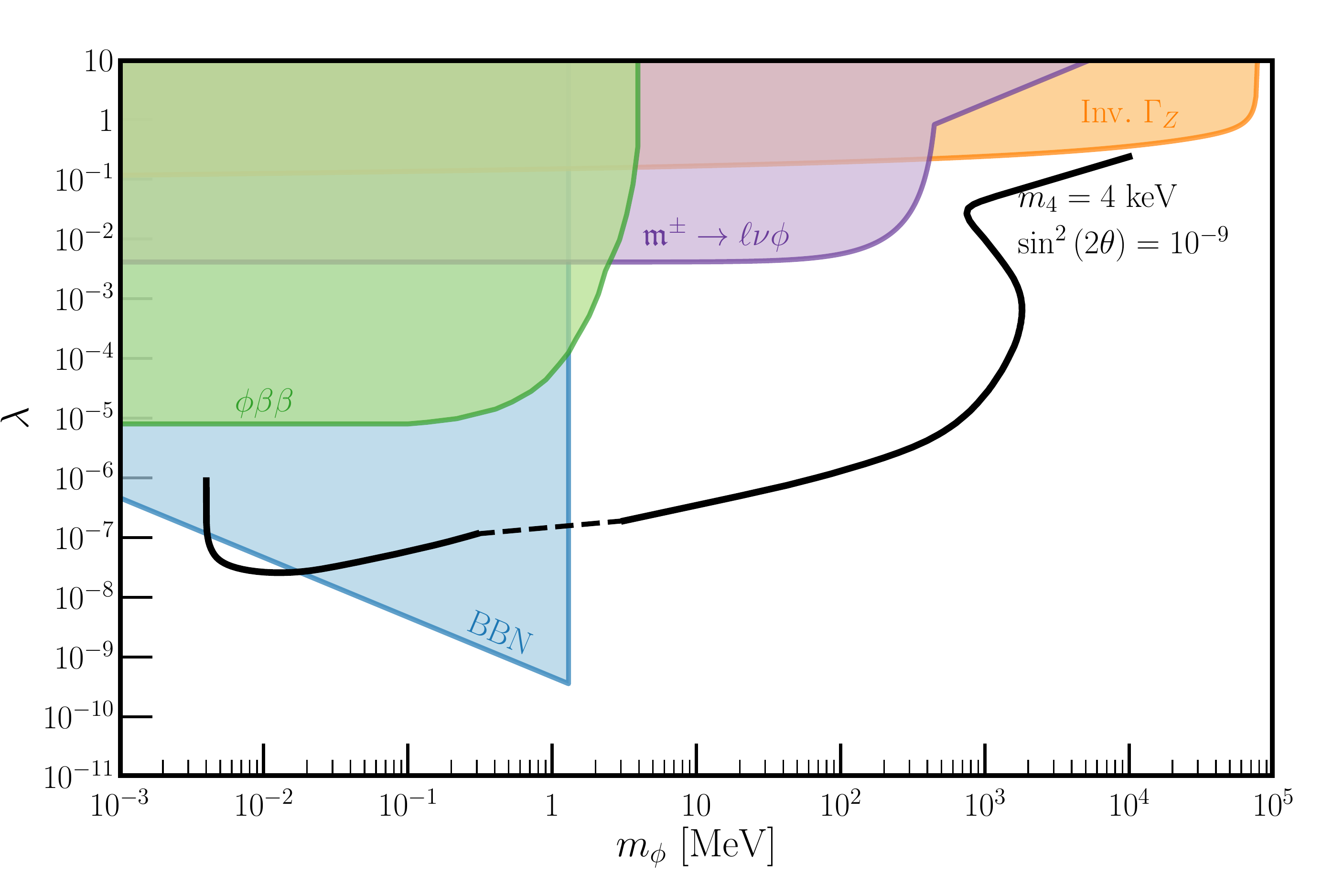}
    \caption{Left panel: dark matter mass/mixing parameter space showing the region predicted by the standard Dodelson-Widrow mechanism (black line), as well as the additional region allowed due to neutrino self-interactions (between the black line and the red, hatched area). Constraints from X-ray telescopes, future sensitivities from KATRIN, and bounds from dwarf galaxies are also shown. Right panel: self-interaction mediator mass/coupling parameter space showing constraints from meson decays (magenta), BBN (blue), and invisible Z-decays (orange) on new neutrino interactions. The black contour, shown for $m_4=4\,{\rm keV}$ and $\sin^22\theta=10^{-9}$, can make up for all the DM. Plots taken from Refs.~\cite{DeGouvea:2019wpf} and \cite{Sen:2021mxl} respectively. 
    }
    \label{fig:DM_parameter_space}
\end{figure}

In the case of a scalar mediator~\cite{DeGouvea:2019wpf}, interaction terms such as $\frac{\lambda_\phi}{2} \nu_a \nu_a \phi + {\rm h.c.}$ are added to the theory, where $\phi$ is a complex scalar with mass $m_\phi$ ranging from keV to GeV.\footnote{For the purpose of generation of DM, only the effective theory is considered. Operators like the one presented here can be further embedded in reasonable UV-complete models~\cite{Berryman:2018ogk, Kelly:2019wow,Blinov:2019gcj}.} The effect of the new interaction is reflected in contributions to the self-interaction rate $\Gamma$ and the thermal potential $V_T$ in Eq.~\eqref{masterequation}. Since the mediator is light relative to the mass of the $Z$ boson, there is typically an epoch during which $T > m_\phi$ and the sign of the new contribution to $V_T$ is opposite to the SM contribution. This may lead to an interference effect as the denominator of $\sin^22\theta_{\rm eff}$ gets suppressed if $|V_{T}^{\rm new}| \sim |V_{T}^{\rm weak}|$ in this period. Additionally, the new interaction could keep the neutrinos in thermal equilibrium with themselves for a longer period than the weak interaction, which also facilitates the production rate of DM. The presence of the new neutrino interaction extends the allowed parameter from a narrow line on the $\sin^22\theta-m_4$ plane to a broad band, as shown in Fig.~\ref{fig:DM_parameter_space} (left panel). Some of the extended parameter space will be challenged by the next generations of X-ray observations but some of the parameter space associated to the smallest mixing angles will remain available even in the absence of an astrophysical discovery. However, these smaller values of $\theta$ require new neutrino interactions that are strong enough to be probed using precision measurement of pion, kaon, and $Z$-boson decays (see the right panel of Fig.~\ref{fig:DM_parameter_space}). These laboratory constraints, as well as upcoming searches at the Deep Underground Neutrino Experiment~\cite{DUNE:2020ypp} and the Forward Physics Facility~\cite{Anchordoqui:2021ghd}, will be discussed in detail in Section~\ref{sec:lab}. At low mediator masses, cosmological surveys provide the strongest constraints and future prospects~\cite{Kelly:2020aks} -- see Section~\ref{sec:cosmo} for more detail on cosmological impacts of this BSM physics.

Self-interactions mediated by a vector boson may include a neutrinophilic interaction, as well as gauged ${L_\mu-L_\tau}$ and ${B-L}$ symmetries~\cite{Kelly:2020pcy}. These scenarios share many features with the scalar-mediated case. The effects on the production mechanism are very similar, leading to an enhancement of the oscillation probability even for very small mixing angles. In the neutrinophilic model, a vector boson $V$ couples only to the active neutrinos. This model leads to efficient production of sterile neutrino DM when the mass of $V$ lies between MeV to GeV and the corresponding coupling to neutrinos ranges from $10^{-6}$ to $10^{-2}$. If the new vector boson $V$ is thermalized in the early Universe and light enough to remain relativistic by the time of BBN, it is required that $m_V \gtrsim 5$ MeV to be compatible with $\Delta N_{\rm eff}$~\cite{Blinov:2019gcj}\footnote{For details of the effects of non-standard neutrino self-interactions on BBN, see~\cite{Grohs:2020xxd}.}.  If $V$ is lighter than $\sim100\,$MeV, it can be produced from neutrino scatterings in the explosion of Supernova 1987A and carry away significant amount of energy, modifying the neutrino emission timescale, which imposes additional constraints for small couplings and mediator masses. There is a prospective signal of this model in DUNE in which the final muon in the process $\nu_\mu n \rightarrow \mu^-p^+V$ carries lower energy than expected since the vector $V$ takes away both energy and transverse momentum. Finally, more constraints come from meson decays, invisible $Z$ boson decay, and invisible Higgs boson decay. The latter constraint turns out to be the most stringent of them and it is expected to be further improved with the high-luminosity run of the LHC~\cite{Cepeda:2019klc}. 

Models with gauge interactions $U(1)_{L_\mu-L_\tau}$ and $U(1)_{B-L}$ are not precisely ``secret'' since other fields in the SM are also charged under these symmetries. They do, however, facilitate the production of sterile-neutrino DM in a similar fashion as the other two cases presented above. The $U(1)_{L_\mu-L_\tau}$ interaction, which was also studied in \cite{Shuve:2014doa}, is more constrained compared to the neutrinophilic case, and the region of parameter space that leads to the expected DM relic abundance can be covered by proposed experiments, including SHiP, NA62, NA64-$\mu$, $M^3$, and DUNE~\cite{Kelly:2020pcy}. The $U(1)_{B-L}$ model is the most strongly constrained among the three vector-mediated models; its parameter space is already narrow and will be fully covered by future high intensity experiments, including BELLE-II, FASER, and LDMX. 

The impact of new interactions among neutrinos and their effects in the dynamics of DM has been also explored in~\cite{Babu:1991at,Dasgupta:2013zpn,Hannestad:2013ana,Mirizzi:2014ama,Cherry:2016jol,Chu:2018gxk}. Ref.~\cite{Benso:2021hhh} studied self-interactions from the effective field theory perspective and included the case in which not all DM is made up of sterile neutrinos. New self-interactions have also been studied in the context of DM in non-standard cosmologies~\cite{Chichiri:2021wvw}. Self-interactions can also be explored in the case of very light vector mediators, $m_V \lesssim$ eV, as done in Ref.~\cite{Alonso-Alvarez:2021pgy}. This case differs from the DW mechanism and the production of sterile neutrino DM comes from the resonant oscillation of a cosmological dark photon condensate.

Going beyond these freeze-in scenarios, neutrino self-interactions can also accommodate connections between the SM and DM in thermal freeze-out scenarios. This can be achieved for DM at the ${\sim}$GeV scales using a variety of interactions among a mediator and DM, all with the interaction between the mediator and neutrinos of the type in Eq.~\eqref{eq:LScalar}~\cite{Kelly:2019wow}. Majoron mediators can allow for light, sub-MeV dark matter, as explored in Ref.~\cite{Berlin:2018ztp}. Furthermore, the Majoron (or a Majoron-like-particle) can act as DM itself, which has been studied, e.g., in Refs.~\cite{Lattanzi:2007ux, Bazzocchi:2008fh, Frigerio:2011in, Queiroz:2014yna, Garcia-Cely:2017oco}.

\subsection{Ultraviolet Completions}\label{sec:theory:UV}
A simple model for neutrino self-interactions involves the so-called `leptonic scalars' $\phi$ that carry a $B-L$ charge of $+2$ but are singlets under the SM gauge group~\cite{Berryman:2018ogk, deGouvea:2019qaz, Dev:2021axj}. They can only couple to the SM fields via higher-dimensional operators, the lowest of which is of dimension six, given in Eq.~\eqref{eq:LHSquaredPhi}, where $\Lambda$ represents
some new physics scale.  After EW symmetry breaking, the operator~\eqref{eq:LHSquaredPhi} yields flavor-dependent NSIs of neutrinos with the leptonic scalar of the form
\begin{equation}
    \lambda_{\alpha \beta} \phi \nu_\alpha \nu_\beta \, ,
    \label{eqn:lambda}
\end{equation}
where $\lambda\sim v^2/\Lambda^2$ and $v$ is the EW vacuum expectation value (VEV). Finally, at energy scales below the mass of $\phi$, this leads to an effective flavor-dependent non-standard neutrino self-interaction of the form $\lambda_{\alpha \beta}\lambda_{\gamma\delta} \nu_\alpha \nu_\beta \nu_\gamma\nu_\delta/m_\phi^2$. 

In this section, we discuss several possible UV-complete models that, after integrating out the heavy degrees of freedom, lead to the effective operator~\eqref{eq:LHSquaredPhi}. 

\paragraph{Self-interactions from Weakly Coupled Physics:} These models are inspired by the tree-level seesaw realizations of the dimension-five Weinberg operator $(LH)(LH)/\Lambda$~\cite{Weinberg:1979sa}, except that all new particles introduced here preserve the $B-L$ symmetry. 
The first model we will discuss is based on Ref.~\cite{Dev:2021axj} and is motivated by the type-II seesaw~\cite{Konetschny:1977bn, Magg:1980ut, Schechter:1980gr, Cheng:1980qt, Mohapatra:1980yp, Lazarides:1980nt} with an $SU(2)_L$-triplet scalar field $\Delta$ with hypercharge $+1$ and $B-L$ charge $+2$. Here the key difference compared to the type-II seesaw is that the neutral component of $\Delta$ does not acquire a VEV, which keeps the custodial symmetry intact. As a consequence, the lepton-number symmetry is not broken and the neutrinos are Dirac-type in this model, with the addition of SM-singlet $B-L=-1$ RH neutrino fields $\nu_{R_i}$. We also add a SM-singlet $B-L=+2$ complex scalar field $\Phi$, whose CP-even and odd components act as the leptonic scalars $\phi$ in the model. The relevant piece of the Yukawa Lagrangian is given by 
\begin{equation}
-\mathcal{L}_Y = y_{\nu,\,\alpha\beta} \overline{L}_\alpha H \nu_{R_\beta} + Y_{\alpha\beta} L_\alpha^{\sf T} C i \sigma_2 \Delta L_\beta + \tilde{y}_{\nu,\,\alpha\beta} \nu_{R_\alpha}^{\sf T} C \nu_{R_\beta} \Phi  + {\rm h.c.} \, ,
\label{eq:LYuk}
\end{equation}
and the scalar potential is given by 
\begin{eqnarray}
\label{V_Delta_phi}
V(H,\Delta,\Phi) =
& & - m^2_H + \frac{\lambda}{4}(H^\dagger H)^2 + M^2_\Delta {\rm Tr}(\Delta^\dagger \Delta) + M^2_\Phi \Phi^\dagger \Phi \nonumber \\
&& + \lambda_1(H^\dagger H){\rm Tr}(\Delta^\dagger \Delta) + \lambda_2 [{\rm Tr}(\Delta^\dagger\Delta)]^2 + \lambda_3 {\rm Tr}[(\Delta^\dagger \Delta)^2] + \lambda_4 (H^\dagger \Delta) (\Delta^\dagger H) \nonumber \\
& & + \lambda_5(\Phi^\dagger \Phi)^2 + \lambda_6 (\Phi^\dagger \Phi)(H^\dagger H) + \lambda_7(\Phi^\dagger \Phi){\rm Tr}(\Delta^\dagger \Delta) 
\nonumber \\
& &
+ \lambda_8(i\Phi H^{\sf T} \sigma_2 \Delta^\dagger H + {\rm h.c.}) \,,
\end{eqnarray}
The effective operator~\eqref{eq:LHSquaredPhi} is generated by integrating out the $\Delta$ field, with the effective coupling~\eqref{eqn:lambda} given by
\begin{equation}
    \lambda_{\alpha\beta} = \sqrt 2 Y_{\alpha\beta}\sin\theta \, ,
\end{equation}
where 
\begin{equation}
\tan 2\theta = \frac{\lambda_8 v^2}{M^2_{\Delta}+v^2 (\lambda_1+\lambda_4 -\lambda_6)/2 - M^2_\Phi} \,,
\label{theta}\end{equation}
is the mixing between the CP-even neutral components of $\Phi$ and $\Delta$ fields. This UV-completion of the leptonic scalar is particularly interesting, as it leads to some distinctive collider signatures~\cite{Dev:2021axj} (see also Section~\ref{subsec:lab:collider}).  

Another possible UV-completion of the operator~\eqref{eq:LHSquaredPhi} is to introduce pairs of vector-like fermions $N_i$ and $N_i^c$ (with $i=1,2,\cdots, n$) which are SM singlets with $B-L$ charges $\mp 1$, respectively. The relevant renormalizable Lagrangian is given by
\begin{align}
    {\cal L} \ \supset \ y_{\alpha i}L_{\alpha} H N_i^c+\lambda_{N,ij}\phi N_i N_j+M_{N,i}N_i N_i^c+{\rm h.c.}
    \label{eq:A3}
\end{align}
This is in a similar vein to the type-I seesaw model~\cite{Minkowski:1977sc, Mohapatra:1979ia, Yanagida:1979as, Gell-Mann:1979vob}, but there are no $B-L$ violating terms here. After integrating out the heavy vector-like fermion fields, we obtain the desired $\lambda_{\alpha\beta}$, with the effective $\lambda$-couplings in Eq.~\eqref{eqn:lambda} given by
\begin{align}
    \lambda_{\alpha\beta} \ = \ v^2 y_{\alpha i}M_{N,i}^{-1}\lambda_{N,ij}M_{N,j}^{-1}y^T_{j\beta} \, .
    \label{eq:A4}
\end{align}
Here the Yukawa couplings $y_{\alpha i}$ also lead to the mixing of the SM neutrinos with the new vector-like fermions, with the mixing angle $\theta\sim yv/M_N$, which is generically constrained to be $\lesssim {\cal O}(0.01)$ for $M_N>v$ from EW precision data~\cite{delAguila:2008pw}. Thus, the $\lambda$-couplings in Eq.~\eqref{eq:A4} are expected to be small $\lesssim {\cal O}(10^{-4})$ in this setup, obliterating their detection prospects~\cite{deGouvea:2019qaz}.

A third possible UV-completion is to replace the SM-singlet vector-like fermions in Eq.~\eqref{eq:A3} by $SU(2)_L$-triplet fermions, as in the type-III seesaw model~\cite{Foot:1988aq}. In this case, the Yukawa Lagrangian is of the form $y_{\alpha i}L_\alpha \sigma^a H N_{ia}$, where $a=1,2,3$ is the $SU(2)_L$ index in the adjoint representation and $\sigma^a$ are the Pauli matrices. After integrating out the heavy $N_i$ fields, the low-energy effective operator takes the form $(L\sigma^a H)(L\sigma_a H)\phi/\Lambda^2$, with the effective $\phi\nu\nu$ coupling~\eqref{eqn:lambda} related to the UV parameters in the same way as in Eq.~\eqref{eq:A4}. Nonetheless, the experimental constraints on $y$ are still applicable in this case, thus ruling out the possibility of large $\lambda_{\alpha\beta}$, but the $SU(2)_L$-triplet fermions might still offer some interesting collider phenomenology, as in type-III seesaw~\cite{Franceschini:2008pz}.

Similar examples of UV-complete models for Majorana neutrinos have been discussed in Ref.~\cite{Blinov:2019gcj}. Using the leptonic scalar field as a portal to the dark sector has been discussed in Ref.~\cite{Kelly:2019wow}, where an additional $Z_2$, $Z_3$, or $U(1)$ symmetry was invoked to stabilize the dark matter.

The effective coupling of $\phi$ to neutrinos as in Eq.~\eqref{eqn:lambda} is similar to that of a Majoron~\cite{Chikashige:1980ui}.  The equivalent coupling $\lambda$ in this case is related to the observed neutrino masses, $\lambda \sim m_\nu/f$, where $f$ is the spontaneous lepton number-breaking scale. However, in order to get sizable couplings $\lambda\sim {\cal O}(1)$ for phenomenological purposes, the lepton number-breaking scale would have to be very low, $f\sim m_\nu\lesssim {\cal O}(1~{\rm eV})$.

\paragraph{Secret Interactions for Active Neutrinos Through Mixing:} Neutrino self-interactions, including those in Eq.~\eqref{eqn:lambda}, can be embedded in UV-complete models through a combination of neutrino mixing and $\nu_s$-philic interactions. For instance, starting from an SU$(2)$-invariant interaction, $\overline{\nu_s} \nu_s \phi$, the secret interaction can leak into the active neutrino sector via neutrino mixing, $(LH)\nu_s$. In the mass basis, 
\begin{equation}
    \mathcal{L} \supset \sum_{i,j} U_{si}^* U_{sj}\, \overline{\nu}_i \nu_j \phi + \dots,
\end{equation}
where $U_{si}$ are the mixing elements of the sterile flavor with the mass eigenstate $i$. In this type of UV completion, self-interactions between active neutrinos can be suppressed in physical observables. For example, meson decays, $M^+ \to \ell^+ \nu \phi$ and $\phi$ emission in $0\nu\beta\beta$ decays are both examples where cancellations akin to the GIM mechanism in $K_L\to \mu^+ \mu^-$ suppress the rate by $m_{4}^2/E^2$, with $E$ the energy scale of the process. As an example, the unitarity of $U$ can be used to show that the amplitude for $W^+ \to \ell_\alpha^+ \nu_j \phi$ is suppressed. Explicitly,
\begin{equation}
\mathcal{M}_{W \to \ell_\alpha \nu_j \phi} \propto \sum_i  U_{\alpha i}^* U_{si} U_{sj}^* \frac{\slashed{p} + m_i}{p^2 + m_i^2} \to  \sum_i  U_{\alpha i}^* U_{si} U_{sj}^* \frac{m_i}{p^2},
\end{equation}
where we assumed the typical momentum exchange to be large, $p^2 \gg m_i^2$. For $p^2 \ll m_4^2$, the rate will be instead suppressed by $1/m_4^2$ and it is still proportional to small mixing factors $|U_{s 4} U_{\alpha 4} U_{sj}|^2$. Because of this cancellation, some of the strongest constraints on the parameter $\lambda_{\alpha}$, such as those coming from meson decays, can be avoided altogether, while interesting astrophysical signatures remain. 

Another interesting aspect of such UV completions is that $\nu_4$ can lead to additional experimental signatures. For instance, the decays in flight of $\nu_4$  can generate effective flavor transitions in neutrino beams~\cite{Palomares-Ruiz:2005zbh,Bai:2015ztj,deGouvea:2019qre,Dentler:2019dhz} as well as lead to apparent neutrino-antineutrino conversion~\cite{Hostert:2020oui}. Finally, it is worth noting that the secret interactions can also leak to the charged lepton and quark sectors via one-loop diagrams, where they can be constrained by searches for long-range forces~\cite{Chauhan:2020mgv,Xu:2020qek}.

\paragraph{Self-interactions from Strong Dynamics:} 
A separate, attractive class of UV-completions for neutrino mass models and \nusi is through strong dynamics. One primary motivation to consider such a class of models is to explain the smallness of the neutrino masses naturally, through their couplings to a strongly coupled sector~\cite{ArkaniHamed:1998pf,vonGersdorff:2008is,Grossman:2010iq,Chacko:2020zze}. The compositeness scale, generated through dimensional transmutation,  can be parametrically lower than the Planck scale. Next, the small neutrino mass can stem from operators with (mass) dimensions greater than four. The composite neutrino framework can connect to dark matter~\cite{Robinson:2012wu,Robinson:2014bma,Hundi:2011et}, to the origin of the baryon asymmetry~\cite{Grossman:2008xb}, and have cosmological imprints~\cite{Okui:2004xn}.  

Compared with weakly coupled neutrino mass theories, the singlet neutrinos are now the hadrons of a new strong force, and the light neutrinos can be partially composite particles. 
Here we follow the recent development~\cite{Chacko:2020zze} to show how would the UV models for composite neutrinos work. At low energy, one wants to realize, e.g., an inverse-seesaw Lagrangian, 
 \begin{equation} 
 \label{L_IR} 
\mathcal{L}_{\rm IR} \supset i\bar{N} \bar{\sigma}^\mu \partial_{\mu} N 
+ i\bar{N}^c \bar{\sigma}^\mu \partial_{\mu} N^c - \left[M_N N^c N + 
\lambda L H N + \frac{\mu^c}{2} \left(N^c\right)^2 + {\rm h.c.} 
\right]~. 
 \end{equation} 
The essential smallness for all the parameters can be naturally explained in this class of UV models and explained below. 
We, in particular, consider the strongly coupled sector to be neutral under the SM gauge groups. Hence, the compositeness scales can lie well below the EW scale. 

We begin with the strong dynamics of a conformal field theory (CFT), deformed by a relevant operator $\mathcal{O}_{\rm S}$. 
When the deformation $\mathcal{O}_{\rm S}$ gets large, it triggers the breaking of the CFT at a scale, denoted by $\Lambda$, which corresponds to the mass scale of the lightest composite particles for composite scenarios without pions or other light composite states. The composite singlet neutrinos with mass $M_N$ of order $\Lambda$ appear in the low-energy Lagrangian. 

The CFT couples to the SM through a neutrino portal
interaction which takes the form
\begin{equation}
\mathcal{L}_{\rm UV} \supset \frac{\hat{\lambda}}{M_{\mathrm{UV}}^{\Delta_{\mathrm{N}} - 3/2}} L H \mathcal{O}_{\rm N} + {\rm h.c.}
\Longrightarrow
\mathcal{L}_{\rm IR} \supset \lambda L H N + {\rm h.c.}
\label{intLHO}
 \end{equation} 
The mass scale $M_{\mathrm{UV}}$ represents the 
UV cutoff of the theory and $\hat{\lambda}$ is an  $\mathcal{O}(1)$ dimensionless 
parameter. Here $\mathcal{O}_{\rm N}$ is a 
primary operator of the CFT and $\Delta_{\mathrm{N}}\geq 3/2$ is its scaling 
dimension, whose validity regime is discussed in detail in Ref.~\cite{Chacko:2020zze}. Note that for $\Delta_{\rm N}=3/2$, the form of the interaction reduces back to the weakly coupled theory.  
At energies of order $\Lambda$, the interaction in Eq.~\eqref{intLHO}
gives rise to the portal term in the low-energy Lagrangian $\mathcal{L}_{\rm IR}$.
Importantly, the portal coupling $\lambda$ is of the order,
 \begin{equation}
\lambda \sim
C_\lambda\,
\hat \lambda
\left(\frac{\Lambda}{M_{\mathrm{UV}}}\right)^{\Delta_{\mathrm{N}}-3/2}\;,
\label{lambdascaling}
 \end{equation}
where the order one multiplicative factor $C_\lambda$ is estimated in Ref.~\cite{Chacko:2020zze}. We can see that this IR parameter is naturally suppressed for large $\Delta_{\mathrm{N}}$.

The Lagrangian also contains a 
small deformation of the CFT, denoted by $\mathcal{O}_{\rm 2N^c}$, which 
explicitly violates lepton number,
 \begin{equation}
\label{Nc2deformation}
\mathcal{L}_{\rm UV} \supset \frac{\hat{\mu}^c}{M_{\mathrm{UV}}^{\Delta_{\rm 2N^c} - 4}}
\mathcal{O}_{\rm 2N^c} + {\rm h.c.}
\Longrightarrow
\mathcal{L}_{\rm IR} \supset \frac{\mu^c}{2} \left(N^c\right)^2 + {\rm h.c.}
 \end{equation} 
Here $\Delta_{\rm 2N^c}$ (that is $\geq 1$ from unitarity requirement) is the scaling dimension of the operator $\mathcal{O}_{\rm 2N^c}$, and $\hat{\mu}^c$ is a dimensionless 
parameter. The limiting case 
of $\Delta_{\rm 2N^c} = 1$ corresponds to a free scalar.
Assuming this deformation carries a lepton number of $(-2)$, at scales of order $\Lambda$ the Lagrangian contains a lepton number violating term.
The mass parameter $\mu^c$ is related to the parameters 
in the ultraviolet theory as 
 \begin{equation} 
\mu^c \sim C_\mu \hat{\mu}^c\, \Lambda \left(\frac{\Lambda}{M_{\mathrm{UV}}} 
\right)^{\Delta_{\rm 2N^c} - 4} ~, 
\label{muscaling} 
 \end{equation} 
where the multiplicative factor $C_\mu$ is estimated in Ref.~\cite{Chacko:2020zze}.

Having all the ingredients needed in the IR, in Eq.~\eqref{L_IR}, generated from the UV theory, we get a 
contribution to the masses of the light neutrinos from the inverse seesaw of order
 \begin{equation} 
\label{eq:numass} 
 {m_\nu} \sim \mu^c \left(\frac{\lambda v_{\rm EW}}{M_N}\right)^2 
\sim \Lambda \left[C_\mu \hat{\mu}^c 
\left(\frac{\Lambda}{M_{\mathrm{UV}}}\right)^{\Delta_{2\mathrm{N^c}}-4} 
\right] 
\left[C_\lambda \hat{\lambda} \left(\frac{v_{\mathrm{EW}}}{\Lambda}\right) 
\left(\frac{\Lambda}{M_{\mathrm{UV}}}\right)^{\Delta_{\mathrm{N}}-3/2} 
\right]^2 \;. 
 \end{equation} 
 The strength of the
lepton-number violation in the composite sector and the partial compositeness are controlled by the first square bracket and second square bracket in Eq.~\eqref{eq:numass}, respectively. Together, these effects generate Majorana neutrino masses. From Eq.~\eqref{eq:numass}, we can see that the scaling dimensions of the operators, $\mathcal{O}_{\rm N}$ and $\mathcal{O}_{\rm 2N^c}$, control the sizes for these effects. In such a way, the scaling dimensions of CFT operators provides a natural explanation for the small neutrino masses. It is worth noting phenomenological differences from other weakly interacting theories. The strong dynamics enables multiple new singlet neutrino productions in a single collision, and in the extreme case, it will undergo a process of showering in the new sector. The resulting spectrum of the accompanied leptons from the charged-current production and the strategy in observing the long-lived singlet neutrinos can enable new opportunities beyond the current searches~\cite{Chacko:2020zze,Liu:2018wte,Liu:2019ayx,Alimena:2019zri,Liu:2020vur,Knapen:2021eip}.

Of particular relevance to the neutrino self-interaction, the Lagrangian (at the scale $\Lambda$) is also expected to contain
four-fermion interactions between the $N$'s. These take the schematic
form,
 \begin{equation}
\mathcal{L}_{\rm IR} \supset - \kappa \frac{\left({\bar{N}\sigma^\mu N}\right)^2}{\Lambda^2} + \kappa^\prime \frac{\left({N^c N}\right)^2}{\Lambda^2} + \ldots \; ,
\label{int4N}
 \end{equation}
 where we have shown two such terms, and $\kappa$ and 
$\kappa^\prime$ are expected to be of order $(4 \pi)^2$. The composite nature of the singlet neutrinos, signified by these renormalizable couplings, call for more explorations. The self-interactions play an important role in the phenomenology of this class of models, particularly in astrophysics and cosmology. More details of this phenomenology are discussed in Ref.~\cite{Chacko:2020zze}.

\section{Cosmological Probes}\label{sec:cosmo}
Neutrino self-interactions can leave an imprint on a variety of cosmological 
observables such as the matter and Cosmic Microwave Background (CMB) power spectra. In this section we summarize the impact of neutrinos and their potential self-interactions on key cosmological epochs, including Big Bang Nucleosynthesis (BBN), the formation of the CMB, and the growth of structure in the Universe. In many cases, the corresponding observables are able to probe unique regions of the neutrino self-interaction parameter space. 
\begin{figure}
    \centering
    \includegraphics[width=0.47\textwidth]{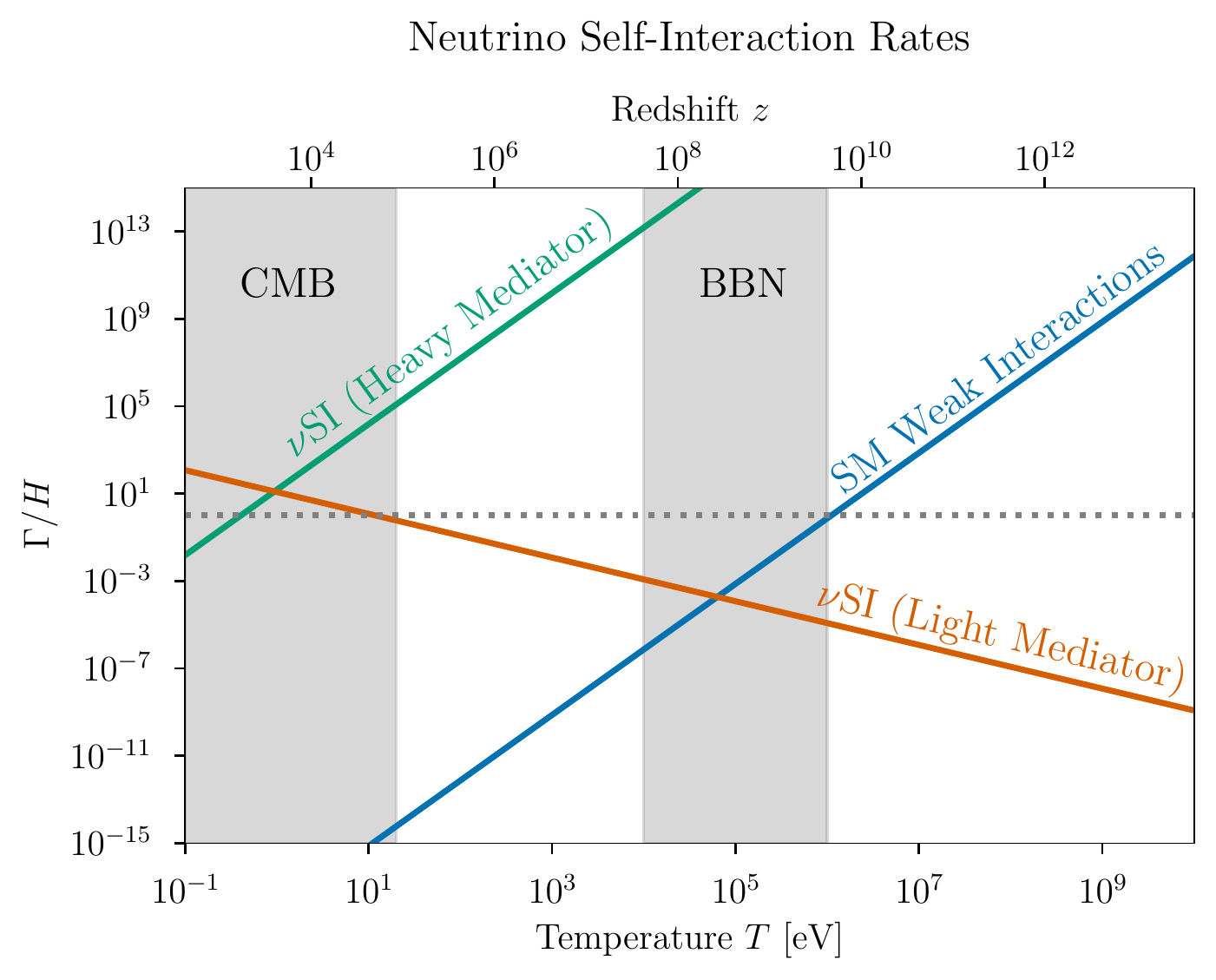}
    \caption{Schematic illustration of neutrino self-interaction (\nusi) rate $\Gamma$ over the Hubble expansion rate $H$ for different \nusi models as a function of temperature (lower horizontal axis) or redshift (upper horizontal axis). The CMB and BBN epochs are highlighted as gray bands. Freeze-out occurs when $\Gamma/H$ falls below 1, indicated by the gray dotted line. For the SM weak interaction this occurs just before BBN, at temperatures $\mathcal{O}(\mathrm{MeV})$. In models with non-standard interactions, the freeze-out of self-interactions can be significantly delayed as in \nusi models with a heavy (compared to the temperatures of interest) mediator. If the mediator is instead light, self-interactions can freeze-out at the usual time, but then become important again through the beyond-SM processes.
    }
    \label{fig:nusi_rate_over_hubble}
\end{figure}

We begin with a brief summary of the role self-interactions in the standard cosmology. 
As the Universe expands, particle momenta redshift and energies decrease.  In a
system of coupled particles in thermal equilibrium, expansion leads directly to
cooling and decreasing temperatures.  The constituent particles have less
energy when they participate in scattering and other reactions.  Weak
interaction cross sections scale with the energies of the incoming particles, so less energy
implies a smaller cross section and subsequent reaction rate.  Those rates eventually become
too small to maintain thermal contact between the neutrinos and the electroweak
plasma. 
Besides gravity, neutrinos experience no other Standard-Model
interactions at a significant level, so the freeze-out of the weak interactions
between the neutrinos and the electroweak plasma is functionally a decoupling
of the neutrinos and the other constituents of the Universe.  In
addition, neutrinos experience weak interactions among themselves at
approximately the same strength as they do with the charged leptons of the
electroweak plasma. The temperature evolution of the rate of weak interactions compared to the expansion rate is shown in Fig.~\ref{fig:nusi_rate_over_hubble}. Therefore, individual neutrinos kinematically decouple
from one another at the same epoch.  In the standard cosmology assuming
SM interactions, the neutrino decoupling epoch occurs on the precipice of BBN when the
plasma temperature has cooled to the MeV scale, roughly when the age of the
Universe is one second.  After decoupling, neutrinos free-stream, initially
moving with ultra-relativistic kinematics and negligible rest masses.  The
free-streaming of neutrinos continues until their rest mass becomes a
significant component of their total energy, at which point neutrinos decouple
from the Hubble flow when the age of the Universe is billions of years. 
Therefore, in the standard cosmology neutrino self-interactions do not play an important role since they are frozen out by BBN and especially during the CMB epoch. However, non-standard self-interactions generated by physics beyond the SM (\nusi) can lead to significantly larger 
scattering rates amongst neutrinos and different time/temperature dependence. Two such examples are shown in Fig.~\ref{fig:nusi_rate_over_hubble} (these are discussed in more detail below).
In these cosmologies key cosmological observables can be modified compared to $\Lambda$CDM.

We start with a description of the influence of non-standard neutrino self-interactions on the CMB in Sec.~\ref{sec:nusi_in_cmb} and the distribution of matter in Sec.~\ref{sec:nusi_in_matter}. Neutrino self-interactions have been proposed as a means of easing the Hubble and $S_8$ tensions, disagreement between late- and early-time measurements of the present expansion rate and matter clustering amplitude, respectively. These anomalies provide an exciting application of \nusi, which we study in Secs.~\ref{sec:H0_tension} and~\ref{sec:nusi_in_s8}. While the combination of the latest cosmological and laboratory data preclude this from completely accounting for the $H_0$ and $S_8$ tensions, they nevertheless provide an interesting study of the complementarity of a wide range of \nusi probes. Such considerations have recently motivated the study of flavour-non-universal \nusi, since laboratory constraints are often strongly flavour dependent; the CMB and BAO constraints on such interactions are described in Sec.~\ref{sec:flav_nonuni}. The rich physics of neutrino decoupling at $\sim$ MeV temperatures is also sensitive to \nusi as described in Sec.~\ref{sec:nusi_in_bbn} through modifications of the neutrino phases-space distributions, which feed into the light element abundances. Finally, in Sec.~\ref{sec:nusi_for_inflation} we point out that \nusi can have important indirect effects on the inference of parameters of inflationary models from CMB observables.

\subsection{Cosmic Microwave Background}
\label{sec:nusi_in_cmb}

In a standard $\Lambda$CDM cosmology at $T\lesssim1$ MeV, free-streaming neutrinos travel supersonically through the photon-baryon plasma at early times, hence gravitationally pulling photon-baryon wave fronts slightly ahead of where they would be in the absence of neutrinos \cite{Bashinsky:2003tk,Baumann:2015rya,Choi:2018gho}. Such free-streaming neutrinos thus imprint a net phase shift in the CMB power spectra towards larger scales (smaller $\ell$), as well as a slight suppression of its amplitude. These effects are present in both CMB temperature and polarization spectra. Free-streaming neutrinos thus lead to a physical size of the photon sound horizon at last scattering $r_*$ that is slightly larger than it would otherwise be. This phase shift is thought to be a robust signature of the presence of free-streaming radiation in the early Universe \cite{Follin:2015hya,Baumann:2017lmt,Choi:2018gho}, and it plays an important role in constraining the abundance of light free-streaming relics in the pre-recombination Universe.
 
\begin{figure}
    \centering
    \includegraphics[width=0.8\textwidth]{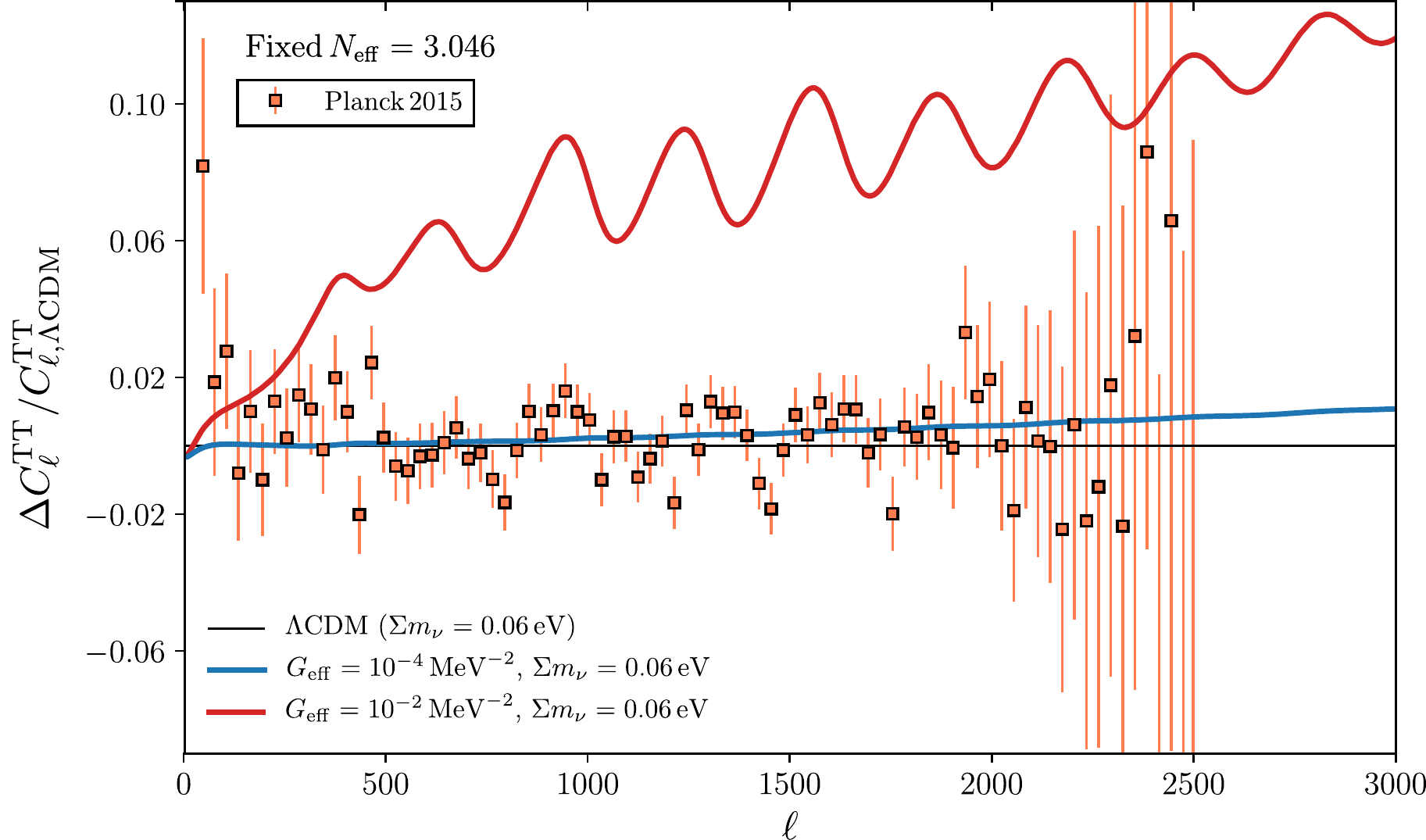}\\\vspace{0.5cm}
    \includegraphics[width=0.8\textwidth]{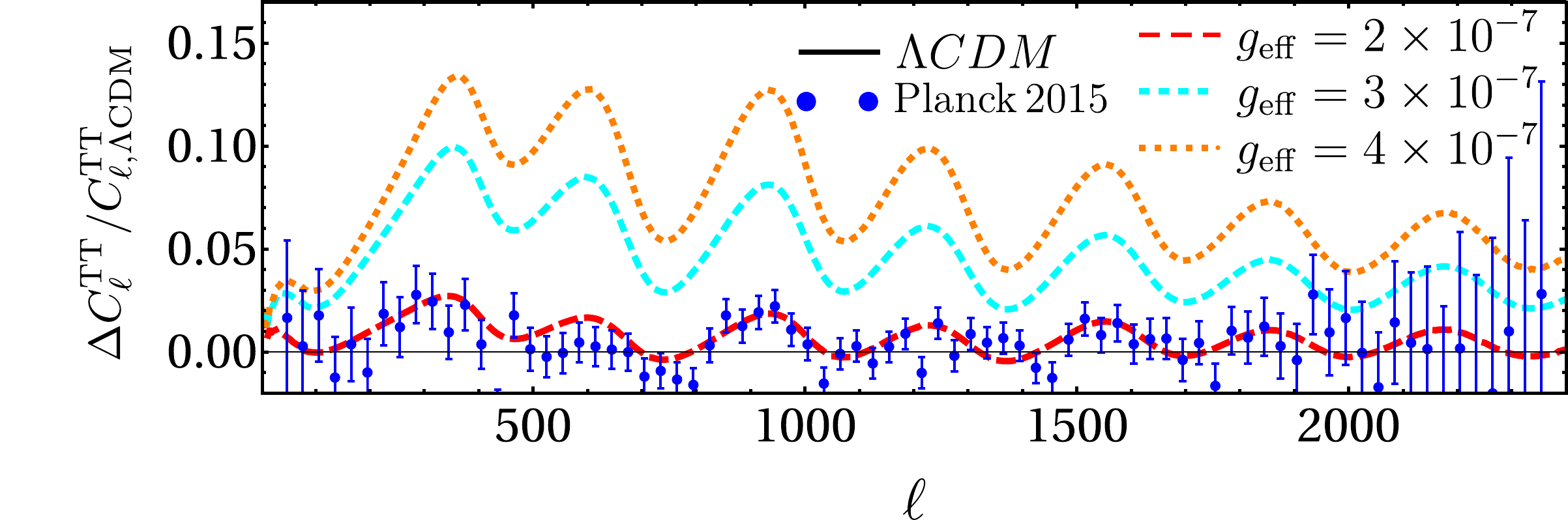}
    \caption{Relative impact of decoupling (upper panel) and recoupling (lower panel) neutrino self-interactions on the CMB power spectrum. 
    Each plot shows the normalized difference of the temperature-temperature power spectrum from the $\Lambda$CDM expectation for several self-interaction model parameters. Observed Planck spectrum residuals are shown by dots with error bars. The impact of decoupling \nusi is greatest at small multipoles, while recoupling affects larger scales/smaller $\ell$. The decoupling (recoupling) scenario corresponds to the ``heavy mediator'' (``light mediator'') model in Fig.~\ref{fig:nusi_rate_over_hubble}. The plots are adapted from Refs.~\cite{Kreisch:2019yzn} and~\cite{Forastieri:2019cuf}. }
    \label{fig:nusi_cmb_impact}
\end{figure}

The presence of neutrino self-interaction at early times (such as those mediated by a massive scalar) delay the epoch at which neutrinos begin to free-stream, as shown schematically in Fig.~\ref{fig:nusi_rate_over_hubble} by the ``Heavy Mediator'' line. Fourier modes of photon-baryon perturbations entering the causal horizon while neutrinos are still tightly coupled will not experience the gravitational tug of supersonic neutrinos and will therefore not receive the associated phase shift and amplitude reduction described above. Compared to the standard $\Lambda$CDM model, neutrino self-interactions thus shift the CMB power spectra peaks towards smaller scales (larger $\ell$) and boost their fluctuation amplitude on angular scales entering the causal horizon prior to the onset of free streaming. These effects are illustrated in the upper panel of Fig.~\ref{fig:nusi_cmb_impact}, which shows the normalized difference between self-interacting neutrinos and $\Lambda$CDM for the massive mediator case; the phase shift and enhanced amplitude of high-$\ell$ modes are evident in the oscillating nature and increasing size of these residuals.
If neutrino self-decoupling is delayed until close to the epoch of recombination, this can lead to a net (although small) reduction of the physical size of the photon sound horizon at last scattering $r_*$. As we briefly discuss in Sec.~\ref{sec:H0_tension} below, this smaller predicted size of the sound horizon is an important ingredient that could help reconcile CMB and late-time measurements of the Hubble constant $H_0$, although other model ingredients beyond neutrino self-interaction are likely necessary to completely eliminate the tension. 

In general, not all neutrino flavors have to interact with the same strength in the early Universe. In fact, the flavor-universal case in which all neutrino species interact with the same coupling strength is tightly constrained by an array of non-cosmological experiments \cite{Blinov:2019gcj,Lyu:2020lps}, leaving flavor-nonuniversal interaction as a more plausible avenue for neutrinos to self-interact at early times. Such nonuniversal interaction will be discussed further in Sec.~\ref{sec:flav_nonuni} below. While cosmological constraints on the possible strength of neutrino self-interaction of course depend on the exact form of the chosen coupling matrix, they tend to have generic features that are nearly always present. Perhaps their most surprising characteristic is the bimodality of the interaction strength posterior distribution. Indeed, CMB and baryon acoustic oscillation (BAO) data, when used to constrain the possible strength of neutrino self-interaction in the early Universe, yield two distinct islands of probability in which interacting neutrino models provide a good fit to them. The first mode, dubbed the moderately interacting (MI) mode, is close in phenomenology to $\Lambda$CDM. There, the onset of neutrino free-streaming is still delayed compared to the SM case, but neutrinos are still largely free-streaming by the time the smallest length scales probed by the CMB are entering the causal horizon. The other statistical mode, dubbed the strongly-interacting (SI) mode, corresponds to neutrinos self-interacting with a strength many orders of magnitude larger than in the SM, resulting in an onset of neutrino free streaming around redshift $z\sim10^4$. That such strong self-interactions are allowed by CMB and BAO data is astonishing and is the result of a multi-parameter degeneracy \cite{Lancaster:2017ksf} between the neutrino interaction strength and the primordial spectrum of fluctuations. The relative statistical weight of each mode depends on the exact data sets and neutrino interaction model used, with Planck CMB data generally disfavoring the SI mode as compared to the MI one for the universal coupling case \cite{Das:2020xke,Brinckmann:2020bcn}, while the nonuniversal case places the modes on a more equal statistical footing (see below).

The previous discussion focused on self-interactions through a massive mediator, i.e., those that can be completely described by a Fermi-type interaction; this leads to delayed decoupling of neutrinos from themselves. Another possibility, realized for a very light mediator, is that neutrinos decouple at $T\simeq 1$ MeV as in the standard scenario, and start free-streaming until a later time, when they cease to do so as they ``self-recouple'' and become collisional again as a result of the self-interactions. The cosmological evolution of the neutrino self-interaction rate in this case (labelled ``Light mediator'') is compared to the previous example in Fig.~\ref{fig:nusi_rate_over_hubble}. The effect on perturbation modes entering the horizon when neutrinos are collisional is the same described above, i.e., these modes will not experience the phase shift towards larger scales and amplitude reduction. Hence the CMB spectra will be boosted and shifted towards smaller scales with respect to the free-streaming case. However, in the recoupling scenario different scales are affected with respect to the case of delayed decoupling: those that enter the horizon after the end of the free-streaming regime. Thus the effects on CMB spectra are seen at multipoles smaller than the one corresponding to the scale that enters the horizon at the time of recoupling. This is illustrated in the bottom panel of Fig.~\ref{fig:nusi_cmb_impact} which shows the normalized shift in the CMB power spectrum due to self-interactions through a light mediator.
The imprint of neutrino perturbations on CMB anisotropies is mostly important when neutrinos make a sizeable contribution to the cosmological energy density, i.e., during the radiation era. Thus the boost and phase shift disappear at multipoles smaller than the one corresponding to matter-radiation equality.

CMB data can be used to constrain the recoupling scenario \cite{Archidiacono:2013dua,Forastieri:2015paa,Forastieri:2019cuf}. Planck 2015 temperature data have been shown to prefer the noninteracting scenario, and constrain, combined with the Planck lensing data, the recoupling redshift $z_\mathrm{rec}$ to be smaller than $3800$ ($95\%$ credible interval) \cite{Forastieri:2019cuf}. This bound is strengthened to $z_\mathrm{rec} < 2300 $ when small-scale polarization is also included. These results can be used to gather information on the coupling of neutrinos to a very light mediator, although the relation between the decoupling redshift and the strength of nonstandard interactions is somewhat model-dependent. In the simple case of interactions between Majorana neutrinos mediated by a light (effectively massless) pseudoscalar, as in Majoron models, the bounds reported imply that the coupling constant $g$ should be $< 7.7 \times 10^{-7}$ ($6.7 \times 10^{-7}$ when small-scale polarization is included).

\subsection{Matter Distribution}\label{sec:nusi_in_matter}

\begin{figure}[t!]
  \centering
\includegraphics[width=0.487\linewidth]{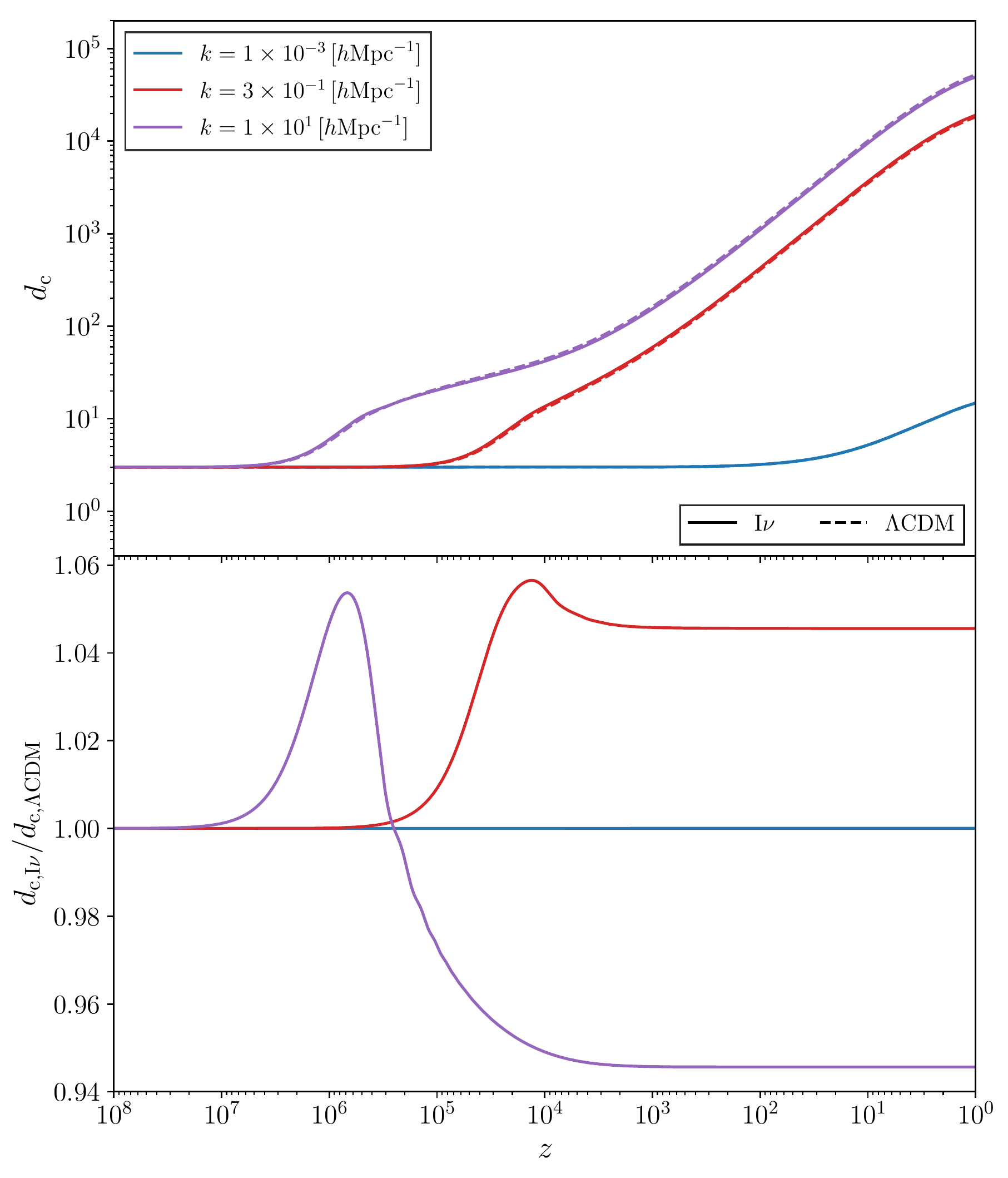}
\caption{The evolution of the gauge invariant dark matter density contrast $d_{\rm c}$ for different $k$ modes as a function of redshift. Solid lines correspond to the interacting neutrino case with a flavor-universal interaction,  whereas dashed lines correspond to the $\Lambda\mathrm{CDM}$ case.  The lower panel shows the ratio of the dark matter fluctuations in the two models. The onset of neutrino free streaming for the interacting neutrino model shown here occurs at $z_{\rm dec,\nu}\simeq 10^4$. Dark matter fluctuations entering the horizon while neutrinos are still tightly coupled decay and appear damped at present relative to $\Lambda \mathrm{CDM}$, while those entering the horizon during neutrino decoupling receive a net boost that persists until the present epoch.}
\label{fig:potentials}
\end{figure}

The growth of matter fluctuations is sensitive to the presence of self-interacting neutrinos through the neutrinos' impact on the two gravitational potentials $\phi$ and $\psi$ in the conformal Newtonian gauge. Indeed, neutrino self-interactions suppress the anisotropic stress of the Universe, leading to $\phi-\psi=0$ before the onset of neutrino free streaming. This contrasts with the $\Lambda$CDM case, for which $\phi = (1 + 2R_\nu/5)\psi$ on large scales at early times for the adiabatic mode \cite{Ma:1995ey}, where $R_\nu$ is the radiation free-streaming fraction. This difference in the evolution of the potentials modifies the gravitational source term driving the growth of matter fluctuations. In the radiation-dominated epoch, the growth of dark matter fluctuation is given by the following solution \cite{Dodelson-Cosmology-2003}:
\begin{equation}\label{eq:sol_growth_cdm} 
d_{\rm c}(k,\tau) = -\frac{9}{2}\phi_{\rm p} + k^2\int_0^\tau d\tau' \tau'\psi(k,\tau')\ln{(\tau'/\tau)},
\end{equation}
 where $d_{\rm c} \equiv \delta_{\rm c} - 3\phi$ in which $\delta_{\rm c} = \delta\rho_{\rm c}/\rho_{\rm c}$ is the standard dark matter energy density contrast in the Newtonian gauge, and where $\phi_{\rm p}$ is the primordial value of $\phi$ on large scales, $k$ is the comoving wavenumber, and $\tau$ is the conformal time. The gauge-invariant variable $d_{\rm c}$ represents the fractional dark matter number density perturbation by unit coordinate volume. At late times, $d_{\rm c}$ is nearly equal to $\delta_{\rm c}$ and it is thus a useful quantity to understand the structure of the matter power spectrum at $z=0$. The integral appearing in Eq.~\eqref{eq:sol_growth_cdm} obtains most of its contribution when $k\tau\sim1$. The changes to the growth of dark matter fluctuations can thus be understood by examining the behavior of the $\psi$ potential at horizon entry, which is particularly sensitive to whether all or some of the neutrinos are self-interacting. 

When modes enter the horizon during the radiation-dominated era, the gravitational potential $\psi$ decays in an oscillatory fashion \cite{Dodelson-Cosmology-2003}. If a fraction of neutrinos are not yet free-streaming, the reduced amount of anisotropic stress implies that $\psi$ starts its oscillatory decaying behavior from a larger amplitude. This boosts the amplitude of the envelope of the decaying oscillations as compared to $\Lambda$CDM, leading to an overall slower decay. While this at first increases the amplitude of dark matter fluctuations at horizon entry as compared to $\Lambda$CDM (see bottom panel of Fig.~\ref{fig:potentials}), the subsequent oscillations of the integrand appearing in Eq.~\eqref{eq:sol_growth_cdm} lead to a net \textit{damping} of the dark matter perturbation amplitude. Another way to think about this is that the slower decay of the potential $\psi$ in the presence of self-interacting neutrinos reduces the horizon-entry boost that dark matter fluctuations experience as compared to $\Lambda$CDM.

For modes entering the horizon at the time of neutrino decoupling, the potential $\psi$ begins decaying from its larger value with $R_\nu < R_{\nu,\Lambda{\rm CDM}}$ but rapidly locks into its standard $\Lambda$CDM evolution due to the onset of neutrino free streaming. This case thus displays the quickest damping of the $\psi$ potential after horizon entry, which leads to a net boost of dark matter fluctuations as compared to $\Lambda$CDM. Indeed, these modes receive a positive contribution near horizon entry from the integral in Eq.~\eqref{eq:sol_growth_cdm}, but without the subsequent extra damping due to the $\psi$ potential quickly converging to its $\Lambda\mathrm{CDM}$ behavior. The evolution of the $k=0.3 \, h/\mathrm{Mpc}$ mode in Fig.~\ref{fig:potentials} displays this behavior. 

Finally, modes entering the horizon well after the onset of neutrino free streaming behave exactly like their $\Lambda\mathrm{CDM}$ counterparts, as illustrated by the $k=10^{-3}\, h^{-1}\mathrm{Mpc}$ mode in Fig.~\ref{fig:potentials}. Taking together the evolution of the different Fourier modes entering before, during, and after neutrino decoupling, we expect the matter power spectrum to have the following properties (at fixed neutrino mass): For large wave numbers entering the horizon while neutrinos are tightly coupled, we expect the matter power spectrum to be suppressed compared to $\Lambda$CDM. As we go to larger scales and approach modes entering the horizon at the onset of free streaming, we expect a ``bump''-like feature displaying an excess of power as compared to $\Lambda$CDM. As we go to even larger scales, the matter power spectrum is expected to asymptote to its standard $\Lambda$CDM value.

\subsection{The Hubble Tension}\label{sec:H0_tension}
The concordance $\Lambda$CDM model provides an excellent fit to many cosmological observations, including the CMB, the distribution of matter at large scales, and primordial abundances of light elements. However, as the sensitivity of current cosmological surveys increases, tiny cracks in this otherwise successful picture start to emerge, mostly in the form of disagreement (or “tension”) between different sets of cosmological and astrophysical data. The most famous example, widely discussed over the past years since the first data release of the Planck satellite, is the so-called “Hubble tension”. Indeed, the value of the Hubble constant $H_0$ inferred from “local” (low-redshift) distance ladder measurements calibrated against Cepheids and the value indirectly inferred from cosmological observations (either CMB-driven or based on an inverse distance ladder calibrated independently of the CMB) within the $\Lambda$CDM framework disagree at the level of $5\sigma$~\cite{Riess:2021jrx, Shah:2021onj, Riess:2020fzl, Bernal:2016gxb, Aubourg:2014yra, DES:2018rjw} (note, however, that other similarly-precise measurements of $H_0$ based on tip-of-the red giant branch calibration are consistent with the CMB-inferred value~\cite{Freedman:2019jwv,Freedman:2020dne,Freedman:2021ahq}). 
Extensive studies of the possible origin of such a disagreement have been pursued in the past years. Leaving aside unaccounted instrumental systematics that might have contaminated the $H_0$ estimates, it is of more interest to this paper the possibility that the $H_0$ tension is signalling the need for going beyond the standard $\Lambda$CDM picture. Indeed, cosmological probes are not directly sensitive to $H_0$. Rather, the $H_0$ estimate can be obtained as a derived quantity when data are analyzed in the context of a specific cosmological model. In particular, $H_0$ enters the calculation of peculiar angular scales that can be measured with cosmological probes, such as the angular scale of the sound horizon:
\begin{equation}
    \theta \sim r_*/d_A(H_0)
\end{equation}
where $r_*$ is the physical scale of the sound horizon and $d_A$ is the angular diameter distance to last scattering, which depends on the late-time expansion history, and therefore on $H_0$.

Different classes of solutions have been proposed~\cite{DiValentino:2020zio,Schoneberg:2021qvd, DiValentino:2021izs, 2021arXiv210101372B, Efstathiou:2021ocp, Knox:2019rjx} that look for departures from the standard $\Lambda$CDM model as a way to reconcile the value of the Hubble constant inferred in these extended or exotic cosmologies with local estimates. Among these solutions, the possibility that neutrinos possess non-standard interactions offer an intriguing alternative. As seen in Sec.~\ref{sec:nusi_in_cmb} and Sec.~\ref{sec:nusi_in_matter}, free-streaming particles at the time of recombination modify the phase of the acoustic oscillations in the baryon-photon fluid. A change in the phase of the oscillations affects the position of the acoustic peaks in the CMB and matter power spectra, from which the angular scale of the sound horizon --- hence, $H_0$ --- is estimated.

In the presence of flavor-universal strong non-standard interactions (SI mode defined in Sec.~\ref{sec:nusi_in_cmb}), the phase shift is altered in such a dramatic way that a rather large departure from the $\Lambda$CDM best fit could be allowed in order to keep the theoretical predictions (mostly represented by the position of the acoustic peaks in the CMB spectra) in agreement with observations. Such a shift in cosmological parameters includes a larger value of the Hubble constant than the estimate obtained within the context of $\Lambda$CDM. Improved constraints of the position of the CMB peaks, and in general on the angular size of the sound horizon, are therefore key to assess the viability of \nusi to ease the Hubble tension. As mentioned in Sec.~\ref{sec:nusi_in_cmb}, when the full suite of CMB (temperature and polarization) and BAO data are combined, the preference for the SI mode --- and its ability to alleviate the $H_0$ tension --- becomes marginal with respect to the MI mode.
The relative weight of the two modes can be altered if one allows for non-universal flavor interactions, though the different cosmological phenomenology with respect to the flavor-universal case prevents the model from successfully tackling the $H_0$ tension. Further discussions can be found in Sec.~\ref{sec:flav_nonuni} below. 

\subsection{\texorpdfstring{$S_8$}{S8} Tension}\label{sec:nusi_in_s8}
Similar to the case of the Hubble parameter, there is a (much milder) tension between late- and early-time inferences of $S_8=\sigma_8 \sqrt{\Omega_m/0.3}$, a parameter that determines the amplitude of linear matter 
fluctuations on $8 h^{-1}$ Mpc scales. 
For example, Planck finds $S_8 = 0.825 \pm 0.011$~\cite{Planck:2018vyg}, while SZ cluster counts~\cite{Planck:2013lkt}, and galaxy surveys like KIDs~\cite{KiDS:2020suj} and DES~\cite{DES:2021wwk} obtain a lower value 0.76-0.78 (with slightly larger uncertainties), leading to a ${\sim}2 \sigma$ discrepancy.

Neutrinos have a variety of indirect effects on the amplitude of matter fluctuations, including modification of the initial amplitude and the evolution of gravitational potentials on certain scales (see Sec.~\ref{sec:nusi_in_matter}), as well as changing the inference of $\Omega_m$, and amplitude and spectral index of primordial curvature fluctuations. In Ref.~\cite{Kreisch:2019yzn} these effects combined to bring the early-time inference into better agreement with late-time measurements. However, the later analyses that included the final Planck likelihood with polarization and BAO data~\cite{Das:2020xke, RoyChoudhury:2020dmd,Brinckmann:2020bcn} found no such preference, and instead recovered $\Lambda$CDM-like values of $S_8$.

\subsection{Non-Universal Interactions}\label{sec:flav_nonuni}
Recent studies of BBN and laboratory experiments constraining neutrino self-interaction have generated interest in the cosmology of flavor-nonuniversal interaction~\cite{Das:2020xke,Brinckmann:2020bcn}. The strongly interacting mode requires a mediator below $\sim$ MeV scale for a coupling strength $g \lesssim 0.1$ in a four-Fermi interaction scenario. However, such MeV-scale mediator can be thermally produced in the early Universe during BBN changing the relic abundance of light elements. Additionally, new decay channels of $K$-meson, $Z$ decay width, and double-beta decays (see Section~\ref{sec:lab}) constrain all of the strongly and moderately interacting modes for flavor-universal case~\cite{Blinov:2019gcj,Lyu:2020lps}. It was shown that laboratory constraints allow strong self-interaction only for tau neutrinos. 

\begin{figure}
    \centering
    \includegraphics[width=0.5\textwidth]{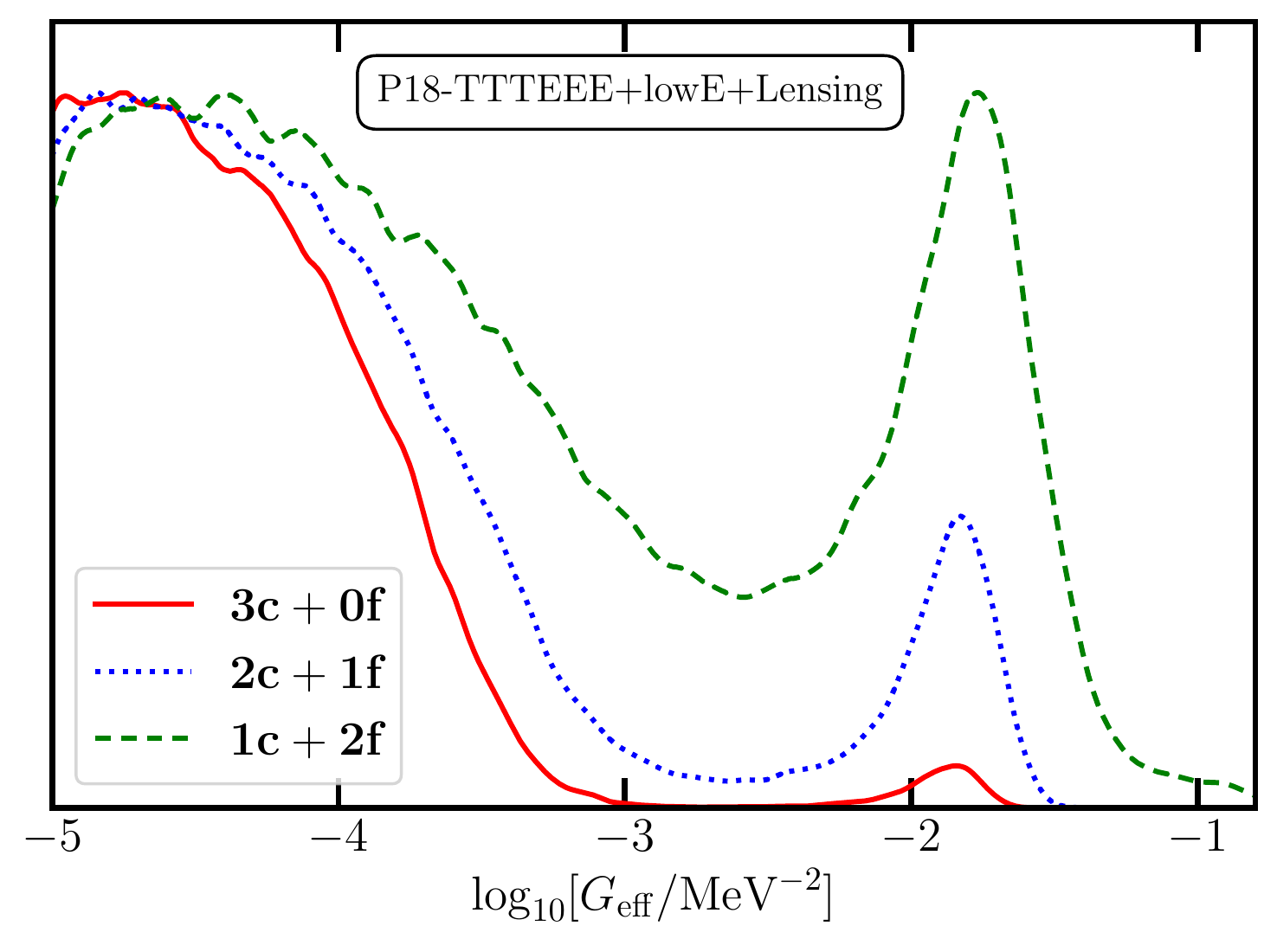}
    \caption{1D posterior of neutrino self-interaction strength from the Bayesian analysis using Planck dataset for both flavor universal and non-universal interaction from Ref.~\cite{Das:2020xke}. The $\mathbf{3c+0f}$ line corresponds to the flavor universal neutrino self-interaction, whereas, $\mathbf{2c+1f}$ and $\mathbf{1c+2f}$ stands for flavor non-universal interactions where respectively two and one neutrino flavors are self-interacting. The significance of the SI mode peak is highly enhanced in flavor non-universal scenario.}
    \label{fig:P18_flv_non_univ}
\end{figure}

Ref.~\cite{Das:2020xke} discusses the implications of the flavor-nonuniversal interaction in the cosmological data assuming massless neutrinos. The coupling between neutrino states is taken to be flavor diagonal. The constraint on the interaction strength from the Bayesian analysis using Planck data~\cite{Planck:2018vyg} is summarised in Fig.~\ref{fig:P18_flv_non_univ}. The significance of the strongly interacting mode discussed in the earlier section, increases dramatically in case of flavor non-universal interaction. In particular, the model with only one neutrino flavor self-interacting provides a similar fit to the Planck data as $\Lambda$CDM. The modification of the CMB spectrum compared to $\Lambda$CDM is milder when only one or two flavor of neutrinos are self-interacting compared to the flavor-universal scenario. This allows for increased flexibility for fitting the CMB spectrum which results in an increased significance for the SI mode.

Notably, the value of interaction strength corresponding to the SI modes does not change appreciably compared to the flavor-universal case. The interaction strength corresponding to the SI mode is such that it keeps neutrinos coupled until just before matter-radiation equality, and thus, affects all the CMB multipole with $\ell \gtrsim 100$. These changes in the CMB spectrum get compensated by variation of $\Lambda$CDM parameters specially the amplitude of the primordial perturbation $A_s$ and the spectral index $n_s$ (and effective number of neutrino degrees of freedom $N_{\rm eff}$ when it is varied). This also explains the existence of a valley region between the SI and MI mode peaks in the 1D posterior. For those values of the interaction strength, \nusi affects only the high-$\ell$ part of the CMB spectrum. In this case, the degeneracy with other parameters fails to fully compensate for those changes, resulting in a poor fit. Since the origin of the SI mode peak is tied to neutrino decoupling around matter-radiation equality --- regardless of the number of interacting flavors, the position of the SI mode peak remains virtually unchanged for flavor-nonuniversal interactions. 

The SI mode corresponds to a higher value of $H_0$, even when $N_{\rm eff}$ is kept fixed, by the virtue of acoustic phase shift of the CMB spectrum induced by the stopping of neutrino free streaming~\cite{Bashinsky:2003tk,Baumann:2015rya,Ghosh:2019tab}, as explained in the earlier section. In the flavor non-universal interaction, since fewer neutrinos are self-interacting, the phase shift compared to $\Lambda$CDM is smaller relative to flavor-universal interaction. Therefore, the increase of $H_0$ in flavor non-universal \nusi is smaller as well. This assessment remains true even when $N_{\rm eff}$ is allowed to vary.  Since SI mode in general allows for higher $H_0$, addition of the local Hubble measurement in the cosmological dataset increases of the significance of the SI mode.

While Ref.~\cite{Das:2020xke} focuses on the study of massless neutrino self-interaction, the case with massive neutrinos is studied in Ref.~\cite{Brinckmann:2020bcn}, 
which considered four cases, introducing a varying neutrino mass sum, a varying total $N_{\mathrm{eff}}$, and a varying interacting fraction of extra relativistic species:
\begin{itemize}
\item Case 1: Varying $N_{\mathrm{eff}}$, varying $\sum m_{\nu}$, all species interacting.\\
The strongly interacting mode is ruled out to high significance compared to a fully free-streaming comparison case when including Planck 2018 polarization data (in agreement with Ref.~\cite{RoyChoudhury:2020dmd}).
\item Case 2: Fixed $N_{\mathrm{eff, free-streaming}} \approx 2$, varying $N_{\mathrm{eff, interacting}}$, varying $\sum m_{\nu}$.\\
The strongly interacting mode persists with high significance compared to weaker interactions (the latter of which are preferred only due to parameter space volume considerations), but interactions are slightly disfavored compared to a fully free-streaming comparison case due to the complexity of the interacting model, despite a similar fit to the data.
\item Case 3: Fixed $N_{\mathrm{eff}} = 3.046$, varying $\sum m_{\nu}$, varying interacting fraction.\\
The strongly interacting mode is marginal and a weaker interaction is preferred, but still slightly disfavored compared to a fully free-streaming comparison case due to model complexity, despite a similar fit to the data. The bound on $N_{\mathrm{eff, interacting}} < 0.79$ at 68\%~CI ($< 2.34$ at 95\%~CI) is dominated by the weakly interacting mode and is much tighter for strongly interacting values, as can be seen by the bound on a species that never decouples $N_{\mathrm{eff, fluid}} < 0.28$ at 68\%~CI ($< 0.50$ at 95\%~CI), where the latter bound is in agreement with~\cite{Blinov:2020hmc}.
\item Case 4: Varying $N_{\mathrm{eff}}$, fixed $\sum m_{\nu}$, varying interacting fraction.\\
The only case where the data show a marginal preference for interactions over the fully free-streaming comparison case. In addition to the usual weakly and strongly interacting modes, an additional mode with a decoupling redshift around $z \approx 1000$ appears, with a slightly disfavored region between this recombination-era decoupling and the usual strongly interacting mode. For this case, the bound $N_{\mathrm{eff, interacting}} < 0.86$ (95\%~CI) remains mostly constant across most of the decoupling redshift/coupling strength parameter space, as illustrated by the bounds for the fluid case only being somewhat tighter at $N_{\mathrm{eff, fluid}} < 0.51$ (95\%~CI), with the latter bound in agreement with~\cite{Blinov:2020hmc}.
\end{itemize}
Common to all of these cases is that they do not help alleviate the $H_0$ tension once Planck 2018 polarization data is included and as such the hints and interacting modes for interacting neutrinos and extra relativistic species have to be considered on their own merits, rather than in the context of cosmological tensions.

In summary, the current cosmological datasets prefer strong flavor-nonuniversal self-interaction over the highly constrained flavor-universal interaction. Moreover, non-universal interaction provides similar fits to the cosmological data as $\Lambda$CDM. Further exploration in this direction, such as theoretical model-building for flavor-nonuniversal interaction, analysis with extended datasets, and estimates of the sensitivity of future experiments are needed to further understand the possible role of flavor-nonuniversal \nusi in cosmology.

\subsection{Big Bang Nucleosynthesis}
\label{sec:nusi_in_bbn}

In the standard picture described in the introduction to this section, neutrinos only experience
gravitational interactions during the period of free-streaming.  If neutrinos
were to experience other, currently unknown interactions, then the standard
picture would no longer capture all of neutrino dynamics~\cite{Blinov:2019gcj, Grohs:2020xxd, Huang:2021dba}.  If these unknown
interactions are confined to the neutrino sector, then the neutrino
distributions can equilibrate with a temperature different from the
photon-baryon fluid.  Instead of a system of decoupled particles each following
their own individual world-lines, the neutrino sector would act as an ideal gas
with a temperature and chemical potential.  In this non-standard picture, weak
decoupling and neutrino decoupling will occur at vastly different scales~\cite{Grohs:2020xxd}.  

The weak decoupling transition is not instantaneous in either the standard or
secret-interaction (\nusi) scenarios~\cite{Grohs:2015eua, Grohs:2015tfy}.  Entropy flows from the electromagnetic
components of the plasma into the neutrino seas over many Hubble times in the
early Universe.  For neutrinos experiencing secret interactions with large
cross sections, the heat and particle number flow as to maintain Fermi-Dirac (FD)
distributions with a temperature $T_{\nu}$ and chemical potential $\mu_\nu$.  The
equations governing the evolution of $T_{\nu}$ and degeneracy parameter
$\eta_{\nu}=\mu_\nu/T_{\nu}$ are
\begin{equation}\label{eq:dtnudt}
  \frac{d T_{\nu}}{dt} = -H T_{\nu} + T_{\nu}\frac{\displaystyle n,_{\eta}\frac{\partial\rho}{\partial t}\biggr|_a
  - 3T_{\nu} n\frac{\partial n}{\partial t}\biggr|_a}
  {\displaystyle 4\rho n,_\eta - 9 T_{\nu} n^2},
\end{equation}
\begin{equation}\label{eq:detanudt}
  \frac{d \eta_{\nu} }{dt} = \frac{\displaystyle 4\rho\frac{\partial n}{\partial t}\biggr|_a
  - 3n\frac{\partial\rho}{\partial t}\biggr|_a}
  {\displaystyle 4\rho n,_\eta - 9 T_{\nu} n^2},
\end{equation}
where $H$ is the Hubble expansion rate, $n$ is the neutrino number density,
$\rho$ is the neutrino energy density, $\partial n/\partial t|_a$ is the number
density added from out-of-equilibrium weak decoupling, $\partial\rho/\partial
t|_a$ is the energy density added, and $n,_\eta$ is the following expression
\begin{equation}\label{eq:n_eta}
  n,_\eta = 3\frac{T_{\nu}^3}{\pi^2}\int d\epsilon\frac{\epsilon}{e^\epsilon + 1}.
\end{equation}
The dummy variable $\epsilon=E_\nu/T_{\rm cm}$ is a dimensionless quantity used to
index the neutrino distributions, $T_{\rm cm}$ is an energy scale which redshifts with
increasing scale factor, ensuring $\epsilon$ is a comoving invariant.  The
above equations are phenomenological and independent on the kind of secret
interaction.  The secret interaction strength is assumed to be large enough to
ensure equilibrium throughout weak decoupling.  In addition, Eqs.\
\eqref{eq:dtnudt}--\eqref{eq:n_eta} only assume one unique temperature and
degeneracy parameter for three flavors of left-handed neutrinos and similarly
three flavors of right-handed anti-neutrinos.  In the case that all three
flavors have unique temperatures and degeneracy parameters, the equations are
similar, with the inclusion of terms for the heat and particle flow between
neutrinos of different flavor.
\begin{figure}[t!]
   \begin{center}
   \includegraphics[scale=0.5]{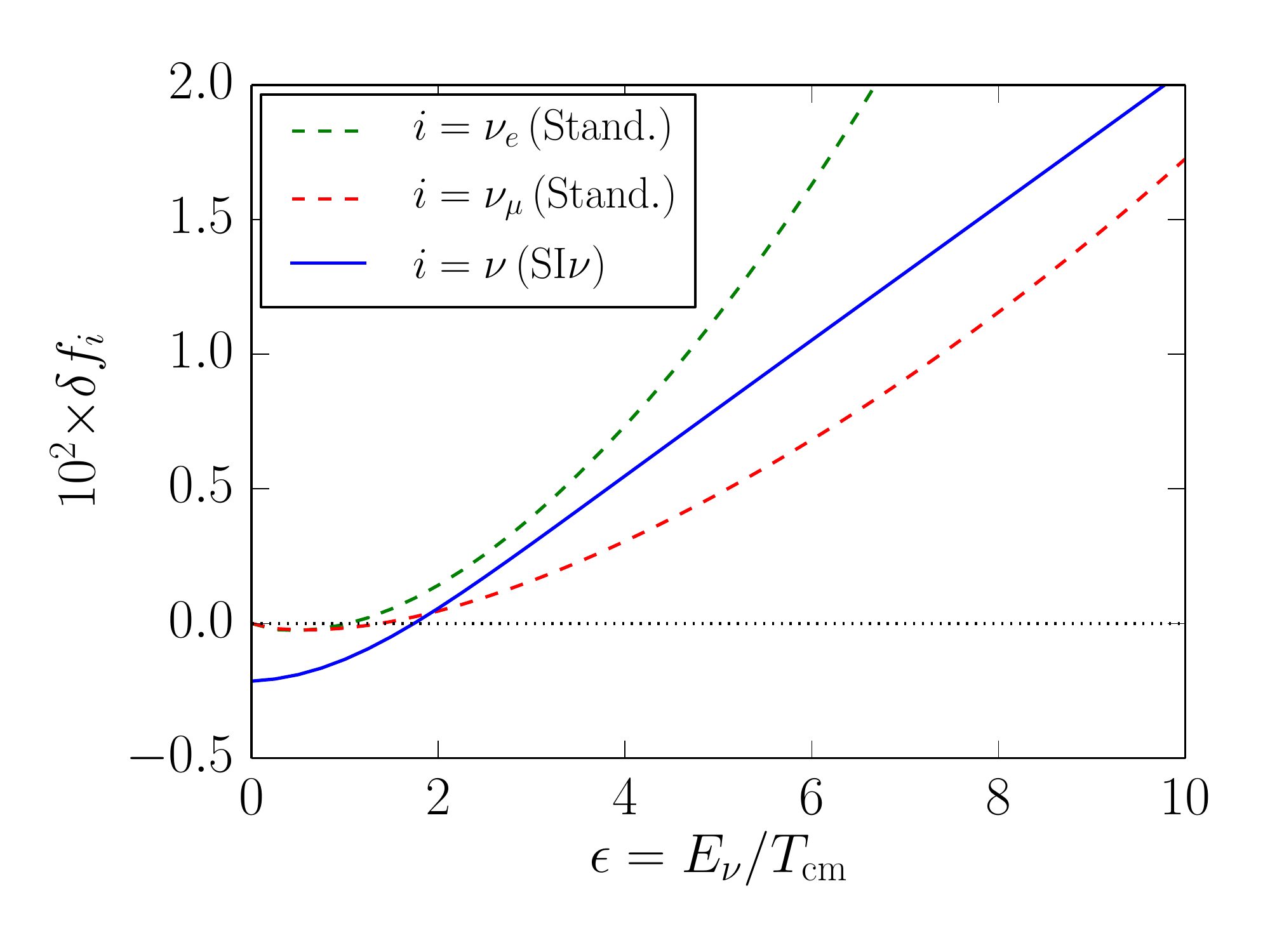}
   \caption{\label{fig:docc} Freeze-out neutrino spectra in the standard
   scenario (dashed) and the self-interaction scenario (solid) plotted against $\epsilon$. In the \nusi model the self-interaction is so large that neutrinos maintain equilibrium distribution until well after weak decoupling. In the standard scenario of Boltzmann transport, the green dashed curve is the
   $e$ flavor and the red dashed curve is either $\mu$ or $\tau$.  The vertical axis
   is the relative change to a non-degenerate FD spectrum. Plot from Ref.~\cite{Grohs:2020xxd}.}
   \end{center}
\end{figure}
Keeping neutrinos coupled to one another slightly changes the neutrino
distributions compared to the result from neutrino transport in the standard
picture.  Fig.~\ref{fig:docc} shows the relative change in the occupation
numbers plotted against $\epsilon$ for the two different physical scenarios at
the conclusion of weak decoupling.  The dashed green and red lines show the
out-of-equilibrium distributions using Boltzmann neutrino transport and
ignoring the effects of neutrino oscillations.  The solid blue line is a
calculation with self-coupled neutrinos.  That distribution also has a FD
spectrum, but with the following parameters:
\begin{align}
  \frac{T_{\nu}}{T_{\rm cm}}-1 &= 2.463\times10^{-3},\\
  \eta_{\nu} &= -4.282\times10^{-3},
\end{align}
showing that the neutrinos have been slightly warmed by weak decoupling.  The
negative degeneracy parameter shows that neutrinos are less numerous than
expected for the given temperature.

With the inclusion of \nusi, there is a larger flow of heat from
the electromagnetic plasma into the neutrino seas than the standard picture.
As the curves in Fig.\ \ref{fig:docc} are all of the same scale, this heat flow
is only slightly larger with the inclusion of secret interactions.  When using
the quantity $N_{\rm eff}$ to parameterize the early-Universe radiation energy density,
the relative change in the secret-interaction scenario compared to the standard
picture is
\begin{equation}
  \delta N_{\rm eff} \simeq 3\times10^{-4},\label{eq:neff_sinu}
\end{equation}
 and well within uncertainties for parameter estimation.  Furthermore, the
change induced in the electron-flavor neutrino (and anti-neutrino) spectrum
changes the integration of the neutron-to-proton interconversion rates.  These
rates set the neutron-to-proton ratio for BBN.  Similar to $N_{\rm eff}$, the secret
interactions only slightly modify the primordial abundances of helium-4 ($Y_{\rm P}$)
and deuterium (${\rm D/H}$)  
\begin{align}
  & \delta Y_{\rm P} \simeq 4\times10^{-4},\label{eq:yp_sinu}\\
  & \delta(  {\rm D/H} ) \simeq 2\times10^{-4},\label{eq:dh_sinu}
\end{align}
and are well within observational precision of these quantities.

The above comparisons are between a secret-interaction scenario and the
standard picture of BBN.  There exist two simple extensions to BBN, namely,
including a dark radiation component~\cite{Mukohyama:1999qx} and a lepton
number asymmetry~\cite{Grohs:2016cuu}.  If there exists a dark-radiation
component (with no coupling to neutrinos through any other interaction besides
gravitation), then secret interactions will only slightly modify the freeze-out
neutrino distributions and resulting cosmological parameters and observables.
The picture is the same for a non-zero lepton number. In this scenario, the
secret interaction must conserve lepton-number for the asymmetry to persist
into BBN.  At that point, neutrinos and anti-neutrinos have similar
temperatures but manifestly different degeneracy parameters.  Nevertheless, the
changes in $N_{\rm eff}$ and the primordial abundances are primarily influenced by the
initial asymmetry conditions and secret interactions add in effects at higher
precision. 

Although neutrinos play a pivotal role in the dynamics of the early Universe,
the requirement of a self-coupled neutrino gas does not fundamentally change
the dynamics of weak decoupling during BBN.  Changes in cosmological parameters
and observables are small and well within experimental precision when including
secret interactions which only maintain equilibrium among the neutrinos.  These
results do not reference a particular phenomenological model for the secret
interactions.  Given that weak decoupling does not provide any meaningful
constraints on secret interactions, the field is open for the consideration of
more detailed models.  For example, if the secret interaction is mediated by an
unknown particle, that particle could exist in appreciable numbers with some
modified form of a thermal distribution.  The particle will have couplings with
neutrinos and could distort the neutrino distributions during weak decoupling,
depending on the mass and coupling strength.  An influx of energy into the
neutrino sector would reverse the flow of entropy in weak decoupling, thereby
changing the dynamics of the electromagnetic plasma and influencing the
primordial abundances.  Other models may require flavor-dependent couplings or
a CP-asymmetry.  In any case, the early Universe provides a unique laboratory for neutrino secret interactions.

\subsection{Inflation}
\label{sec:nusi_for_inflation}
\begin{figure}
\centering
\includegraphics[width=0.47\textwidth]{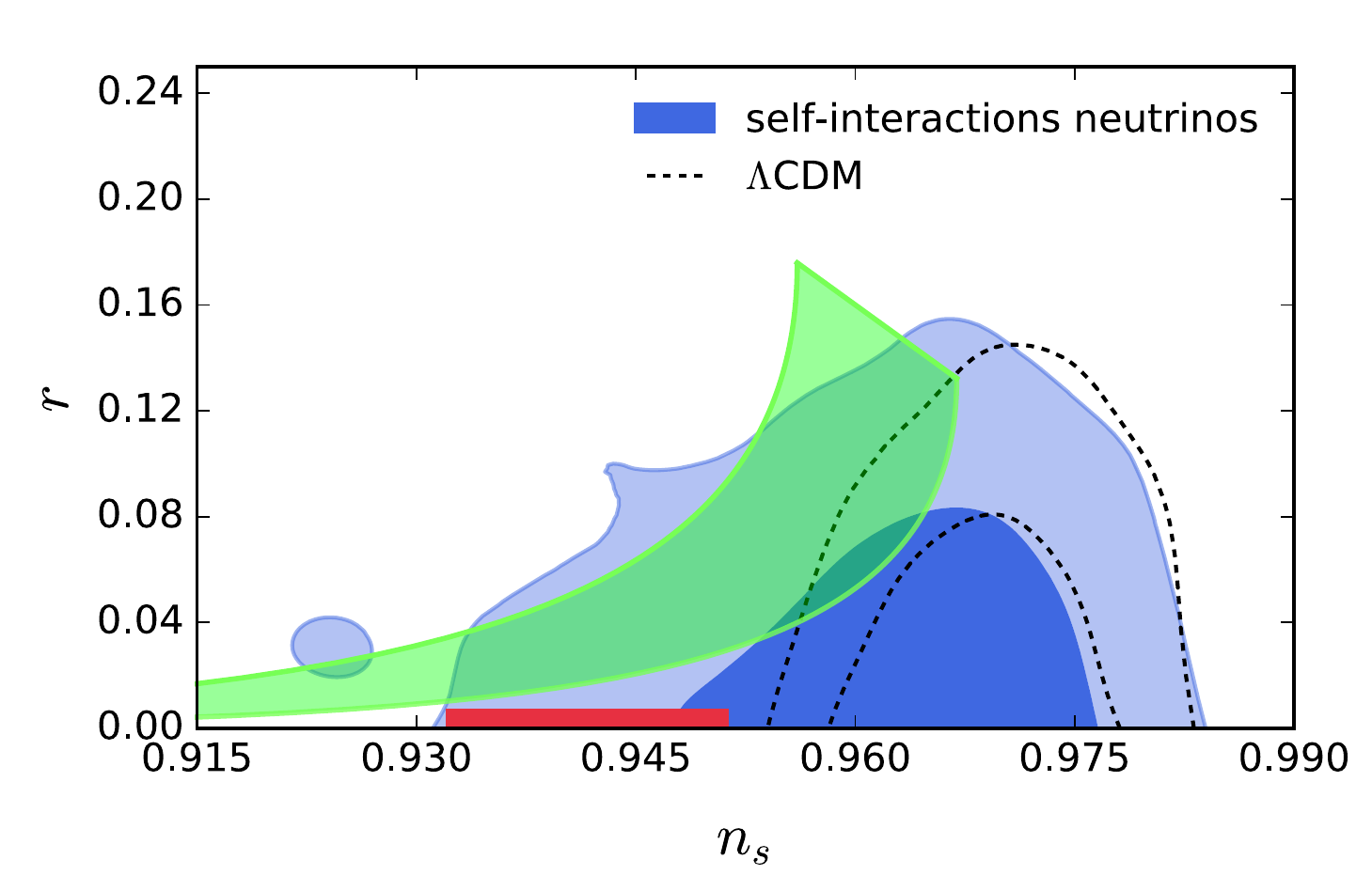}
\includegraphics[width=0.47\textwidth]{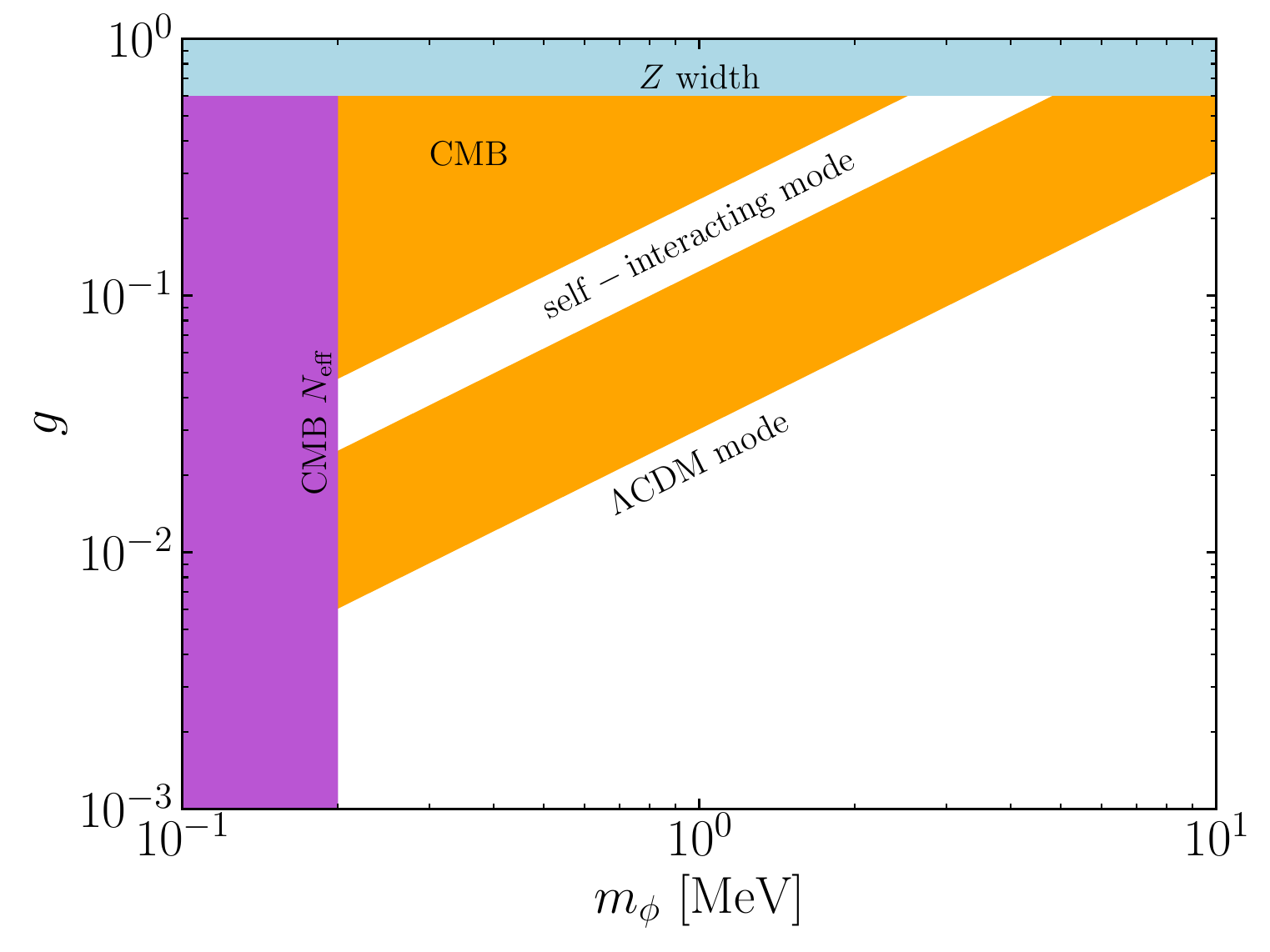}
\caption{Left: Comparison of the predictions of Natural Inflation (green) and Coleman-Weinberg inflation (red) with the 68\% (dark blue) and 95\% (light blue) confidence regions for the interacting neutrino mode. Dashed lines mark the confidence regions obtained within the standard $\Lambda$CDM framework assuming free-streaming neutrinos ($G_{\mathrm{eff}}=0$). 
Right: The constraints on a neutrino self interaction; the region that relaxes the inflation constraints is shown in white labeled ``self-interacting mode.''
The orange region is disfavored from CMB data \cite{Barenboim:2019tux}, the purple region is disfavored from BBN data \cite{Huang:2017egl}, and the blue region is disfavored from measurements of the $Z$ width \cite{Bilenky:1992xn,Laha:2013xua}.
Both figures are from Ref.~\cite{Barenboim:2019tux}.}
\label{fig:rvsns} 
\end{figure}

Two well-motivated inflation models are Natural Inflation \cite{Freese:1990rb} (employing a shift-symmetric inflaton with a naturally flat potential) and Coleman-Weinberg inflation \cite{Linde:1981mu,Albrecht:1982mp} (which makes use of a guaranteed 1-loop contribution to the inflaton potential).
Natural inflation generally predicts non-zero and measurable tensor modes in the CMB, while Coleman-Weinberg inflation predicts no tensor modes, but a somewhat low spectral index, $n_s$.
Past constraints on $n_s$ and the ratio of tensor to scalar modes, $r$, have put these models of inflation in moderate tension with the data \cite{Barenboim:2013wra}.
These models are only disfavored in the vanilla $\Lambda$CDM model.
In Ref.~\cite{Barenboim:2019tux}, it was shown that these constraints can be significantly relaxed in the presence of a new neutrino self-interaction which allows both of these models at $<1\,\sigma$ again; see the left panel of Fig.~\ref{fig:rvsns}. This is because in the presence of non-standard \nusi, the posterior distributions of $r$ and $n_s$ are significantly different from those in $\Lambda$CDM, to compensate for the modifications of the power spectrum described in Sec.~\ref{sec:nusi_in_cmb}.

This scenario works for a new scalar interaction with $g\sim0.1$ and $m_\phi\sim1$ MeV; see the right panel of Fig.~\ref{fig:rvsns}.
In addition, this preferred region overlaps with the parameter space which could also partially alleviate the $H_0$ tension (see Section~\ref{sec:H0_tension} and Ref.~\cite{Kreisch:2019yzn}).

The parameters relevant to relax the constraints on inflation models can be tested elsewhere.
Mediators in the range $m_\phi\in[0.2,5]$ MeV are exactly the region of interest that IceCube is sensitive to via high-energy neutrinos scattering off the C$\nu$B if the astrophysical uncertainties can be overcome \cite{Barenboim:2019tux}.
This interaction is ruled out by kaon decay measurements \cite{Blinov:2019gcj} unless the interaction is only in the tau neutrino sector where it is still viable.

\section{Astrophysical Probes}\label{sec:astro}
Neutrinos emitted by Galactic and extragalactic astrophysical sources provide tests of BSM neutrino self-interactions that are complementary to cosmological and laboratory-based searches. Their probing power stems primarily from their very long baselines, of tens of kpc for Galactic neutrinos and of Mpc--Gpc for extragalactic neutrinos.  While propagating across these vast distances, astrophysical neutrinos may have a significant chance of scattering off the background of low-energy ($\sim$0.1~meV) relic neutrinos, even if the neutrino-neutrino coupling strength is feeble.  The scattering may affect the energy spectrum, flavor composition, and arrival times/directions of the astrophysical neutrinos in characteristic and potentially detectable ways.  Notably, if the neutrino self-interaction is resonant, it may introduce dips in the astrophysical neutrino energy spectrum around the resonance energy, and a pile-up of neutrinos at lower energies. 

Previous works have studied the effects of BSM self-interactions in neutrinos from core-collapse supernovae (SNe) and in high-energy extragalactic astrophysical neutrinos.   Neutrinos from core-collapse supernovae, with energies of up to a few tens of MeV, are sensitive to neutrino self-interactions via mediators with keV-scale masses, when they occur during propagation~\cite{Kolb:1987qy,Farzan:2014gza,Dighe:2017sur,Shalgar:2019rqe}, or MeV-scale masses, when they occur in the SN core and affect the explosion mechanism~\cite{Shalgar:2019rqe} and flavor conversions in the core~\cite{Dighe:2017sur}.  High-energy extragalactic neutrinos, with energies of TeV--PeV, are sensitive to MeV-scale mediator masses~\cite{Ioka:2014kca,Ng:2014pca,Ibe:2014pja,Kamada:2015era,DiFranzo:2015qea,Kelly:2018tyg,Murase:2019xqi,Bustamante:2020mep}.  In both cases, the effects of BSM self-interactions may be detectable in the flux of neutrinos from a single astrophysical source, or in the diffuse flux from a population of sources.  

Studying the effect of BSM self-interactions on astrophysical neutrinos today is timely, in preparation for the imminent detection of the next Galactic core-collapse SN, the discovery of the diffuse supernova neutrino background, the detection of more TeV--PeV neutrinos, and the discovery of EeV cosmogenic neutrinos in existing and envisioned neutrino telescopes.

We separate the discussion of these phenomena based on observations of SNe (Section~\ref{sec:astro:SNe}) and those from high/ultra-high energy astrophysical neutrinos (Section~\ref{subsec:astro:UHENeutrinos}). Additionally, in Section~\ref{sec:astro:snsi} we discuss how many of these searches are modified when \textit{both} \nusi and sterile neutrinos (which also self-interact) are added to the SM.  Figure~\ref{fig:nusi_constraints} summarizes some of the \nusi constraints coming from astrophysics.

\subsection{Core-Collapse Supernovae}\label{sec:astro:SNe}
Core-collapse supernovae (CCSNe) are host to a huge swath of neutrinos streaming out from the core, eventually leading to cooling of the progenitor. It is estimated that almost 99$\%$ of the binding energy of the SN is released in the form of neutrinos. The huge density of emitted neutrinos naturally makes a SN an ideal astrophysical laboratory to test non-standard \nusi. Bounds on self-interactions from SN neutrinos can be divided broadly into two categories: (i) direct bounds arising from the observation of $\mathcal{O}(30)$ neutrinos from SN1987A over a period of approximately $10\,$s~\cite{Bionta:1987qt,Hirata:1988ad, Bratton:1988ww, Alekseev:1988gp}, and (ii) changes in the SN neutrino flux and spectra due to introduction of new physics in the neutrino sector, as compared to those predicted by simulations.

Our current favored understanding of the explosion mechanism of a SN follows the neutrino-driven delayed explosion scenario~\cite{Janka:2017vcp,Janka:2006fh}, where neutrinos emitted from the SN core transfer enough energy to the stalled shock wave to cause a successful explosion. Large non-standard neutrino self-interactions in the SN core of the type, $\nu \nu \rightarrow \nu \nu$, can give rise to $2\nu \rightarrow 4\nu$ interactions at the next-to-leading order~\cite{Shalgar:2019rqe}. This could result in a net reduction of the energy transferred by the neutrinos to the stalled shockwave, thereby halting the explosion completely. The very fact that we have observed neutrinos from SN1987A can be used to constrain stronger-than-weak non-standard neutrino-interactions in all flavors. If the new physics mediating neutrino self-interactions is lepton-number violating, e.g., like a Majoron model~\cite{Gelmini:1980re}, then as the neutrinos get trapped in the core, number-conserving and number-changing processes can lead to a thermal equilibrium among the neutrinos, whereas chemical equilibrium with the charged leptons is obtained through weak interactions only. If the new interactions are stronger than weak interactions, this can lead to the thermal equilibrium happening on time scales much shorter than weak interactions. This can generate extra entropy in the SN core, thereby leading to a core-bounce at sub-nuclear densities. This will affect the neutrino spectra, and can be used to constrain neutrino self-interactions~\cite{Fuller:1988ega}. Furthermore, depending on the mass, the mediator particle can also be produced on-shell within the SN core~\cite{Kachelriess:2000qc,Farzan:2002wx}. If the mediator mass and coupling to the neutrinos are in the right ballpark values, it can cause the SN to cool faster than what is expected. This leads to additional bounds on these couplings~\cite{Berryman:2018ogk}. This argument has been used to constrain all kinds of new light scale physics, coupled to neutrinos (e.g., see Ref.~\cite{Escudero:2019gzq}).  

Flavor evolution of neutrinos within a SN is driven mostly by their self-interactions, and hence is highly sensitive to new physics in that sector. Deep inside the SN envelope, the neutrino density is high enough that neutrinos feel a potential due to their interactions with the ambient neutrinos. Such neutrino-neutrino self-interactions in a dense neutrino gas can lead to non-linear collective flavor conversions, thereby influencing their flavor evolution, and hence the emitted spectra~\cite{Duan:2006jv}. It was demonstrated using simple toy models that secret self-interactions of neutrinos, stronger than weak interactions, can drastically alter our understanding of neutrino flavor evolution~\cite{Blennow:2008er, Das:2017iuj}. However, the impact of such collective oscillations is still uncertain due to the possibility of flavor decoherence, particularly caused due to neutrinos moving in different trajectories (known in the community as multi-angle effects)~\cite{Chakraborty:2011nf}. Recent theoretical developments have also shown that neutrino non-standard self-interactions can also cause fast flavor conversions of neutrinos --- a phenomenon where the neutrino ensemble undergoes rapid flavor conversions just outside the neutrinosphere~\cite{Dighe:2017sur}. 
This results in altering the flavor content of the neutrinos, and consequently the neutron-proton ratio inside the SN. As a consequence, these secret interactions can have direct consequences for the SN explosion mechanism, as well as nucleosynthesis of heavy elements. This is a topic of intense investigation, and the final word is yet to be determined.

\begin{figure*}[t!]
 \centering
 \includegraphics[width=0.6\textwidth]{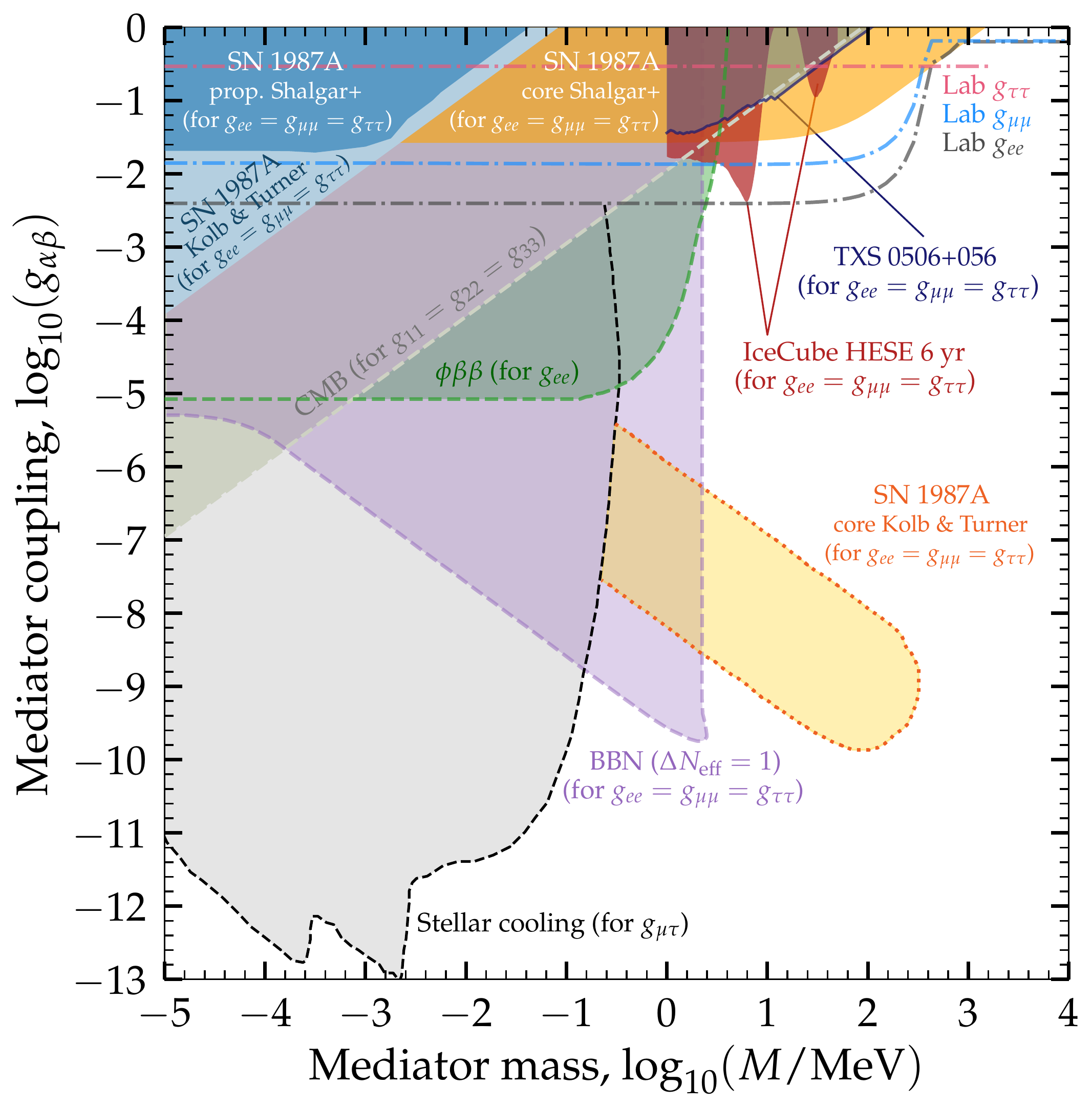}
 \caption{\label{fig:nusi_constraints}Limits on the coupling strength of neutrino self-interactions, $g_{\alpha\beta}$, as a function of the mediator mass, $M$.  We show limits from the propagation of neutrinos from SN 1987A~\cite{Kolb:1987qy, Shalgar:2019rqe}, from inside the SN 1987A core~\cite{Kolb:1987qy, Shalgar:2019rqe}, CMB~\cite{Archidiacono:2013dua} (see also Ref.~\cite{Escudero:2019gvw}), BBN~\cite{Blinov:2019gcj}, laboratory measurements of particle decays~\cite{Berryman:2018ogk}, double beta decay ($\phi\beta\beta$)~\cite{Brune:2018sab}, stellar cooling~\cite{Escudero:2019gzq}, IceCube High Energy Starting Events (HESE)~\cite{Bustamante:2020mep}, and a high-energy neutrino detected by IceCube from the blazar TXS 0506+056~\cite{Kelly:2018tyg}. Figure modified from Ref.~\cite{Bustamante:2020mep}.}
\end{figure*}

Finally, \nusi can also manifest themselves through the interactions of the SN neutrinos with the cosmic neutrino background (C$\nu$B). During propagation, neutrinos from a SN can scatter with the C$\nu$B and lose energy, and/or get deflected causing a time delay. Knowing the distance at which the SN occurred (e.g., SN1987A occurred roughly at a distance of 50 kpc in the Large Magellanic Cloud), one can estimate the time delay, and change in spectral shape due to these scatterings to put tight constraints on \nusi~\cite{Kolb:1987qy, Shalgar:2019rqe,Murase:2019xqi}. The same operator giving rise to \nusi can also cause neutrinos to decay if the mediator is light enough. This can cause spectral distortions, which can be used to put bounds on such couplings, and consequently on \nusi~\cite{Ando:2004qe, deGouvea:2019goq}. Refs.~\cite{Farzan:2014gza, Jeong:2018yts} explored the possibility of sterile neutrinos undergoing secret interactions on the diffuse supernova neutrino background. A collection of various limits on the \nusi coupling, as discussed, is depicted in Fig.\,\ref{fig:nusi_constraints}. The possible observation of neutrinos from a CCSN at similar/closer distances to SN1987A with current/next-generation neutrino experiments promises exciting prospects for extending these searches across this parameter space.

\subsection{High-Energy and Ultra-High-Energy Neutrinos}\label{subsec:astro:UHENeutrinos}
Complementary to neutrinos emitted in SNe, many astrophysical sources (across a vast span of distances from Earth) exist that produce neutrinos across orders of magnitudes of energies. With the advent of gigaton-scale neutrino telescopes (such as IceCube~\cite{IceCube:2013cdw} and ANTARES~\cite{ANTARES:2011hfw}), more and more neutrinos from these sources are being observed and being identified as extragalactic. Even with these first observations, fundamental properties of neutrinos are capable of being explored with unprecedented precision. Moreover, future neutrino telescope proposals, including Baikal-GVD~\cite{Baikal-GVD:2019fko}, KM3NeT~\cite{KM3Net:2016zxf}, P-ONE~\cite{P-ONE:2020ljt}, TAMBO~\cite{Romero-Wolf:2020pzh}, IceCube-Gen2~\cite{IceCube-Gen2:2020qha}, and their combination~\cite{Schumacher:2021hhm}, are capable of even deeper understanding of neutrino properties. By measuring the properties of these astrophysical neutrinos~\cite{Song:2020nfh}, we have the opportunity to combine laboratory- and astrophysics-based studies of neutrinos to search for physics beyond the standard model, including the possibility of \nusi.

\subsubsection{Absorption Effects and Neutrino Telescopes}
Astrophysical neutrinos probe the interactions that occur during propagation and detection, in addition to the source physics. Since the cosmic neutrino background (C$\nu$B) permeates the Universe, neutrino self-interactions can lead to distortions in the spectrum as ``signal" neutrinos scatter from C$\nu$B neutrinos. This idea goes back to SN 1987A, where Kolb \& Turner placed limits on neutrino self-interactions \cite{Kolb:1987qy}. Later work built on this, considering sources such as the diffuse supernova neutrino background (DSNB) and high-energy astrophysical neutrinos, and a variety of neutrino self-interaction scenarios \cite{Baker:2006gm,Ng:2014pca,Ibe:2014pja,Blum:2014ewa,Jeong:2018yts,Barenboim:2019tux,Shalgar:2019rqe,Bustamante:2020mep,Creque-Sarbinowski:2020qhz,Esteban:2021tub}.

Qualitatively, the effect of self-interactions on a spectrum can be understood by considering a collision between a signal neutrino and a nonrelativistic C$\nu$B neutrino. The result of the scattering will be two outgoing neutrinos, generally at higher energy than the C$\nu$B neutrino and lower energy than the signal neutrino. This can be described as removing neutrinos from the spectrum at energies where the scattering cross section is relatively large and injecting neutrinos at a lower energy. In practice, the accumulation of neutrinos at the low-energy end of the spectrum is often below the detector threshold, so the higher-energy dips are more phenomenologically significant. The exact form of the resulting spectrum depends on the details of the neutrino self-interactions and the source. For example, with a massive scalar mediator there is a resonant energy that depends on the mediator mass, and this sets the scale for the location of dips in the spectrum. For sources distributed over cosmological distances, redshift of neutrino energies during propagation is important.

\begin{figure}[hbtp]
    \centering
    \includegraphics[width=0.5\textwidth]{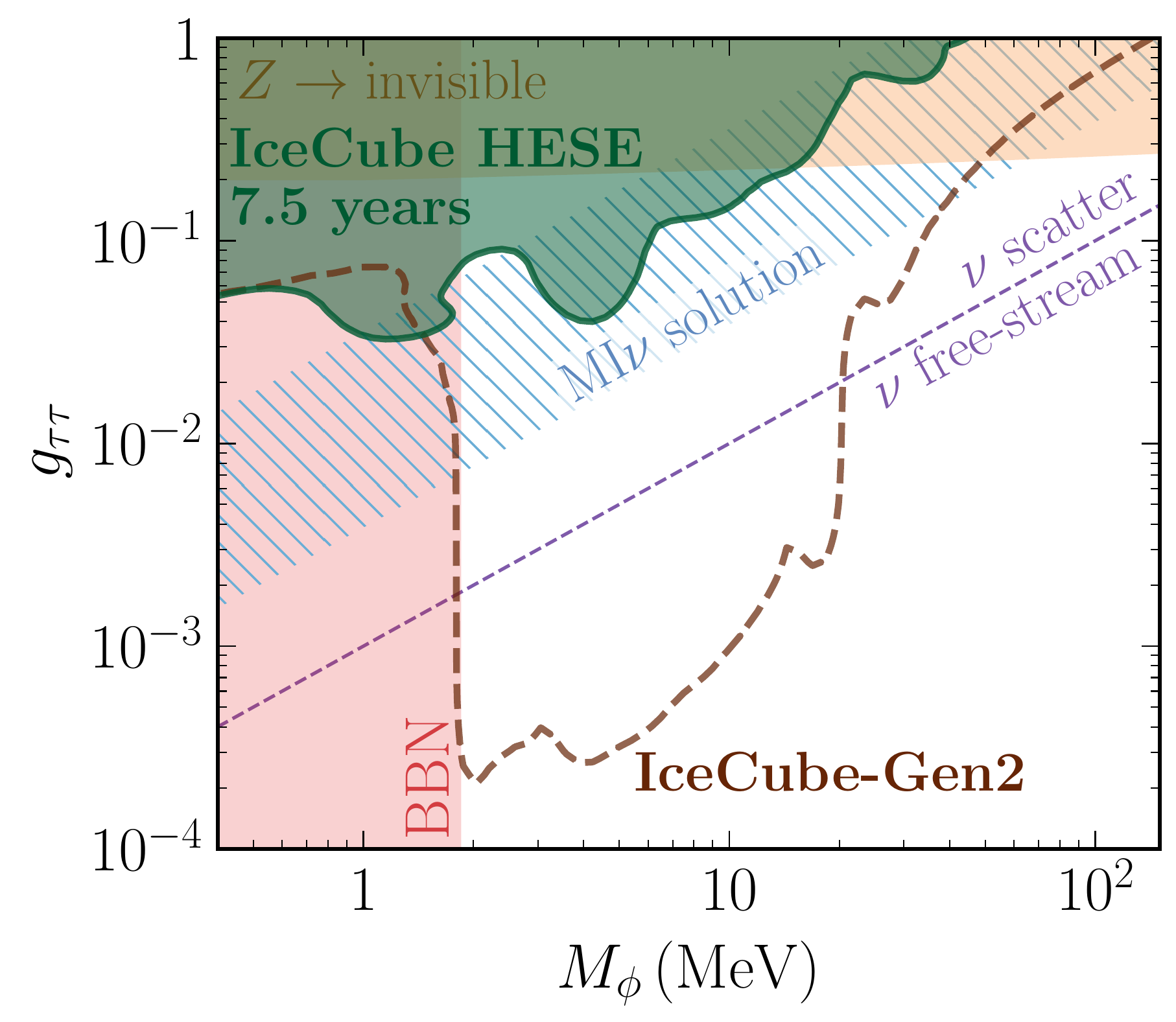}
    \caption{Present constraints on neutrino self-interactions from the IceCube HESE diffuse neutrino flux [solid green], together with the Gen2 sensitivity [dashed]. Also shown is the Moderately Interacting neutrino mode (labelled as ``MI$\nu$ solution'') and, more generically, the scale above which neutrinos self-scatter in the Early Universe at times relevant for CMB observations. The constraints here assume self-interactions only in the $\nu_\tau$ sector to avoid strong constraints from $K$ decay, but comparable sensitivity should apply to all flavors. Figure from Ref.~\cite{Esteban:2021tub}.}
    \label{fig:bounds_Gen2}
\end{figure}

In addition to mediator mass, the position of dips, particularly their separation in energy, depends on neutrino masses. Since the C$\nu$B neutrinos are non-relativistic, the center-of-mass energy of the HE/C$\nu$B scattering is $s = 2E_\nu m_\nu$, where $E_\nu$ is the HE$\nu$ energy. For resonant scattering with a neutrinophilic mediator $\phi$, this resonance occurs when $s = m_\phi^2$ or $E_\nu = m_\phi^2/(2m_\nu)$. The cosmological bound of $\sum m_{\nu} \lesssim 0.12$~eV, together with the measured neutrino squared mass splitting, disfavors neutrino mass degeneracy. As such there should be at least two observable dips in the neutrino spectrum at IceCube, given the mass of two eigenstates ($\nu_1$, $\nu_2$) are close to each other. In case of normal ordering (NO, $m_1< m_2 < m_3$, i.e., one heavy and two light states) the mass of the lightest state is greater than that compared to Inverted Ordering (IO, $m_3< m_1 < m_2$, i.e., one light and two heavy states). Therefore,  keeping $\sum m_{\nu} \sim 0.12$~eV, the dips in case of NO are a factor of $\mathcal{O}(1)$ apart in energy, whereas for IO the dips are a factor of $\mathcal{O}(100)$ apart in energy~\cite{Esteban:2021tub}. As NO is slightly favored over IO, Fig.~\ref{fig:bounds_Gen2} depicts the \nusi constraints considering NO. 

Detecting high-energy astrophysical neutrinos with IceCube opened a novel window into exploring \nusi. There exist, however, astrophysical uncertainties related to the unknown normalization and spectrum of the primary neutrino flux. Two phenomenological approaches have been suggested in the literature to overcome them. On the one hand, astrophysical acceleration mechanisms typically predict a power-law spectrum. Self-interactions, specially if they are resonantly enhanced at a particular energy, would introduce spectral deviations such as dips or bumps that can be looked for. For recent analyses in this direction using IceCube data, see Refs.~\cite{Bustamante:2020mep,Esteban:2021tub}.

On the other hand, if high-energy neutrinos can be associated to an astrophysical source, and the luminosity of this source is inferred by other means, one can estimate the primary neutrino luminosity. If it is consistent with observations, self-interactions leading to a mean free path smaller than the distance to the source can be excluded. Despite relying on astrophysical assumptions, this method can yield competitive constraints with rather small statistics, see Ref.~\cite{Kelly:2018tyg}. 

IceCube data available up to now has provided unique constraints that, unfortunately, are typically weaker than laboratory probes. The future, however, is very promising. Fig.~\ref{fig:bounds_Gen2} shows the reach of the proposed IceCube-Gen2~\cite{IceCube-Gen2:2020qha} neutrino observatory. The impressive statistics of this detector, combined with the high-energy reach that will robustly determine the underlying spectrum, will increase the current sensitivity of IceCube by three orders of magnitude. It will even overcome laboratory constraints for a large mediator mass range. Furthermore, the sensitivity in that figure is only based on analyzing the all-flavor diffuse flux. Exploring flavor effects can bring about interesting complementarity with flavor-dependent self-interactions. And, in addition, Gen2 is expected to detect many neutrino point sources~\cite{IceCube-Gen2:2020qha}. As discussed above, this will open a complementary channel.

This program will be boosted by any detection of Ultra-High Energy neutrinos, with energies of 100~PeV and higher. Although very challenging to detect, their higher energy will generically imply probing even heavier mediators. Altogether, future astrophysical neutrino data will robustly explore uncharted territory, where any hint will be cross-checked with complementary probes.

\subsubsection{Neutrino Echoes}
\begin{figure}[hbtp]
\centering
\includegraphics[width=0.54\textwidth]{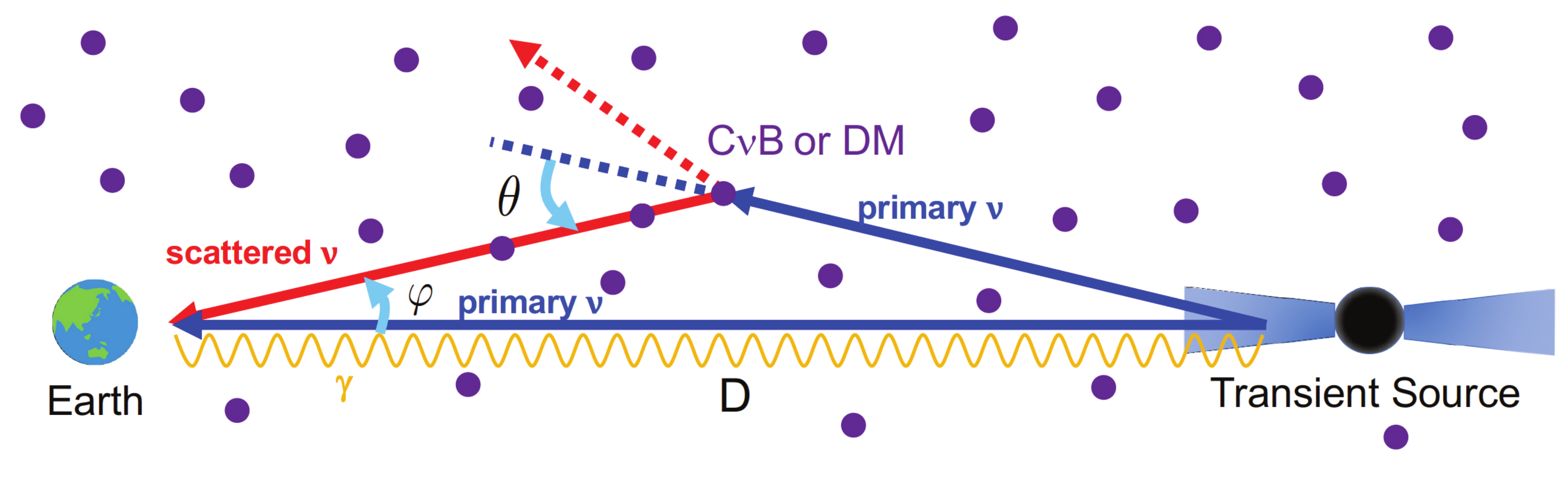}  
\includegraphics[width=0.45\textwidth]{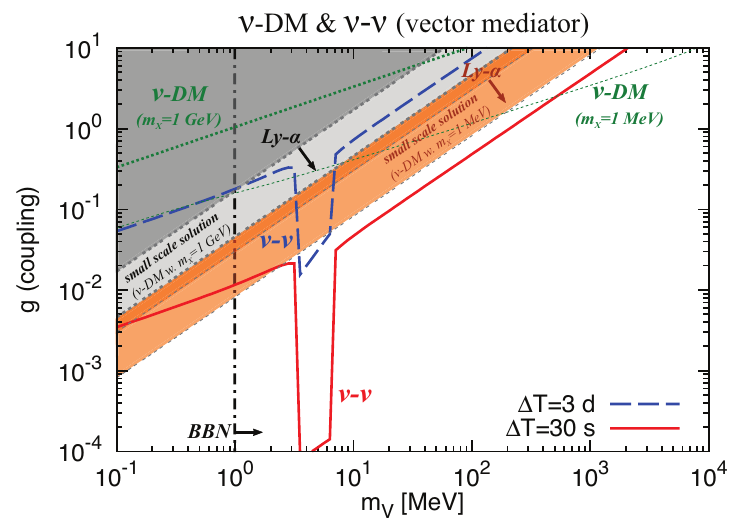}   
    \caption{(Left panel) Schematic picture of neutrino “echoes” induced by
BSM interactions, reproduced from~\cite{Murase:2019xqi}. (Right panel) Constraints on a vector model of neutrino self-interactions in the coupling versus mediator mass plane.  The transient distance and neutrino
mass are $D =$ 3 Gpc and $m_{\nu}$ = 0.1 eV, respectively. }
    \label{fig:echo}
\end{figure}

In addition to spectral and flavor distortions of the astrophysical neutrino flux~\cite{Ioka:2014kca, Ng:2014pca,Cherry:2014xra,Shoemaker:2015qul,Cherry:2016jol}, neutrino scattering on the C$\nu$B can also produce distinct temporal signatures~\cite{Murase:2019xqi}. As depicted in Fig.~\ref{fig:echo}, these neutrino ``echoes'' are produced via the lengthened path scattered neutrinos follow en route to the Earth. Neutrino-bright multimessenger transient sources will then allow for the sensitivity to a time delay induced by the neutrino self-interaction.  

\subsection{Sterile Neutrino Secret Interactions}\label{sec:astro:snsi}
Neutrino secret interactions may also couple active neutrinos of the Standard Model with additional sterile (non-weakly-interacting) species. Indeed, sterile neutrinos have been proposed at various energy scales, from the eV to the Grand Unification scale at $10^{15}$~GeV. Such additional species are motivated either by theoretical expectations, for example in the case of the seesaw mechanism and leptogenesis  with a heavy sterile neutrino, or by experimental observation, as light sterile neutrinos at masses of  eV and keV which could explain neutrino oscillation anomalies and dark matter, respectively.

Secret interactions between active and sterile neutrinos can be probed by similar means as active-active secret interactions. Let us assume a coupling between active and sterile neutrinos~\cite{Fiorillo:2020jvy,Fiorillo:2020zzj}
\begin{equation}\label{eq:secretSI}
\mathcal{L}_{\mathrm{SI}}=\sum_{\alpha} \lambda_{\alpha} \bar{\nu}_{\alpha} \gamma_{5} \nu_{s} \varphi,
\end{equation}
where $\alpha$ runs over the three active flavors, $\varphi$ is a (pseudoscalar) mediator, $\nu_s$ is the sterile neutrino field, and $\lambda_\alpha$ are three couplings. 

A first set of constraints that must be satisfied come from laboratory experiments. The new interaction leads to additional decay channels for the kaon of the form $K\to\nu_s \ell \varphi$ and $K\to \nu_s \ell \bar{\nu}_{\ell'} \nu_s$. As shown in Ref.~\cite{Fiorillo:2020zzj}, the measurements of the kaon decay rate can be used to constrain the couplings $\lambda_\alpha$ to the level of $\lambda\sim 0.1-1$. The constraints do not apply when the masses of the mediator and of the sterile neutrino become of the order of 100~MeV, since the decays become kinematically forbidden. Furthermore, the choice $\lambda_\tau\neq 0,\;\lambda_e=\lambda_\mu=0$ cannot be constrained by this method, since the decays $K\to \nu_s \tau \varphi$, $K\to\nu_s \tau \bar{\nu}_{\ell'} \nu_s$ are always kinematically forbidden. 

A second way to constrain the secret interaction is by looking at dense environments of neutrinos, such as the primordial plasma and  the supernovae. As shown in Ref.~\cite{Fiorillo:2020zzj}, the most important constraints come from cosmology. The presence of an additional sterile relativistic species in equilibrium with the plasma at a temperature of order $T\sim 1$~MeV increases the effective number of relativistic degrees of freedom, jeopardizing BBN. Therefore, the newly added species $\varphi$ and $\nu_s$ must have masses larger than about $10$~MeV.\footnote{Sterile neutrinos at lighter mass scales can still be present if they are not in equilibrium with the plasma at BBN: for example, keV sterile neutrinos are a candidate for dark matter if they are produced via oscillation of active neutrinos before BBN.} These masses are also safe from constraints from supernovae.

\begin{figure}
    \centering
    \includegraphics[width=0.45\textwidth]{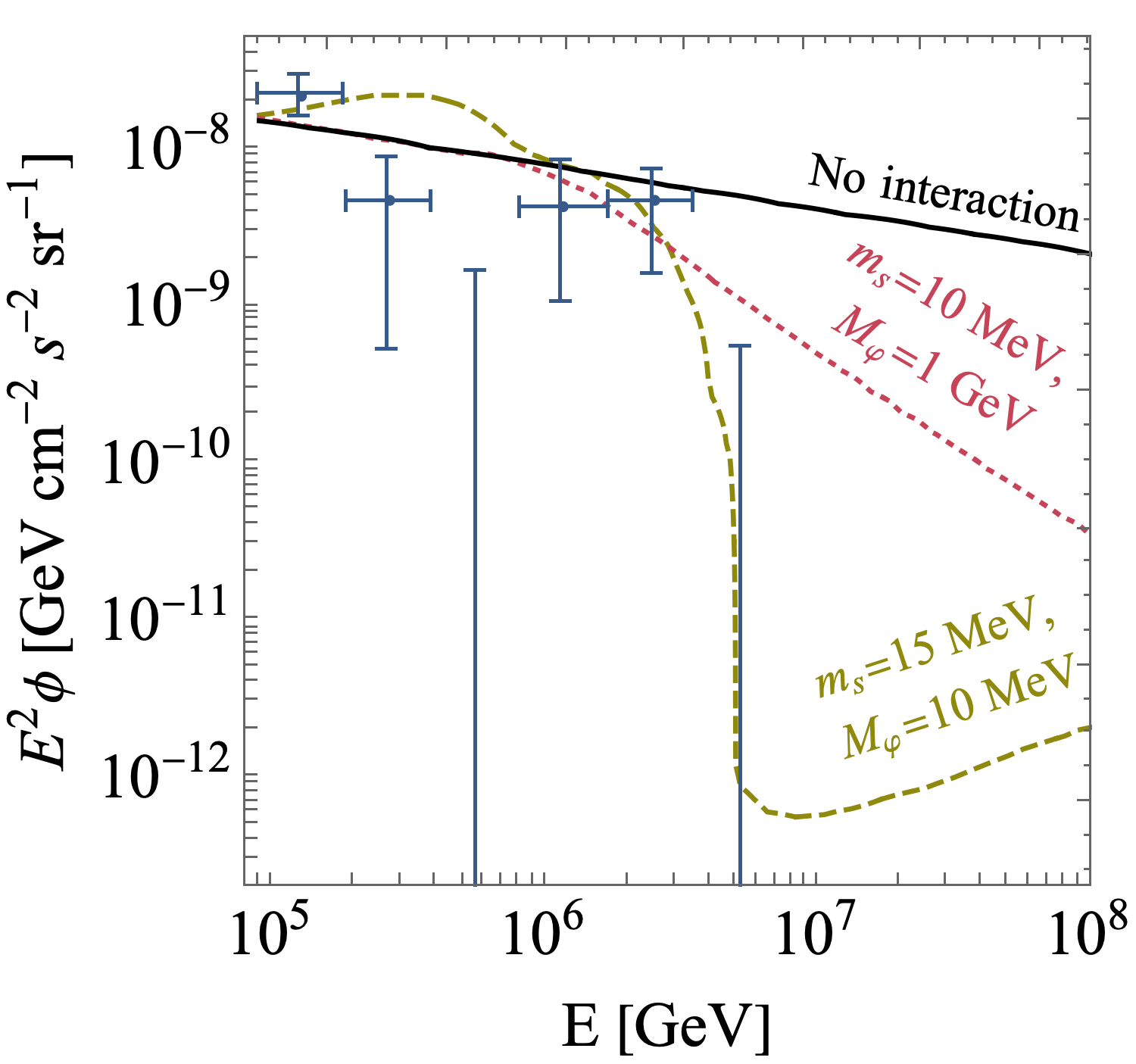}
    \includegraphics[width=0.46\textwidth]{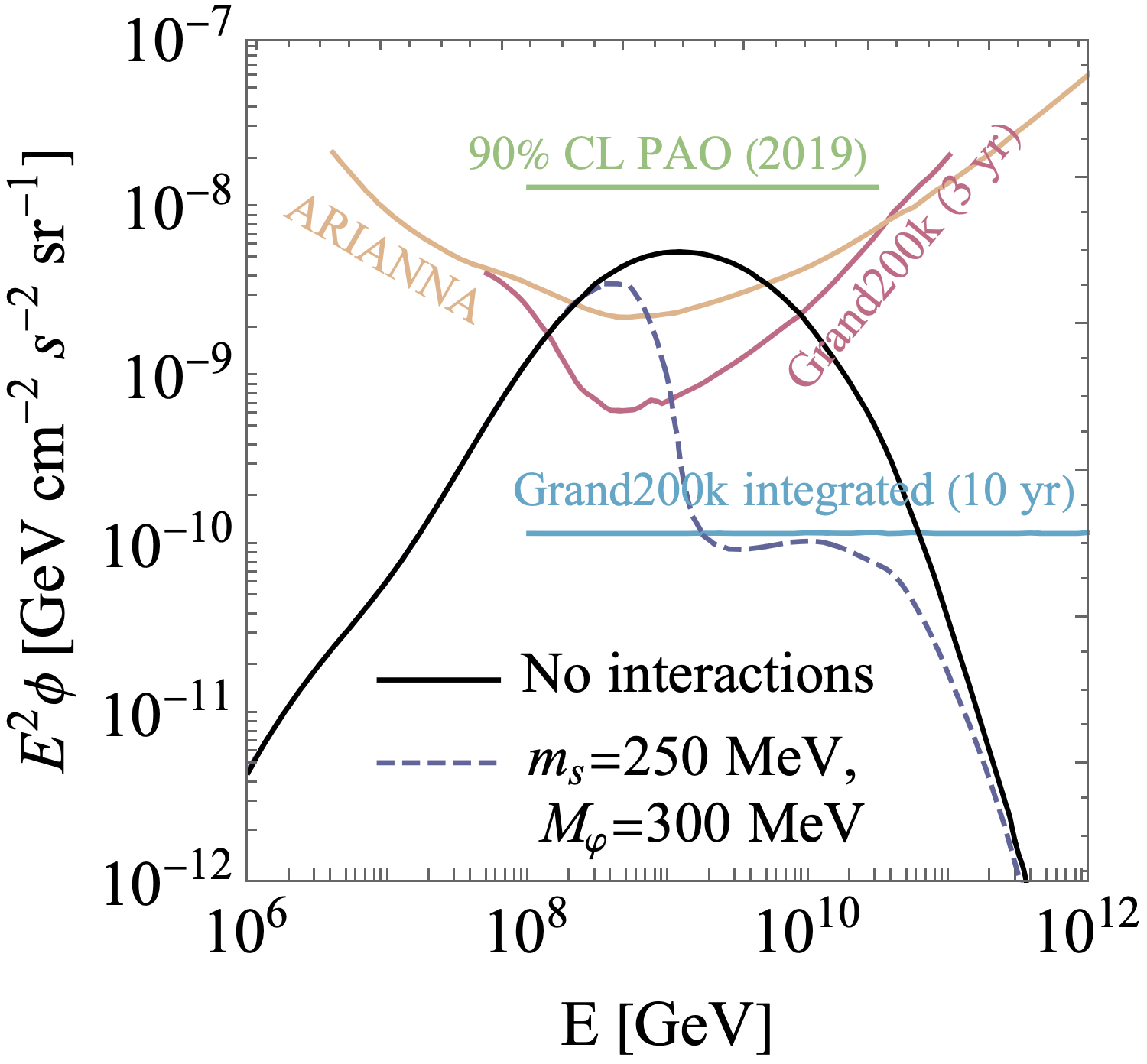}
    \caption{Effects of active-sterile secret interactions on astrophysical neutrino fluxes. (Left panel) Spectral distortion of astrophysical neutrino power-law flux in the IceCube energy range. For the dashed line only the tau neutrino is coupled to the interaction; for the dotted line all three flavors couple with the same strength. The parameters chosen are shown in figure. (Right panel) Spectral distortion of cosmogenic neutrino flux in the ultra-high energy range. The curve is obtained in a simplified 1+1 flavor model. The upper bound of Pierre Auger Observatory, and the sensitivities of ARIANNA and GRAND200k are shown as well. Figures adapted from Refs.~\cite{Fiorillo:2020jvy,Fiorillo:2020zzj}}
    \label{fig:sterilesecret}
\end{figure}

Finally, active-sterile secret interactions can be constrained by observation of high-energy and ultra-high-energy neutrinos. Secret interactions cause the neutrinos, propagating to the Earth, to collide with the C$\nu$B via the reaction $\nu_a\nu_a\to\nu_s\nu_s$. The cross-section peaks at an active neutrino energy $E_\nu\sim m_\varphi^2/m_a\sim 10^8\;\mathrm{GeV} (m_\varphi/100\;\mathrm{MeV})^2$, where $m_a$ is the mass scale of the active neutrinos, assumed to be $0.1\;\mathrm{eV}$. Sterile neutrinos are produced collinearly with the active neutrinos, and in their subsequent propagation can regenerate the high-energy active neutrino flux via the reaction with the C$\nu$B $\nu_s\nu_a\to\nu_a\nu_s$. Furthermore, if $m_s>m_\varphi$, sterile neutrinos can also directly regenerate active neutrinos via the decay $\nu_s\to\nu_a\varphi$. The non-standard propagation causes mainly two effects: a depletion of the flux at an energy $E_\nu\sim 10^8\;\mathrm{GeV} (m_\varphi/100\;\mathrm{MeV})^2$, and a small pile-up of neutrinos at lower energies, caused by the regeneration process. Therefore, the astrophysical neutrino spectrum is distorted by the interaction. The energy scale at which these effects are expected depends on the mass of the mediator. For $m_\varphi\simeq 10$~MeV, the distortion of the spectrum happens around $1$~PeV, in an energy region which can already be probed by IceCube. In order to give rise to significantly large effects, couplings of order 1 are necessary. Such couplings are allowed by meson decay bounds only if the secret interaction couples the tau neutrino only, as mentioned above. This effect is shown for benchmark values of the parameters in the left panel of Fig.~\ref{fig:sterilesecret}. For larger masses of the mediator, the spectrum distortion naturally shifts at ultra-high energies, in the EeV region. A guaranteed population in this region are cosmogenic neutrinos, produced by cosmic-ray $p\gamma$ collisions with the Cosmic Microwave Background. We show the impact of secret interactions on the cosmogenic flux, assuming cosmic-rays are purely made of protons, in the right panel of Fig.~\ref{fig:sterilesecret}, showing that it is in principle detectable by proposed ultra-high energy neutrino radio telescopes such as GRAND~\cite{GRAND:2018iaj}.

\section{Laboratory Probes}\label{sec:lab}
In Section~\ref{sec:theory}, we discussed how new neutrino self-interactions require the existence of a new mediator. Depending on the mass of such a particle, it may be produced and potentially studied in a variety of laboratory environments. This section details several ways to search for the physical effects of neutrinophilic mediators in the laboratory and summarizes the current experimental landscape of these searches, as well as some future prospects.

We discuss the following searches: production of neutrinophilic mediators in double-beta decay processes (Section~\ref{subsec:lab:doublebeta}); new, rare meson/charged lepton/Higgs boson decay processes (Section~\ref{subsec:lab:raredecay}); scattering-related production of the new mediator in neutrino experiments (Section~\ref{subsec:lab:scattering}); and production of mediators in collider environments such as the LHC (Section~\ref{subsec:lab:collider}).

\subsection{Double-Beta Decays}\label{subsec:lab:doublebeta}
For realizations of the neutrino self-interaction like in Eq.~\eqref{eqn:lambda} in which the mediator $\phi$ carries lepton number, new effects may be present in searches for neutrinoless double-beta decay. In these experiments a nucleus spontaneously undergoes the decay process, converting two neutrons into protons simultaneously, and emitting two electrons. If neutrinos are Majorana fermions, then this process may occur without any outgoing neutrinos; the final-state electrons therefore have a specific total energy that is distinct from the broad energy distribution in neutrino-full double-beta decay.

If the interaction of Eq.~\eqref{eqn:lambda} exists, and the mass of $\phi$ is below the $Q$-value of the nuclear double-beta transition, then $\phi$ may be emitted in this decay in lieu of the two neutrinos. In this case, the signal electrons will have a distribution instead of a well-defined value; however, the spectrum is distinct from that in standard double-beta decay. This effect has been studied in detail in Refs.~\cite{Burgess:1992dt, Burgess:1993xh, Gando:2012pj, Agostini:2015nwa, Blum:2018ljv, Cepedello:2018zvr, Brune:2018sab}, and provides constraints\footnote{Because double-beta decay searches always involve outgoing electrons, this type of search for new mediators can only constrain the coupling of $\phi$ to electron flavor, $\lambda_{ee}$.} on the order of $\lambda_{ee} \lesssim 10^{-5}$ for mediators lighter than ${\sim}$MeV.

Even if the new neutrino self-interaction mediator does not carry lepton number, and if the new mediator is heavier than the $Q$-value of the double-beta decay transition in question, new-physics effects can still be manifest in these searches. Ref.~\cite{Deppisch:2020sqh} studied this scenario in detail, finding that constraints from double-beta decay searches are competitive with searches for self-interactions in cosmology.

\subsection{Rare Decay Processes}\label{subsec:lab:raredecay}
Charged pions, kaons, and $D$ mesons have significant leptonic branching ratios. The two-body decays into a single charged lepton and a neutrino are particularly important because their rates are chirality-suppressed. In models with neutrino self-interactions, the final-state neutrinos can radiate the \nusi mediator if kinematically allowed. This opens up the possibility of three-body decays of the meson $m$, e.g., $m^- \to \ell_\alpha^- \nu_\beta \phi$; while these decays suffer from the three-body phase-space suppression, they are not chirality-suppressed and therefore can be important. The decay rate for these new channels should not exceed limits on the processes including a single (detectable) charged lepton in the final state, e.g., $m^- \to \ell^-_\alpha \overline{\nu}_\alpha \nu \overline{\nu}$. Constraints on the couplings of an $L$-charged scalar were derived in this way in Ref.~\cite{Berryman:2018ogk}; these limits are shown in gray ($g_{ee}$), blue ($g_{\mu\mu}$) and pink ($g_{\tau\tau}$) in Fig.~\ref{fig:nusi_constraints}. Similar limits were previously reported for fixed $m_\phi = 1$ keV in Ref.~\cite{Lessa:2007up}.

We should note that limits from meson decays also exist in the scenarios where the boson is heavier than the decaying meson. This means that, in the relevant semi-leptonic meson-decay Feynman diagram, two further neutrino lines need to be attached to the boson line in order to have the process kinematically open. Then, one has four-body decay into a charged lepton and three neutrinos. While such rate is coupling- and phase-space-suppressed with respect to the above three-body decay realization, it appears relevant for, e.g., neutrinophilic dark matter searches \cite{Kelly:2019wow}. 

This document has largely focused on $\nu$SI induced by new scalars, but we briefly remark on the possible existence of new vector interactions. These generate (lepton-number-conserving) meson decays such as $m^- \to \ell^-_\alpha \overline{\nu}_\beta V.$ These decays have been studied in, e.g., Refs.~\cite{Laha:2013xua, Bakhti:2017jhm}; we also note the studies of meson decays in the context of gauged $L_\mu-L_\tau$ in Refs.~\cite{Ibe:2016dir, Dror:2020fbh}. These constraints, while nontrivial, are generally not competitive with constraints from other sources; see, e.g., Ref.~\cite{Ilten:2018crw,Bauer:2018onh}.

Additional \nusi will also generate new contributions to the decays of the Higgs and $Z$ bosons. In Ref.~\cite{Berryman:2018ogk}, the decays $H \to \nu_\alpha \nu_\beta \phi$ and $Z \to \nu_\alpha \nu_\beta \phi$ were used to derive constraints $|\lambda_{\alpha\beta}| \lesssim 0.7$ and $|\lambda_{\alpha\beta}| < 0.5(1+\delta_{\alpha\beta})$, respectively, for a sub-GeV $L$-charged scalar using measurements of invisible decay widths. Ref.~\cite{Brdar:2020nbj} refined the limit for $Z \to$ invisible by including rescattering contributions to $Z \to \nu_\alpha \overline{\nu}_\beta$ enabled by the new scalar. The constraint from $H\to$ invisible is dominant for a GeV-scale new scalar; below this, both $H$ and $Z$ decays are subdominant to meson decays. Ref.~\cite{Brdar:2020nbj} also derived a constraint from four-body $\tau$ decays, i.e., $\tau^- \to \ell^-_\alpha \overline{\nu}_\beta \overline{\nu}_\gamma \phi$, though this turns out to be subdominant to meson decays, as well.

Accelerator neutrino experiments rely on meson decays to produce beams of neutrinos and antineutrinos. If present, then the meson decays discussed in this section will manifest as nonstandard components of the beam. To wit, these neutrinos will be less energetic than their two-body-decay counterparts; in the case of the $L$-charged scalar, this can lead to nonstandard flavor structure \cite{Jones:2019tow} and the appearance of antineutrinos in a neutrino beam (and vice versa). In a real-world environment, however, it would be difficult to identify these contributions: pure beams of (anti)neutrinos cannot be produced, so the nonstandard component would be overwhelmed by this background. In the next subsection, however, we discuss how the same interactions can appear in the detector at such an experiment, and the additional handles that can be used to identify such a process. We conclude by noting that the new force mediator may decay back into neutrinos or, in the case of vector mediators, charged leptons within the detector volume at such an experiment, which can lead to an observable signal \cite{Bakhti:2018avv, Berryman:2019dme}.

\subsection{Neutrino Scattering}\label{subsec:lab:scattering}
\begin{figure}[h]
    \vspace{0.5cm}
    \centering\includegraphics[width=0.618\linewidth]{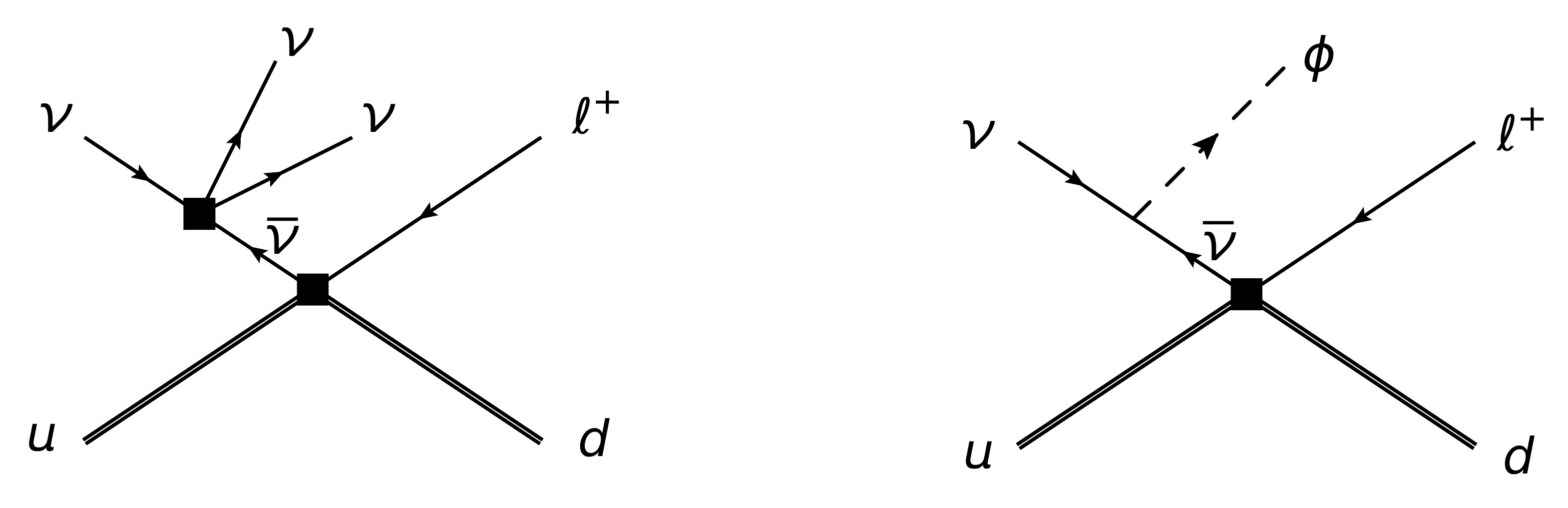}
    \caption{Feynman diagrams of charged-current process with initial-state neutrino radiation via contact operator (left) or light mediator (right). Time goes from left to right. Arrows indicate flow of lepton number.}\label{fig:mono-nu-diagram}
\end{figure}

Neutrino self-interactions, if sufficiently strong, could manifest in neutrino scattering processes. Let us consider two scenarios where \nusi occurs through an effective operator $(\bar\nu\nu)(\bar\nu\nu)$ or via the exchange of a light mediator $\phi$. One could then imagine the radiation of two neutrinos or a $\phi$ in any scattering processes triggered by a neutrino. It is most useful to consider this occurring in association with a charged-current neutrino scattering process where all the final-state particles are visible except for the radiated two neutrinos or $\phi$. The processes are depicted by the Feynman diagrams shown in Fig.~\ref{fig:mono-nu-diagram}.

This class of processes was first considered in Ref.~\cite{Bardin:1970wq} in the context of the Standard Model and then in models containing a light Majoron~\cite{Barger:1981vd}. Two classes of signatures to probe this model were identified:
\begin{description}
    \item [Wrong-Sign Leptons] Already in Refs.~\cite{Bardin:1970wq, Barger:1981vd} it has been noticed that the final-state charged lepton can be of ``wrong sign'', i.e., carry the opposite lepton number to that of the initial-state neutrino. It can serve as a striking signal in experiments with a relatively pure neutrino beam and a magnetized detector with charge identification capability. Indeed, Ref.~\cite{Berryman:2018ogk} considered a published search for CPT violation performed by the MINOS experiment~\cite{MINOS:2012ozn} and reinterpreted its results to constrain the $\phi\nu\nu$ coupling as a function of $\phi$ mass.
    \item [Mono-Neutrino] If a neutrino detector cannot identify charge, one has to resort to the other handle --- the missing transverse momentum with respect to the incoming neutrino direction. One can therefore search for neutrino charged-current interaction final states along with missing transverse momentum, which was dubbed the ``mono-neutrino'' signature in Refs.~\cite{Kelly:2019wow, Kelly:2021mcd}. 
\end{description} 

\begin{figure}[t]
    \centering\includegraphics[width=0.80\linewidth]{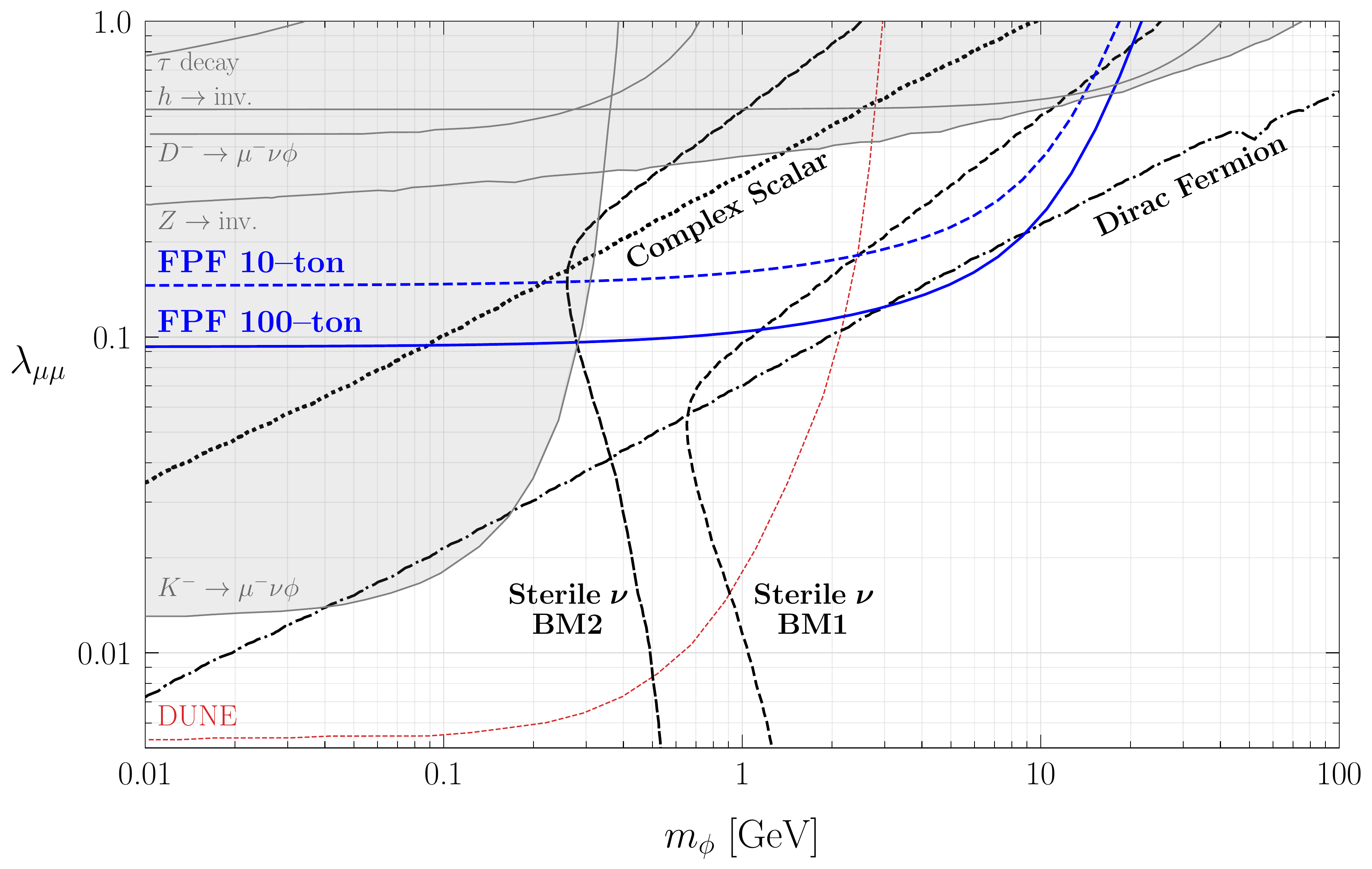}
    \caption{Parameter space of neutrinophilic scalar spanned by its coupling to muon neutrinos $\lambda_{\mu\mu}$ and its mass $m_\phi$. The region excluded by previous measurements is shown in gray-shaded, while the projected reach of DUNE and the FPF are shown as red and blue colored lines, respectively. Various DM target lines are shown in black. See text and Refs.~\cite{Kelly:2019wow, Kelly:2021mcd} for details.} \label{fig:mono-nu-reach}
\end{figure}

For concreteness, let us consider the benchmark model of a massive scalar $\phi$ with couplings to muon neutrinos as described by the low-energy Lagrangian $\mathcal{L} \sim \frac{1}{2}\lambda_{\mu\mu} \nu_\mu \nu_\mu \phi$. The corresponding parameter space is shown in  Fig.~\ref{fig:mono-nu-reach}. The gray-shaded regions have been excluded by previous measurements including laboratory-based constraints detailed in Section~\ref{subsec:lab:raredecay} --- from precision measurements of $\tau$-lepton decays as derived in Ref.~\cite{Brdar:2020nbj}; constraints from precision measurements of $D$-meson and kaon decays as derived in Ref.~\cite{Pasquini:2015fjv, Berryman:2018ogk}; and constraints from the measurements of the $Z$-boson and Higgs invisible decay width at high-energy colliders as derived in Ref.~\cite{Berryman:2018ogk, Brdar:2020nbj}. In addition, we also show the projected sensitivity from future experiments as colored lines:
\begin{description}
    \item [DUNE:] The Deep Underground Neutrino Experiment, or DUNE, is a long-baseline neutrino facility that is currently under construction in the US. Here an intense beam of neutrinos with energies of a few GeV is created at Fermilab and send to a far detector located roughly 1300~km, with the goal of measuring the neutrino oscillation parameters with a high precision. In addition, DUNE contains a liquid argon near detector, which will observe a large number of neutrino interactions. As proposed in Refs.~\cite{Berryman:2018ogk,Kelly:2019wow}, this setup can be used to probe neutrinophilic scalars with masses of up to a few GeV. Since the DUNE near detector cannot identify charge of the outgoing lepton, the proposed analysis strategy focuses on the identification of the ``mono-neutrino'' signature. The projected reach is presented as a red dashed line in Fig.~\ref{fig:mono-nu-reach}. 
    \item [FPF at the LHC:] As the collider experiment with the highest beam energy, the LHC is also the source of the highest-energy neutrinos produced in a laboratory environment. Indeed, at its interactions points, the LHC produces an intense and strongly collimated beam of up to multi-TeV-energy neutrinos in the forward direction. Two dedicated LHC neutrino experiments, FASER$\nu$~\cite{FASER:2019dxq, FASER:2020gpr} and SND@LHC~\cite{Ahdida:2020evc, Ahdida:2750060}, will start their operation in 2022 and observe thousands of high-energy neutrino interactions within a few years. To further increase the neutrino event rate, larger LHC neutrino experiments for the high-luminosity LHC era have been proposed as part of the Forward Physics Facility (or FPF)~\cite{Anchordoqui:2021ghd}. The proposed detectors include the emulsion-based neutrino detector FASER$\nu$, the electronic neutrino detector AdvSND, and the liquid-argon neutrino detector FLArE~\cite{Batell:2021blf}. These detectors would be able to detect about a million interactions of TeV neutrinos, providing an opportunity to produce and probe neutrinophilic scalars with larger masses. Although FPF neutrino experiments will have charge identification capabilities, the sensitivity of the ``wrong-sign lepton'' signature is diluted due to a similar number of neutrinos and anti-neutrinos in the LHC neutrino beam. Therefore, at the FPF the ``mono-neutrino'' signature seems more promising. The corresponding sensitivity reaches of this analysis strategy have been obtained in Ref.~\cite{Kelly:2021mcd} and are shown as blue lines in Fig.~\ref{fig:mono-nu-reach}. 
\end{description}
It is worth noting that the corresponding reaches in the parameter space are sensitive to the hadronic energy resolution of the upcoming neutrino detectors. Precision measurements of hadronic energies in this neutrino scattering allow for further separation of the signal (with large missing transverse momentum) and the SM background (with zero or very small missing transverse momentum)~\cite{Kelly:2021mcd}.

In the same figure, we also show the dark matter targets as black curves of various styles. This includes the sterile neutrino dark matter production via neutrino self-interaction mediated by $\phi$~\cite{DeGouvea:2019wpf,Kelly:2020aks,Kelly:2020pcy} and the thermal freeze out fermionic and scalar dark matter candidates that couples to $\phi$~\cite{Kelly:2019wow}, as discussed in Section~\ref{sec:theory:DM}. The benchmark values for the mass and coupling parameters are given in Ref.~\cite{Kelly:2021mcd}. As we can see, future measurements with DUNE and FPF experiments will be able to probe a large fraction of well-motivated parameter space of such a neutrinophilic mediator. 

\subsection{Collider Probes}\label{subsec:lab:collider}
The leptonic scalar mediators for neutrino self-interaction discussed in Section~\ref{sec:theory} can be effectively probed at high-energy colliders using their direct production and subsequent decay into neutrinos, thus giving rise to characteristic missing transverse energy signatures~\cite{deGouvea:2019qaz, Dev:2021axj}. The LHC and future colliders could potentially extend the reach to higher masses, complementary to the low-energy laboratory probes and the astrophysical and cosmological observations discussed above.  

\begin{figure}[t!]
  \centering
  \includegraphics[height=0.35\textwidth]{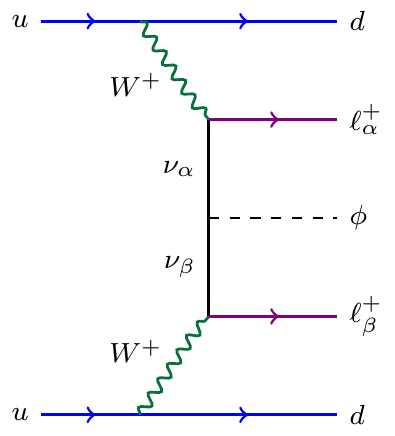}
  \caption{Representative Feynman diagram for the production of leptonic scalar $\phi$ at the LHC.}
  \label{fig:VBFdiagram}
\end{figure}

Given the interaction in Eq.~\eqref{eqn:lambda}, the leptonic scalar $\phi$ can be produced at colliders by radiation off a neutrino leg. In particular, it can be produced in vector-boson fusion (VBF) processes (see Fig.~\ref{fig:VBFdiagram}), leading to the characteristic signal of same-sign dileptons, two forward jets in opposite hemispheres, and missing transverse energy, i.e., 
\begin{equation}
    pp \to \ell_\alpha^\pm \ell_\beta^\pm jj + E_T^{\rm miss} \quad (\alpha,\, \beta = e,\, \mu, \tau) \, .
    \label{eqn:VBF}
\end{equation}
Although this process by itself may not uniquely distinguish any specific UV-complete model, such as those discussed in Section~\ref{sec:theory}, it can be used in conjunction with additional signals arising in specific UV-completions (see below) to probe the leptonic scalar at the LHC.

\begin{figure}[t!]
  \centering
     \includegraphics[height=0.5\textwidth]{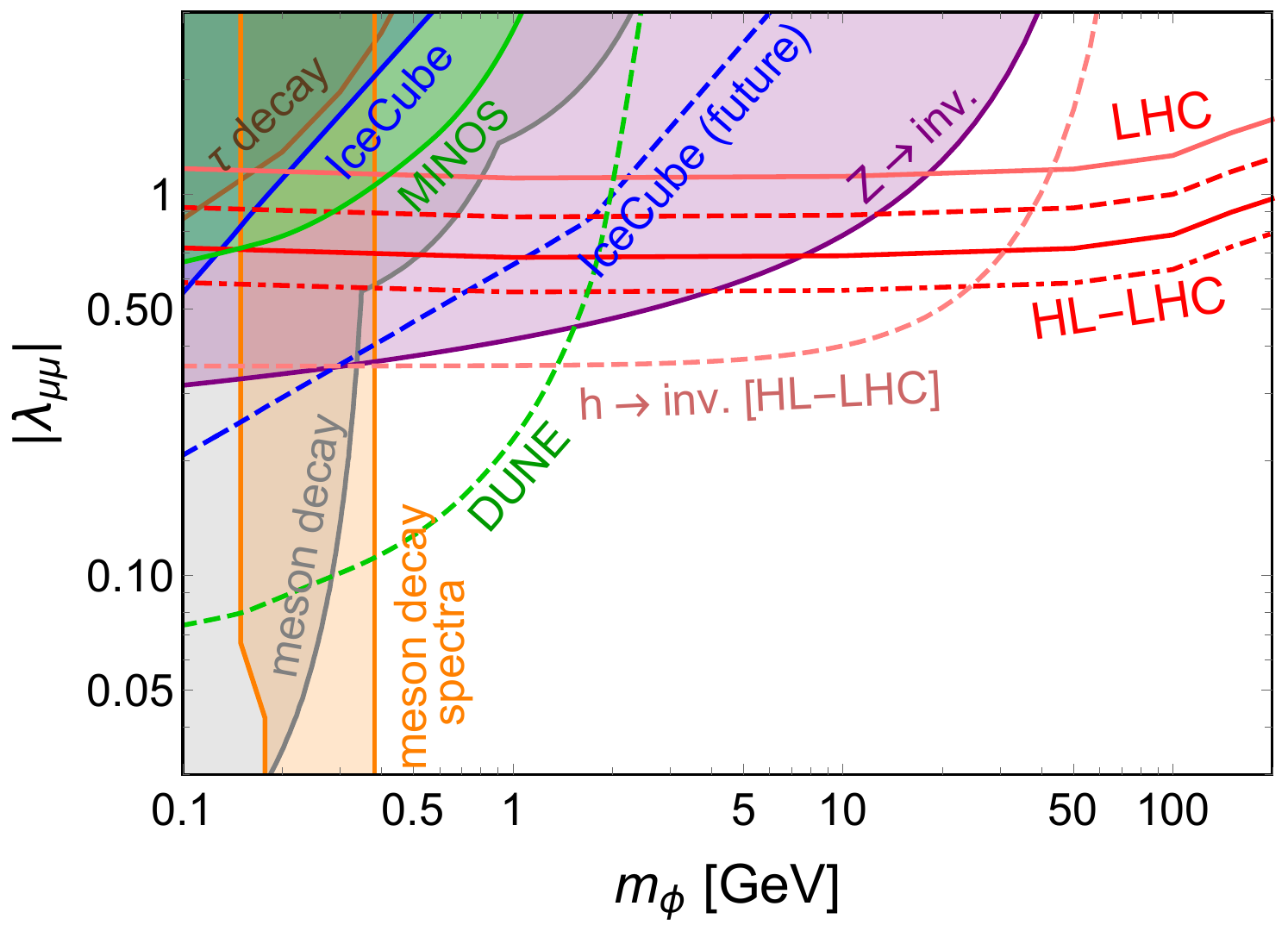}
  \caption{Projected sensitivity for the coupling $|\lambda_{\mu\mu}|$ as a function of the leptonic scalar mass $m_\phi$ at 14 TeV LHC with luminosity of 300 fb$^{-1}$ (solid thin red line) and HL-LHC with 3 ab$^{-1}$ and with systematic errors of 0\% (dot-dashed thick red line),  10\% (solid thick red line) and 20\% (dashed thick red line). Also shown are the low-energy limits from meson decay (gray), $\tau$ decay (brown), heavy neutrino searches in meson decay spectra (orange), invisible $Z$ decay (purple) and the prospect of invisible SM Higgs decay at HL-LHC (dashed pink), the current IceCube limits on neutrino--neutrino interactions (blue) and future prospect (dashed blue), as well as the MINOS limit (green) and future prospect at DUNE (dashed green). All the shaded regions are excluded. From Ref.~\cite{deGouvea:2019qaz}.}
  \label{fig:LHC}
\end{figure}

The 95\% C.L. LHC and HL-LHC sensitivities to the coupling $|\lambda_{\mu\mu}|$ using the process in Eq.~\eqref{eqn:VBF} are shown in Fig.~\ref{fig:LHC}. The corresponding limits for $|\lambda_{ee}|$ and $|\lambda_{e\mu}|$ can be found in Ref.~\cite{deGouvea:2019qaz}. The dot-dashed thick red line is for the most optimistic case at the 14 TeV HL-LHC with 3 ab$^{-1}$ integrated luminosity and without any systematic error. With a realistic 10\% (20\%) systematic error, the sensitivities at the HL-LHC are slightly weaker, denoted by the solid (dashed) thick red lines.
This implies that our leptonic scalar signals are rather robust against the systematic uncertainties on the background determination.
For comparison,  we also show the prospects at the 14 TeV LHC  with only 300 fb$^{-1}$ integrated luminosity, which is achievable in the upcoming run within a few years. Since the difference between the LHC prospects with 0\%, 10\% and 20\% systematic uncertainties is not appreciable, we show only the prospects with 10\% systematic error as the thin red line in Fig.~\ref{fig:LHC}. 

The prospects of $\lambda_{\alpha\beta}$ at the LHC and HL-LHC are largely complementary to the low-energy constraints discussed above. To see it more clearly, we show in Fig.~\ref{fig:LHC} the limits from meson decays (gray), $\tau$ decays (brown), heavy neutrino searches in two-body meson decay spectra (orange), the invisible $Z$ decay (purple), neutrino-matter scattering at MINOS (green), 
and IceCube limits on new neutrino--neutrino interactions (blue). All the shaded regions are excluded. Also shown are the prospects of invisible SM Higgs decay at HL-LHC by the dashed pink lines and the prospects at IceCube-Gen2 by the dashed blue lines. We have also shown the prospect from DUNE by dashed green line. One can see that the HL-LHC prospects exceed all the existing limits when the scalar mass $m_\phi \gtrsim 10$ GeV. At a future 100-TeV $pp$ collider, further improvement in the sensitivity is expected. We would like to emphasize that this is a direct probe of scalar-mediated neutrino self-interactions at high-energy colliders.

There can be additional collider signals in specific UV-complete models, such as those discussed in Section~\ref{sec:theory}. For example, in the type-II seesaw-motivated model of Ref.~\cite{Dev:2021axj}, the leptonic scalar can be produced either from the doubly-charged scalar $H^{\pm\pm} \to W^\pm W^\pm \phi$ or from the singly-charged scalar $H^\pm \to W^\pm \phi$. As the leptonic scalar $\phi$ decays exclusively into neutrinos, these new channels will lead to same-sign dilepton plus  missing transverse energy plus jets signal at the hadron colliders, which is different from the standard type-II seesaw.  We find that the mass of doubly-charged scalars in the small and large Yukawa coupling scenarios can be probed up to respectively 800 GeV and 1.1 TeV at $2\sigma$ significance, corresponding to a $95\%$ confidence level, in the new channels at the HL-LHC with integrated luminosity of 3 ab$^{-1}$, and can be improved up to 3.8 TeV and 4 TeV respectively at future 100 TeV colliders with luminosity of 30 ab$^{-1}$.
We also find that since in the large Yukawa coupling case, the missing energy is completely from the leptonic scalar, its mass can be determined with an accuracy of about $10\%$ at the HL-LHC.

\section{Outlook \& Conclusions}\label{sec:conclusions}
This white paper provides an overview of the current status and future prospects of exploring beyond-the-Standard-Model interactions among  neutrinos. As we have demonstrated, these new-physics scenarios have a range of motivations and have the possibility of addressing a number of important questions in particle physics and cosmology. This includes the unknown generation of neutrino masses, explanations of the origin of dark matter in the Universe, and relieving tensions in different measurements of cosmological parameters.

Depending on whether or not the new interactions couple predominantly to neutrinos (and therefore are neutrinophilic), the prospects for searches for these new interactions, and their associated mediators, can be very difficult (compared against searches, for instance, for dark photons or dark Higgs bosons). We have discussed a wide range of mediator masses in this white paper, ranging from below the keV scale to hundreds of GeV. This parameter space is ripe for exploration and for novel ideas in how to test these interactions.

In order to thoroughly test this parameter space, probes of all types are necessary. They include (i) cosmological probes using the cosmic microwave background, large- and small-scale structures, Big-Bang nucleosynthesis, and even imprints from inflation; (ii) astrophysical probes using high-energy cosmic neutrinos and supernovae; and (iii) laboratory probes using double-beta decay, meson rare decays, accelerator neutrino experiments, as well as high-energy colliders like the LHC. In this white paper, we have summarized the current status of these three categories of searches, and how they provide complementary information when searching for neutrino self-interactions. 

Additionally, many upcoming experimental endeavors in the short- and long-term will provide excellent capabilities for searching for new neutrino self-interactions. We have highlighted a number of these prospects that can and will guide us through the next generation of experiments. We conclude that this is a very exciting time for the prospects of discovering neutrino self-interactions, and our understanding of this phenomenon will soon be improved substantially.

\section*{Acknowledgements}
NB was supported in part by NSERC, Canada.
The work of F.K. is supported by the Deutsche Forschungsgemeinschaft under Germany’s Excellence Strategy - EXC 2121 Quantum Universe - 390833306.
W.T. is supported by National Science Foundation - Grant No. PHY-2013052.
The work of P.S.B.D. is supported in part by the U.S. Department  of  Energy  under  Grant  No.  DE-SC0017987.
The work of B.D. is supported in part by the U.S.~Department of Energy under grant No. DE-SC0010813.
The work of Z.L. is supported in part by the U.S.~Department of Energy under grant No. DE-SC0022345. 
The work of Yongchao Zhang is supported by the National Natural Science Foundation of China under Grant No.\  12175039, the 2021 Jiangsu Shuangchuang (Mass Innovation and Entrepreneurship) Talent Program No.\ JSSCBS20210144, and the ``Fundamental Research Funds for the Central Universities''. AD was supported by the U.S. Department of Energy under contract number DE-AC02-76SF00515.  The work of MB is supported by the {\sc Villum Fonden} under project no.~29388.
Y.Z. is supported by the Arthur B. McDonald Canadian Astroparticle Physics Research Institute.


\providecommand{\href}[2]{#2}\begingroup\raggedright\endgroup

\end{document}